\documentclass{pazha2}
\usepackage{graphicx}
\usepackage[figuresright]{rotating}
\usepackage{multirow}
\usepackage{color}
\definecolor{darkblue}{rgb}{0,0,0.9}

\def\*{$^{*}$}
\def\aa{$^{\mbox{\small a}}$}
\def\bb{$^{\mbox{\small b}}$}
\def\cc{$^{\mbox{\small c}}$}
\def\dd{$^{\mbox{\small d}}$}

\begin{document}
\journalinfo{to be}{2025}{51}{11-12}{655}{629}{660}[691]
\sloppypar

\title{\bf X-ray and Gamma-ray Emission from \protect\\ a High-Temperature
  Plasma and the Spectra of Accreting Black Holes} 
\year=2025
\author{
S. A. Grebenev\email{grebenev@cosmos.ru}\addres{1}
\addrestext{1}{Space Research Institute, Russian Academy of Sciences, Moscow, 117997 Russia}
}

\shortauthor{GREBENEV}
\shorttitle{X-RAY AND GAMMA-RAY EMISSION FROM A HIGH-TEMPERATURE PLASMA} 

\submitted{January 21, 2025}
\revised{April 22, 2025}
\accepted{August 19, 2025}

\begin{abstract}
\noindent
We present the results of our numerical computations of the
broadband radiation spectra forming in a layer of
high-temperature ($kT_{\rm e}\sim 50$ keV) semitransparent (with
a Thomson optical depth $\tau_{\rm T}\sim1-3$) plasma with an
electron density $N_{\rm e}\sim
10^{17}-10^{19}\ \mbox{cm}^{-3}$, typical for the accretion disk
regions surrounding a black hole in X-ray binaries. In our
computations we took into account the bremsstrahlung processes
of the production (and absorption) of photons and their
subsequent Comptonization. The intrinsic radiation from such a
high-temperature plasma is shown to be enough to explain the
X-ray spectra observed in the low (hard) state of Galactic black
hole candidates and X-ray novae. No commonly assumed additional
soft (with energies $h\nu\la 1$ keV) photons to maintain
Comptonization are required; moreover, their presence would lead
to severe distortions of the spectrum compared to the observed
one or would require a very fine tuning of plasma parameters. In
the hard X-ray range the forming power-law radiation spectrum
with a photon index $\alpha\sim1.4-1.7$, which has an
exponential cutoff at energies $h\nu\ga 100$ keV, exceeds
considerably the bremsstrahlung flux that might be expected from
such a plasma layer in the limit of an infinitely small optical
depth. This is a result of the multiple inverse Compton
scattering of bremsstrahlung photons. It is important that,
according to our computations, the power-law radiation spectrum
for such a high-temperature plasma should extend in an
invariable form downward along the energy axis to the
ultraviolet, optical, and infrared ranges ($h\nu\sim 1-3$~eV).
At energies $h\nu\la 1$ eV the optical depth for bremsstrahlung
absorption grows rapidly and the radiation spectrum becomes the
Rayleigh-Jeans one. To explain the steeper $\alpha\sim2.1-2.5$
X-ray spectra observed from accreting black holes in their high
(soft or two-component) state, it is indeed necessary that a
large number of soft photons additional to the intrinsic plasma
bremsstrahlung photons enter a hot cloud. Such photons could be
emitted by the surface of an outer dense and cold accretion disk
whose inner edge during these states characterized by a powerful
soft component in the X-ray spectrum approaches the black hole
maximally closely. The optical and infrared emission from
systems in these states is associated precisely with the
emission from the outer disk, whereas during their low state it
can be produced directly in the hot central disk region bloated
by instabilities. Under favorable conditions (disk size and
inclination) the low-frequency emission from this region can
noticeably exceed in flux and luminosity the emission from the
outer cold accretion disk regions.\\

\noindent
{\bf DOI:} 10.1134/S106377372670009X

\keywords{\sl Compton scattering, Comptonization, recoil effect,
  Doppler effect, bremsstrahlung emission and absorption, disk
  accretion, black holes, X-ray sources.}
\end{abstract}

\section{INTRODUCTION}
\noindent
The hard X-ray emission from accreting black holes in binary
systems (quasi-stationary sources like Cygnus~\mbox{X-1} or
transient ones --- X-ray novae) is believed to be produced
through the multiple inverse Compton scattering (Comptonization)
of soft X-ray and ultraviolet photons by high-temperature
($kT_{\rm e}\sim 25$--50 keV) electrons in the central accretion
disk region. Calculations (Eardley et al. 1975, 1978; Shapiro et
al. 1976; Sunyaev and Truemper 1979; Sunyaev and Titarchuk 1980;
Pozdnyakov et al. 1982; Grebenev et al. 1993; Zdziarski et
al. 1996, 2020) show that Comptonization in a semitransparent
($\tau_{\rm T}\sim1$) rarefied plasma bloated by instabilities
can indeed give rise to a characteristic power-law X-ray
spectrum with an exponential cutoff at high, $h\nu\ga kT_{\rm
  e}$ energies observed from black holes during their hard state 
(Sunyaev and Truemper 1979; Sunyaev et al. 1988, 1991; Grebenev
et al. 1993; Grove et al. 1998).

The low-frequency photons required for Comptonization can be
supplied by the outer cold ($kT_{\rm e}\la 1$ keV) disk
region. It is there that the soft blackbody component (Shakura
and Sunyaev 1973) dominating in the X-ray spectrum of accreting
black holes during their soft and two-component states is formed
(Makishima et al. 1986; Grebenev et al. 1991, 1197a; Gilfanov et
al. 1993; Tanaka and Shibazaki 1996; Remillard and McClintock
2006; Done et al. 2007; Belloni 2010). At this time, the
boundary of the cold disk region approaches the black hole
maximally closely. However, during the hard state of black holes
of interest to us, when the central hot disk zone expands
greatly, the contribution of the outer disk zone to the overall
radiation spectrum drops (along with its surface
temperature). The solid angle at which the central zone is seen
from the cold disk also decreases; accordingly, the number of
soft photons capable of entering it is reduced (Garcia and
Kallman 2010; Gilfanov 2010). The change in the size of the hot
zone directly manifests itself as a change in the frequency of
the quasi-periodic X-ray flux oscillations recorded in the power
spectrum of black hole candidates (see, e.g., Mereminskiy et
al. 2018).

An alternative model suggests that synchrotron radiation from
high-temperature thermal (Wardzinski and Zdziarski 2000; Dexter
et al. 2021) or hard nonthermal (Poutanen and Vurm 2009;
Veledina et al. 2011; Poutanen and Veledina 2014) electrons, a
certain fraction of which can be present in the hot disk zone,
supplies soft photons for Comptonization. The observations of a
very hard power-law component in the radiation spectrum of
Cyg~X-1, the best-known source containing a black hole, point to
the possible existence of nonthermal electrons (see, e.g.,
McConnell et al. 1994, 2002; Ling et al. 1997; Ibragimov et
al. 2005; Laurent et al. 2011; but, at the same time, Jourdain
et al. 2012). However, even if such a hard component exists, it
can be associated with the relativistic jets observed in many
accreting black holes (Mirabel and Rodriguez 1999; Gallo et
al. 2014; Russell et al. 2014; Fender and Gallo 2014; Kantzas et
al. 2021), i.e., be formed in a place geometrically separated
from the accretion disk. Synchrotron photons inevitably appear
in the advection-dominated accretion flow (ADAF) models as well
(Narayan et al. 1998; Yan and Narayan 2014; Liu et
al. 2025). However, such flows, as a rule, are more transparent
for Thomson scattering and the Comptonization efficiency in them
can be limited.\\ [-2mm]

These models share the belief that: 
\begin{enumerate}
  \item a powerful external source of low-frequency photons with
    energies $\la 1$ keV entering the high-temperature disk
    region is needed for efficient Comptonization and the
    formation of a hard radiation spectrum typical for accreting
    black holes;

 \item the intrinsic thermal radiation from a high-temperature
   plasma is too weak and is unable to explain the observed
   X-ray and gamma-ray emission from accreting black holes.
\end{enumerate}
Accordingly, the escape from a high-temperature plasma cloud of
photons emitted by a built-in or external (cloud-irradiating)
source of soft X-ray emission is commonly considered when
computing the hard X-ray spectra of accreting black holes. The
Monte Carlo method (Pozdnyakov et al. 1983), which considers the
Compton scattering of such photons by taking into account all of
the relativistic corrections important for the scattering in
such a hot ($kT_{\rm e}\ga 50$ keV) medium, is widely used. The
production of intrinsic photons in the plasma through
the bremsstrahlung processes is neglected in such
computations.

In this paper we check the validity of this approach and the
underlying assertions. For a wide set of parameters of the layer
of semitransparent hot plasma we found a numerical solution of
the complete Kompaneets (1957) equation that takes into account 
both Comptonization and bremsstrahlung processes occurring in
the plasma. The computations are based on the computer code
developed by Grebenev and Sunyaev (2002) to investigate the
radiation spectra forming in a layer of accreting matter
spreading over the surface of a neutron star with a weak
magnetic field. The code was adapted to the current problem,
although the computations were performed without applying the
relativistic corrections to the scattering cross section that
are important for such a hot plasma. For this reason, the
estimates of the plasma parameters can slightly differ from the
true ones, but, in general, this simplification does not change
the main conclusions of this paper. The results of completely
relativistic computations will be presented in a separate paper.

In this paper we compute the broadband spectra of the intrinsic
high-temperature plasma radiation in this layer and compare them
with the X-ray spectra observed from a number of widely known
accreting black holes in their hard state. We show that no
additional soft photons are needed for the formation of their
canonical hard spectrum with a photon index
$\alpha\simeq1.6$. Moreover, the penetration of external
low-frequency photons into the plasma layer generally leads to a
softening of its radiation spectrum, i.e., to an increase in the
photon index to $\alpha\simeq2.0$ or higher. The successful
description of the observed radiation spectra for black holes
within the model of Comptonization of external soft photons is
associated with the total neglect of the thermal
(bremsstrahlung) radiation from the plasma in the central hot
accretion disk regions that contributes substantially to its
spectrum.

\section*{FORMATION OF THE RADIATION SPECTRUM}
\noindent
In a hot and fairly rarefied plasma the radiation spectra are
formed under the combined action of bremsstrahlung processes
(emission and absorption) and Compton scattering.
\begin{figure*}[!t]
\centering
\includegraphics[width=0.70\linewidth]{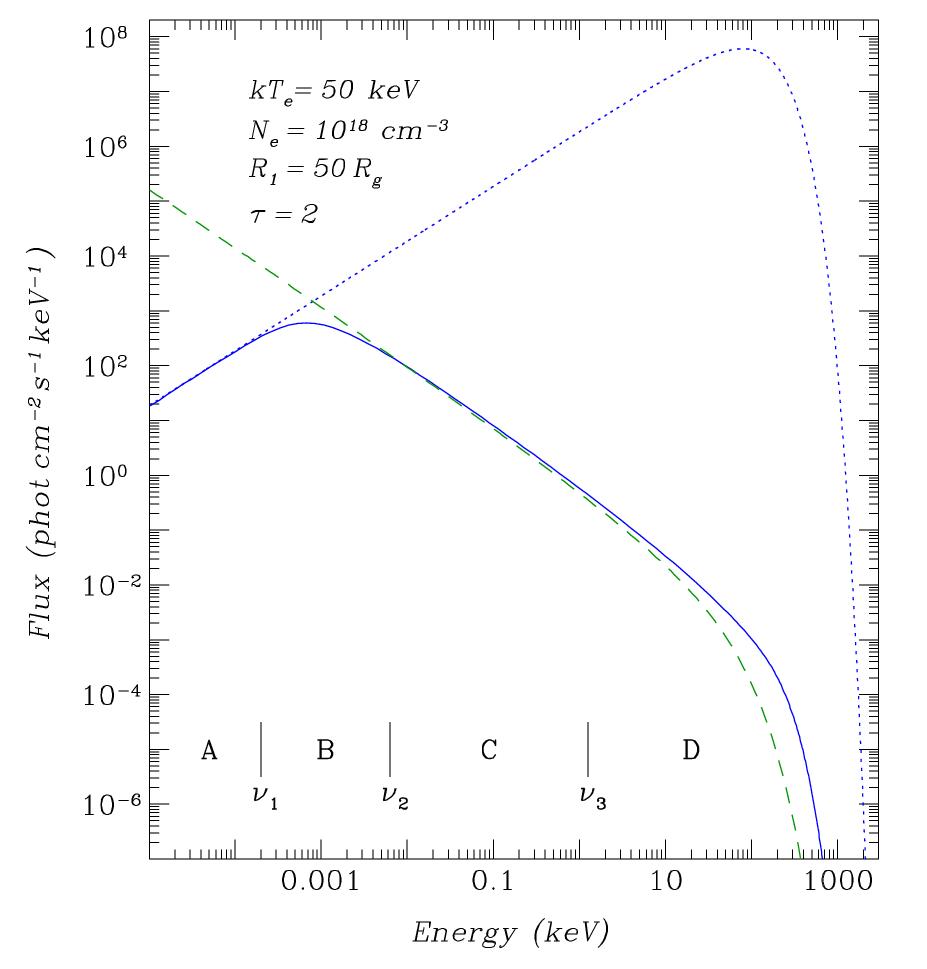}
\caption{\rm Broadband (from optical to X-ray) photon spectrum
  for a layer of high-temperature plasma with the electron
  temperature $kT_{\rm e}=50$ keV, the transverse optical depth
  for Thomson scattering $\tau_{\rm T}=2$, and the electron
  density $N_{\rm e}= 10^{18}\ \mbox{cm}^{-3}$ (solid line). The
  dashed line indicates the bremsstrahlung spectrum for this
  plasma layer; the dotted line indicates the Planck radiation
  spectrum that could be formed in a very dense and optically
  thick plasma layer with the same temperature.\label{fig:ff}}
\end{figure*}

As an example, the solid line in Fig.~\ref{fig:ff} indicates the
photon spectrum of the radiation $F_{\nu}(\nu)$ emerging from a
homogeneous layer of plasma with the same temperature of its
electrons and ions $kT_{\rm e}=50$ keV, the electron density
$n_{\rm e}=10^{18}\ \mbox{cm}^{-3}$, and the transverse optical
depth for Thomson scattering $\tau_{\rm T}=2$. The choice of
such a density is explained in Appendix~I. The plasma is assumed
to be composed of pure hydrogen; it is fully ionized at the
temperatures under consideration. By discussing a homogeneous
plasma layer, it is easier to understand the formation of its
radiation spectrum, although in a real disk the temperature, the
density, and even the optical depth, of course, depend on the
radius. More realistic cases will be considered below.

The normalization of the spectrum suggests that the radiating
surface of the layer is perpendicular to the line of sight (the
inclination is $i=0$\deg) and has an area $S\simeq 7.0\times
10^{16}\ \mbox{cm}^2$ (the area of one side of the hot accretion
disk region); the source itself is at the distance $d=2.5$
kpc. Since $S=\pi (R_1^2-R_0^2),$ where $R_1$ is the outer
radius of the high-temperature disk region and $R_0\ll R_1$ is
its inner radius (the radius of the innermost stable circular
orbit), the value of $S$ being used corresponds to $R_1\simeq
1500$~km. For a Schwarzschild black hole $R_0=3\ R_{\rm g}$, 
where $R_{\rm g}=2GM\,/c^2\simeq 3 (M/M_{\odot})$ km is its
gravitational radius.  Accordingly, for a black hole with the
mass $M=10\ M_\odot$ the radius $R_1=50\ R_{\rm g}$, i.e.,
despite the assumption about the hard state of the source, the
high-temperature region remains fairly compact. Note that the
half-thickness of the hot disk zone under the assumption of a
uniform density distribution is $H=\tau_{\rm T}/(2\,\sigma_{\rm
  T} N_{\rm e})\simeq 15.0$ km. The details of our computations
of the presented spectrum will be described below.

The bremsstrahlung absorption is efficient only at $h\nu\ll
kT_{\rm e}$, since the absorption cross section decreases
rapidly with frequency (as $\nu^{-3}$). At frequencies where it
dominates in the opacity, an intensive production and absorption
of photons occur, providing a thermal equilibrium in the plasma
and giving rise to a Planck (or, more precisely, Rayleigh-Jeans)
radiation spectrum. In Fig.~\ref{fig:ff} the corresponding
region of frequencies (or energies $h\nu$) is denoted by 
the letter A. The dotted line indicates the Planck radiation
spectrum for the same temperature and area of the radiating
surface, while the dashed (green) line indicates the
bremsstrahlung spectrum for a layer of optically thin plasma
with the same parameters. In region A the computed radiation
spectrum coincides with the Planck one and has nothing in common
with the thermal spectrum of an optically thin plasma.

Starting with some frequency $\nu_1,$ the bremsstrahlung
absorption and Thomson scattering cross sections become equal
and then the optical depth for absorption become less than unity
(see Appendix~II). Scattering lengthens the path of photons in
the plasma, increasing the probability of their bremsstrahlung
absorption (Zeldovich and Shakura 1969; Illarionov and Sunyaev
1972; Felten and Rees 1972; Sunyaev and Shakura
1974). Accordingly, in this frequency range (region B in
Fig.~\ref{fig:ff}) the real spectrum of the radiation emerging
from the plasma layer deviates from the Rayleigh-Jeans one and passes
below the bremsstrahlung spectrum of an optically thin plasma.

At even higher frequencies $\nu\ga\nu_2$ (region C) the
bremsstrahlung absorption ceases to play any prominent role even
if the lengthening of the photon path is taken into account, and
the radiation emerging from the plasma layer has a spectrum
almost coincident with the bremsstrahlung spectrum of an
optically thin plasma. Because of their multiple scatterings by
electrons (Comptonization), the photons in this region increase
their frequency due to the Doppler effect (in each scattering,
on average, by $\Delta\nu/\nu\sim kT_{\rm e}/m_{\rm
  e}c^2$). However, this actually manifests itself only in
region D (at frequencies $\nu\ga\nu_3$). A photon flux is formed
in this region upward along the frequency axis into the zone
$h\nu\sim 3 kT_{\rm e},$ where the Doppler effect becomes
inefficient and, besides, the recoil effect $\Delta\nu/\nu\sim
-h\nu/m_{\rm e}c^2$ preventing a further growth of the photon
frequency comes into play. The combined action of these two
effects gives rise a Wien component $F_{\nu}\sim \nu^2
\exp{(-h\nu/kT_{\rm e})}$ in the radiation spectrum of the
plasma layer (Zeldovich and Sunyaev 1969; Illarionov and Sunyaev
1972, 1975). In Fig.~\ref{fig:ff} the Wien component (a
significant excess of hard X-ray photons compared to the spectrum
of an optically thin plasma) appears at energies $h\nu \ga 30$
keV.

The pattern and role of Comptonization in the formation of the
spectrum and, in the long run, the production efficiency of hard
X-ray photons are largely determined by the outward diffusion of
photons. 

\section*{COMPUTATION OF THE RADIATION SPECTRUM}\label{sec:komp}
\noindent
In the plane-parallel case, the radiative transfer equation
corresponding to the problem under consideration can be written
as (Kompaneets 1957; Grebenev and Sunyaev 2002)
\begin{equation}\label{eq:komp}
\frac{1}{3}\frac{\partial}{\partial \tau} \left(
\frac{\alpha_{\rm T}}{\alpha_{\rm T}+\alpha_{\rm ff}}\frac{\partial
U_{\nu}}{\partial \tau}\right)= 
\frac{\alpha_{\rm ff}}{\alpha_{\rm T}}
\left(U_{\nu}-B_{\nu}\right) -
\end{equation}
$$ -\frac{kT_e}{m_{\rm e}\,c^2}\ x\frac{\partial}{ \partial
  x}\left(x\frac{\partial U_{\nu}}{\partial x}
-3U_{\nu}+xU_{\nu}\right).$$ Here and below, we use the notation
$x=h\nu/kT_{\rm e}$. The quantity $U_{\nu}(\tau)$ is the angle-averaged radiation
intensity at an optical depth for Thomson scattering $\tau$ and
$$B_{\nu}(T_{\rm e})= \frac{2 h\nu^3}{c^2}
\left(e^{\,h\nu/kT_{\rm e}}-1\right)^{-1}$$ is the blackbody
radiation intensity corresponding to the plasma temperature. The
term on the left and the first term on the right in
Eq.~(\ref{eq:komp}) describe the spatial diffusion of photons
and their bremsstrahlung production and absorption,
respectively, while the second term on the right describes the
Comptonization process in the nonrelativistic
limit.\footnote{The relativistic corrections to
  Eq.~(\ref{eq:komp}) were discussed and used in the
  calculations of Cooper (1971), Arons (1971), Illarionov and
  Sunyaev (1972), Itoh et al. (1998), Challinor and Lasenby
  (1998), Sazonov and Sunyaev (2000), and Grebenev and Sunyaev
  (2019). All of these were the studies of Compton scattering in
  an infinite medium (the early Universe or the hot gas of galaxy
  clusters). Scattering in a bounded volume requires corrections
  to the transport cross sections like those deduced by Grebenev
  and Sunyaev (1987) for the scattering of hard photons in a
  cold medium, but also with allowance made for the high
  electron velocities.}. We neglected the term that describes
the induced scattering and the production of photons in the
double Compton effect.

The opacities for bremsstrahlung absorption and
Thomson scattering appearing in the equation can
be written with a sufficient accuracy in the limit of a
purely hydrogen ($N_{\rm e}=\rho/m_{\rm p}$) plasma as (Allen 1973;
Lang 1974) 
$$ \alpha_{\rm ff}=
21.2\  N_{\rm e} T_{\rm e}^{-7/2}\ x^{-3} \left(1-e^{-x}\right)
g(x)\ \ \mbox{K}^{7/2} \mbox{cm}^5\ \mbox{g}^{-1}
$$ and $\alpha_{\rm T}\simeq
0.4\ \mbox{cm}^2\ \mbox{g}^{-1}$. The Gaunt factor $g(x)$ is
expressed via the Macdonald function $K_0$ and its asymptotics
\begin{equation}\label{eq:gaunt}
  g(x)=\frac{\sqrt{3}}{\pi}
  e^{x/2}K_0\left(\frac{x}{2}\right) \simeq 0.55\times
\end{equation}
$$
\left \{ 
\begin{array}{l@{\,}l}
\!\!e^{x/2}\ \left[\, \ln(6.1/x)\,
(1+x^2/16)-0.98\,\right],& \mbox{at}\ x < 1.7,\\ [1mm]
\!\!(2/x)^{0.5}\left[1.25-(0.312-0.24/x)/x\right],& \mbox{at}\ x
\geq 1.7.\\ 
\end{array}\right. 
$$ Note that the term $(\alpha_{\rm ff}/\alpha_{\rm T})\,B_{\nu}$
on the right-hand side of Eq.~(\ref{eq:komp}) describes the
plasma emissivity for the bremsstrahlung processes. Accordingly, the flux of
bremsstrahlung photons from the hot disk region in
a direction perpendicular to its plane is 
\begin{equation}\label{eq:spec-brem}
  B_{\rm ff}(x)=B_0\ \frac{g(x)\ e^{-x}}{x}\,
  \ \left(\frac{S}{d^2}\right).
\end{equation}
Here, $B_0\,=\,8.9\times10^{23}\,\tau_{\rm
  T}\ kT_{*}^{-3/2}\ N_{18}\ \mbox{\rm phot. cm}^{-2}$ $\mbox{\rm
  s}^{-1}$  $\mbox{\rm keV}^{-1},$ $kT_*=kT_{\rm e}/50\ \mbox{\rm
  keV},$ and $N_{18}=N_{\rm e}/10^{18}\ \mbox{\rm cm}^{-3}.$

The boundary conditions for Eq.~(\ref{eq:komp}) were specified
in a standard form for the diffusion approximation --- at
the outer boundary of the layer (at $\tau=0$) as
\begin{equation}
\left.\left(\frac{\alpha_{\rm T}}{\alpha_{\rm T}+\alpha_{\rm ff}}
\frac{\partial U_{\nu}}{\partial \tau}
-\frac{3}{2}U_{\nu}\right)\right|_{\tau=0}=0, 
\end{equation}
and at the lower boundary (at $\tau=\tau_{\rm T}$) as 
\begin{equation}\label{eq:bclow}
\left.\left(\frac{\alpha_{\rm T}}{\alpha_{\rm T}+\alpha_{\rm ff}}
\frac{\partial U_{\nu}}{\partial \tau}+
\frac{3}{2}U_{\nu}\right)\right|_{\tau=\tau_{\rm T}}=0. 
\end{equation}
Along the frequency axis the boundary conditions were taken in
the form
\begin{equation}\label{eq:bcnu}
U_{\nu}|_{\nu=\nu_{\rm min}}=B_{\nu}(T_{\rm e})
\end{equation}
\begin{equation}
\left. \left(x\frac{\partial U_{\nu}}{x}-3U_{\nu}+
xU_{\nu}\right)\right|_{\nu =\nu_{\rm max}}=0, 
\end{equation}
implying that at the frequency $\nu_{\rm min}$ a local Planck
(Rayleigh-Jeans) radiation spectrum is formed over the entire
layer (at all optical depths), while at the frequency $\nu_{\rm
  max}$ the photon flux along the frequency axis due to
Comptonization is absent. To ensure that these conditions are
used properly, we adopted $h\nu_{\rm min} = 0.01$ eV $\ll
kT_{\rm e}$ and $h\nu_{\rm max} = 1300$ keV $\gg kT_{\rm e}$,
i.e., considered the problem in a very wide energy range. The
division in Thomson optical depth $\tau$ was chosen in
coordination with $h\nu_{\rm min}$ so that in the outer regions
of the layer the step in total optical depth for absorption
$\tau_{\rm tot}=\tau\,(1+\alpha_{\rm ff}/\alpha_{\rm T})$ did
not exceed unity. For our computations we used a
quasi-logarithmic grid with 240 grid points along the $\tau$
axis (the logarithmic one at $\tau\leq 1$ and the linear one at
greater depths if $\tau_{\rm T}>1$) and 250 grid points along
the energy axis (the logarithmic one at $h\nu\la kT_{\rm e}$ and
the linear one at high energies). This choice of the grid
ensures the stability of the solution, which was important due
to the presence of an alternate coefficient at the minor term
($\sim U_{\nu}$) of the Kompaneets operator (see
Eq.~(\ref{eq:komp})). System (\ref{eq:komp})--(\ref{eq:bcnu})
was solved by the matrix sweep method (the Feautrier method, see
Mihalas 1978).

As mentioned above, all our computations of the radiation field
were performed by assuming a one-dimensional geometry of the
problem. Moreover, the plasma temperature, $T_{\rm e}(\tau)$,
and density, $N_{\rm e}(\tau)$, distributions inside the
accretion disk are assumed to be uniform and independent of
$\tau$. Note that Grebenev and Sunyaev (2002) for a layer of
matter spreading over the surface of a neutron star and Lapidus
et al. (1986), London et al. (1986), and Ebisuzaki and Nomoto
(1986) for exponential atmospheres of X-ray bursters previously
showed that the method being used allows the problems of
radiation escape from a medium with highly nonuniform
temperature and density distributions to be successfully solved.
\begin{figure}[!p]

  \hspace{-5mm}\includegraphics[width=1.1\linewidth]{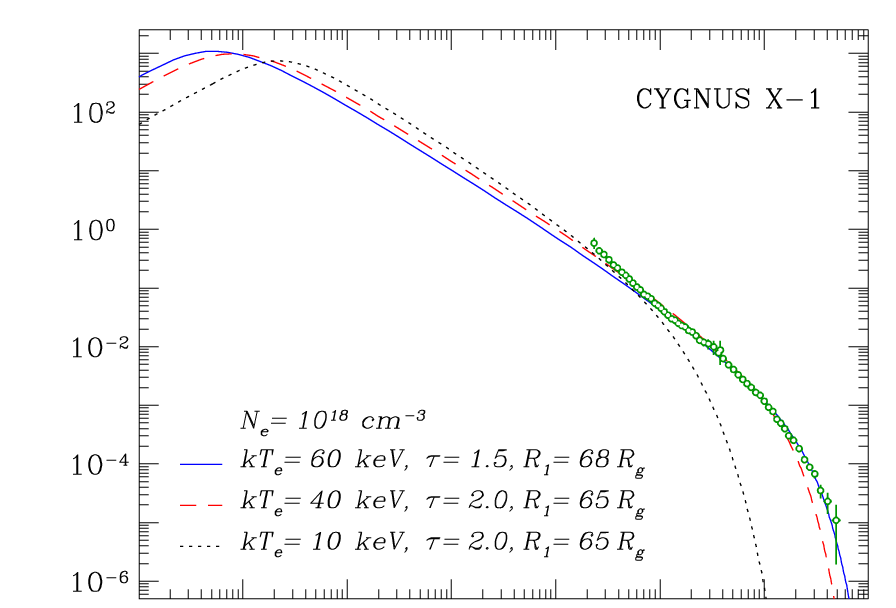}\\ [-4mm]
  
  \hspace{-5mm}\includegraphics[width=1.1\linewidth]{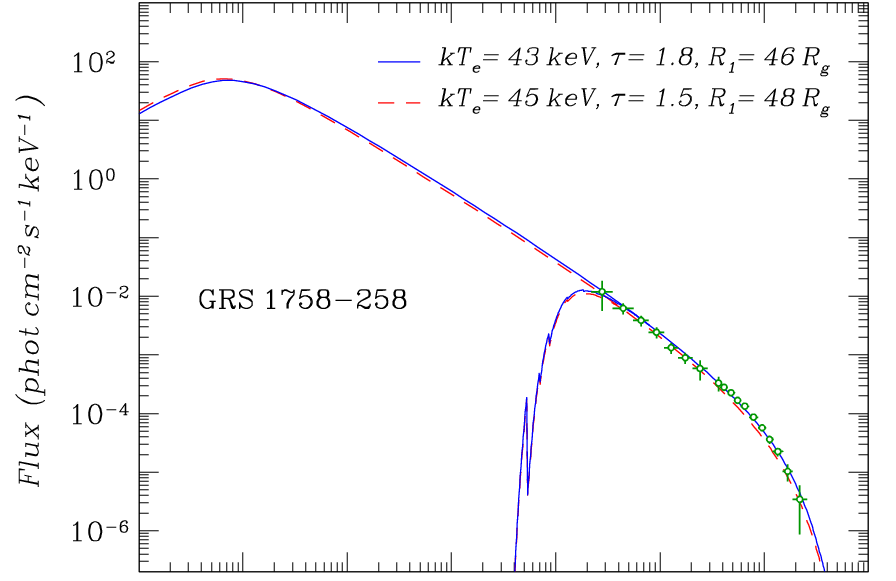}\\ [-4mm]
  
  \hspace{-5mm}\includegraphics[width=1.1\linewidth]{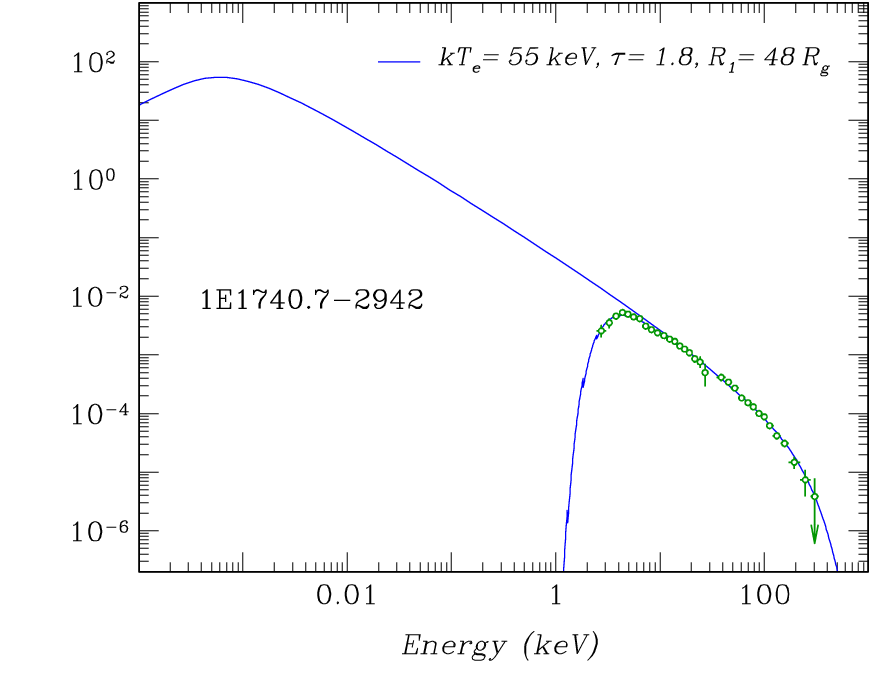}\\ [-4mm]
\caption{\rm The same photon spectra as those in
  Fig.~\ref{fig:ff}, but for a disk with slightly different
  outer radius $R_1$, temperature $kT_{\rm e},$ and transverse
  Thomson optical depth $\tau_{\rm T}$ (specified in the figure);
  the data in comparison with the broadband X-ray spectra of the
  sources Cyg X-1, GRS 1758-258, and 1E\,1740.7-2942 measured
  by the GRANAT observatory. The electron density in the disk
  was everywhere assumed to be $N_{\rm e} = 
  10^{18}\ \mbox{cm}^{-3}$.\label{fig:cygx1}}
\end{figure}

\section*{COMPARISON WITH OBSERVATIONS}
\noindent
The photon spectrum of a hot plasma layer presented in
Fig.~\ref{fig:ff} was obtained through our computations as a
combination $F_{\nu}(\nu)=2\,U_{\nu}(0)/h\nu \times S/d^2$. The
distance to the source was assumed to be $d=2.5$ kpc, in
accordance with the traditional assumption about the distance to
the best-known black hole candidate, the source Cyg X-1. Let us
compare the results of X-ray observations of this object with
our computations. The green dots on the upper panel of
Fig.~\ref{fig:cygx1} indicate the canonical spectrum of the
source (in the $\gamma_2$ state, according to the classification
of Ling et al. (1987)) measured in March 1990 in the wide energy
range 2--600 keV by the \mbox{ART-P} and SIGMA telescopes
onboard the GRANAT orbital astrophysical observatory (Grebenev
et al. 1993). The solid and dashed lines indicate the results of
our numerical computations with slightly differing (specified in
the figure) parameters of the plasma layer.

The observations are seen to be in excellent agreement with the
computations. We had only to increase slightly the disk area
relative to the original assumption (to increase the outer
radius by 30\% to $\sim65\ R_{\rm g}$) and either to lower the
plasma temperature to $kT_{\rm e}=40$ keV (red dashed line), or
to increase the temperature to $kT_{\rm e} = 60$ keV, while
simultaneously decreasing the transverse optical thickness of
the disk to $\tau_{\rm T}=1.5$ (blue solid line). The blue and
red lines fit slightly better the hardest part of the spectrum
with an exponential cutoff ($h\nu\ga 200$ keV) and the part of
the spectrum in the X-ray ($h\nu\la 100$ keV) range,
respectively. The desire to fit the spectrum at high, $h\nu\sim
m_{\rm e}c^2$, energies as accurately as possible may be
excessive, since our computations are based on the
nonrelativistic radiative transfer equations and applying the
corresponding corrections should lead to a harder spectrum.

Note that in the X-ray range 2--60 keV the observed spectrum of
the source can be satisfactorily fitted by a simple power law
$F_{\nu}\sim \nu^{-\alpha}$ with a photon index $\alpha\simeq
1.57\pm0.01$ and that the entire spectrum in the range 2--600
keV was successfully fitted by Grebenev et al. (1993) in their
original paper within the model of Comptonization (the inverse
Compton scattering of seed photons) in a flat layer with $kT_{\rm
  e}\simeq 57$ keV and an optical depth $\tau_{\rm T}\simeq
2.0$. The computations were performed by the Monte Carlo
method. Note also that although the attempts to fit the spectrum
using the approximate analytical solution of the Kompaneets
equation (without including any bremsstrahlung processes)
proposed by Shapiro et al. (1976) and, in a more general form,
by Sunyaev and Titarchuk (1980) led to similar parameters of the
layer, they generally turned out to be not so successful, since
they failed to reproduce the hardest part of the
spectrum. Grebenev et al. (1993) explained this failure by the
neglect of the relativistic effects in the analytical solution
compared to the completely relativistic Monte Carlo
computations. Another reason could be a small transverse optical
depth of the disk, whereas the analytical solution was obtained
in the diffusion approximation suggesting a large optical depth
($\tau_{\rm T}\gg 1$). Curiously, our computations, which are
also based on the nonrelativistic Kompaneets equation to
describe the motion of photons upward along the energy axis and
the diffusion approximation to describe their spatial
propagation (but take into account the bremsstrahlung processes
and do not assume the presence of external seed photons),
nevertheless, describe the experimental spectrum noticeably
better than does the mentioned analytical solution.

There is slight disagreement with the observations only at very
low energies $h\nu\la 10$ keV. It could have an instrumental
origin, since the efficiency of the \mbox{ART-P} X-ray telescope
rapidly dropped below $\sim 4$ keV, or a natural --- physical
one. In particular, it can be explained by a decrease of the
plasma temperature in the hot disk region as one recedes from
the black hole. As an illustration, the black dotted line in
Fig.~\ref{fig:cygx1} indicates the spectrum of the radiation
emitted by a plasma layer with the temperature $kT_{\rm e}=10$
keV. It can be seen that the contribution of this radiation
component can explain the observed excess of the source's
spectrum at these energies. There can also be other natural
explanations related to the Comptonization of low-energy
photons, synchrotron ones or those entering the high-temperature
region of the disk from its cold outer regions. The efficiency
of this process will be discussed below in the paper.

The two lower panels in Fig.~\ref{fig:cygx1} show the results of
our computations in comparison with the broadband X-ray photon
spectra of two more known black hole candidates, GRS~1758-258
and 1E\,1740.7-2942. The observations were carried out by the
GRANAT observatory in September and October 1990 (Grebenev et
al. 1995, 1997b). The source GRS 1758-258 is very similar in
spectral shape to \mbox{Cyg~X-1}, but it is characterized by a
slightly earlier high-energy cutoff. The spectrum of
1E\,1740.7-2942, which is flatter and harder, is fitted in the
range 2--60 keV by a power law with a photon index $\alpha\simeq
1.35\pm 0.09$. Both sources are located near the Galactic
center, at a distance $d\simeq 8.0$ kpc, which is taken into
account in the figure. At low energies their spectra are
distorted by interstellar absorption with an optical depth
corresponding to a hydrogen column density $N_{\rm H}\simeq
1\times 10^{22}\ \mbox{cm}^{-2}$ (for GRS~1758-258) and $\simeq
8\times 10^{22}\ \mbox{cm}^{-2}$ (for 1E\,1740.7-2942). The
computed curves are presented with and without the correction
for absorption.

Figure (its lower panels) again demonstrate surprisingly good
agreement of our computations with the observations. We only
again to slightly change the outer radius of the disk (by $\sim
\color{blue}10\color{black}$\%), the electron temperature,
and/or its transverse optical depth.

The middle panel for GRS~1758-258 presents two computed curves
that fit the data equally well. The curve that corresponds to a
larger optical depth, but a lower temperature is bent slightly
more strongly and fits better the shape of the exponential
cutoff in the spectrum. The second (flatter) curve fits the data
in the standard X-ray range slightly better. In general,
however, the differences are hardly visible.

Of course, such good agreement of the model with the
observations at almost invariable parameters of the plasma layer
(the hot disk region) cannot be accidental. It indicates that
the formation of the spectrum of accreting black holes in their
hard state is reproduced by the model correctly.

The GRANAT observations of black holes were used for their
comparison with our computations because of the very wide
coverage of the X-ray range typical for them. It is difficult to
advance into the range of lower energies, since, as the lower
panels in Fig.~\ref{fig:cygx1} show, interstellar absorption
distorts catastrophically the radiation spectrum of most
sources, cutting it off at energies below $h\nu\la 1$
keV. Nevertheless, in the concluding sections of the paper we
will present the spectra of some of the sources for which it is
possible to check the validity of the model at low energies as
well.
\begin{figure*}[!t]
 \centering
 \includegraphics[width=0.59\linewidth]{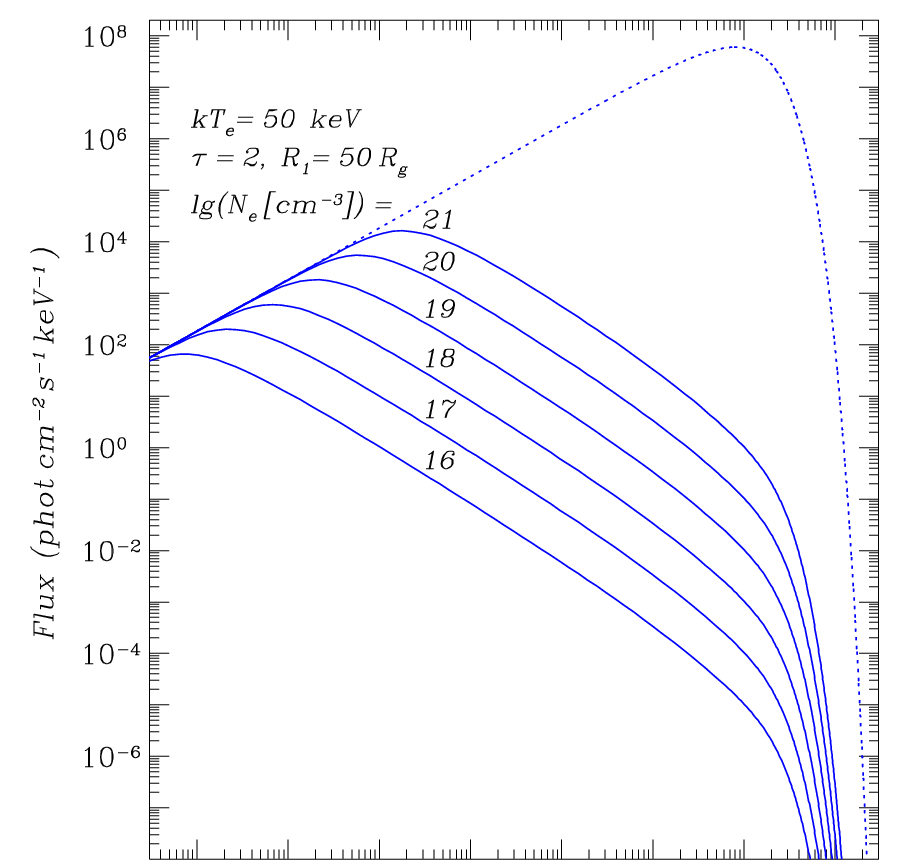}\\
 \includegraphics[width=0.59\linewidth]{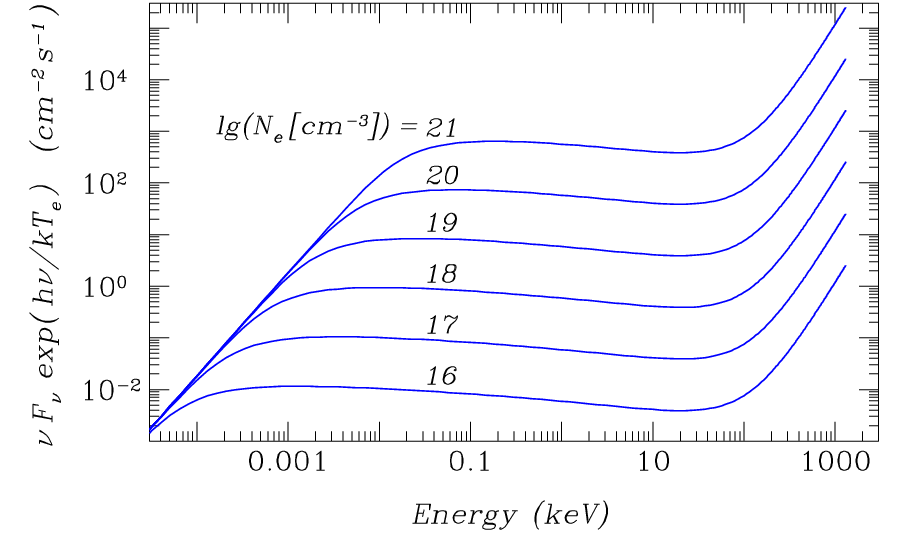}
\caption{\rm The {\sl upper panel\/} shows the photon spectra,
  as in Fig.~\ref{fig:ff}, but forming in a plasma layer with
  different electron densities (in the range $N_{\rm
    e}=10^{16}-10^{21}\ \mbox{cm}^{-3}$). The temperature, the
  transverse optical depth, and the outer radius of the disk
  remain invariable: $kT_{\rm e}=50$ keV, $\tau_{\rm T}=2,$ and
  $R_1=50\ R_{\rm g}$. The distance to the source is $d=2.5$
  kpc. The dotted line indicates the Planck radiation spectrum
  that should have been emitted by a very dense and optically
  thick plasma layer of the same temperature. The {\sl lower
    panel\/} shows the same spectra, but multiplied by
  $\nu\exp{(h\nu/kT_{\rm e})}$. \label{fig:den}}
\end{figure*}

\section*{THE DEPENDENCE ON PLASMA PARAMETERS}
\noindent
The above examples show that the shape and normalization of the
radiation spectra for the hot layer depend on its parameters and
the plasma parameters $N_{\rm e},\ kT_{\rm
  e},\ \mbox{and}\ \tau_{\rm T}$ in a complex way. Spectra with
a similar shape can be obtained by slightly different
methods. Consider this dependence in more detail.

Figure~\ref{fig:den} (upper panel) shows the photon spectra of
the radiation emerging from a plasma layer with the same
$kT_{\rm e}$, $\tau_{\rm T}$, $R_1,$ and $d$, as those in
Fig.~\ref{fig:ff}, but with a different electron density
changing from $N_{\rm e}=10^{16}$ to $10^{21}\ \mbox{\rm
  cm}^{-3}$. It can be seen that at energies exceeding $h\nu_1$,
i.e., where the layer is optically thin for bremsstrahlung
absorption, the difference between the spectra is reduced to a
change in its normalization; the spectral shape does not change
noticeably. This is also true for the most interesting region of
the exponential cutoff, as shown by the lower panel of the figure
that presents the same spectra, but multiplied by
$\nu\exp{(h\nu/kT_{\rm e})}$. The spectra are almost parallel to
one another, differing in normalization by an order of
magnitude. This would be expected, since the spectrum is formed
through the Comptonization of bremsstrahlung photons, with the
intensity of their emission being proportional to $2H\,N_{\rm
  e}^2\sim \tau_{\rm T}\, N_{\rm e}\sim N_{\rm e}$, where $H$ is
the disk half-thickness. The spectra in the figure were obtained
at an electron density $N_{\rm e}$ changing each time by an
order of magnitude; their normalization changes accordingly.

Significantly, the energy $h\nu_1$ itself depends on the
electron density (and in a critical way) and it determines the
transition of the spectrum to the Rayleigh-Jeans
form. Figure~\ref{fig:den} shows that the slope of the photon
spectrum for the hot accretion disk region in the optical and
near-infrared ranges (0.9--3.3 eV) remains the same as that in
the hard range ($\alpha\sim1.6$) at a plasma density $N_{\rm
  e}\la 10^{18}\ \mbox{\rm cm}^{-3},$ becomes almost flat
($\alpha\sim 0$) at $N_{\rm e}\sim 10^{19}\ \mbox{\rm cm}^{-3},$
and acquires the slope $\alpha = -1$ corresponding to the
Rayleigh-Jeans spectrum $F_{\nu}\sim (2\nu/c^2)\, kT_{\rm e}$
only at $N_{\rm e}\ga 10^{20}\ \mbox{\rm cm}^{-3}$. Below we
will show that in the hard state of accreting black holes
precisely this spectrum of the hot disk region determines their
observed optical and infrared emission.

\begin{figure*}[t]
  \centering
  \includegraphics[width=0.70\linewidth]{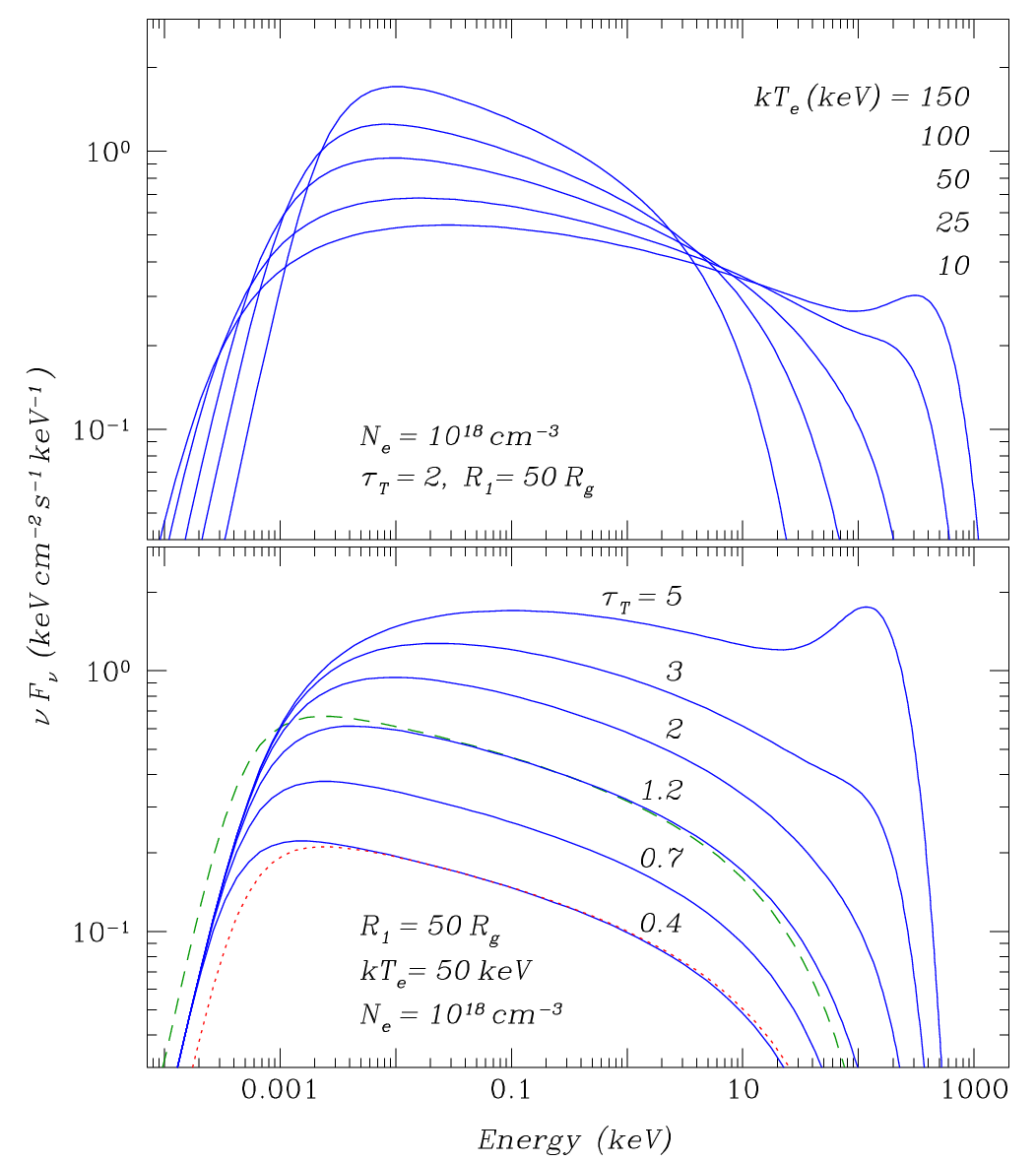}
  \caption{\rm The radiation spectra, as in Figs.~\ref{fig:ff}
    and \ref{fig:den}, but in the form $\nu F_{\nu}$. The upper
    and lower panels show the evolution of the spectral shape
    with increasing electron temperature $kT_{\rm e}$ and with
    changing transverse optical depth for Thomson scattering
    $\tau_{\rm T}$ respectively (the dotted red and dashed green
    lines indicate the spectrum for $\tau_{\rm T}=0.7$ shifted
    in normalization for comparison with the spectra for
    $\tau_{\rm T}=0.4\ \mbox{\rm and}\ 1.2,$ respectively). The
    remaining disk parameters were taken to be the same as those
    in Fig.~\ref{fig:ff}; the distance to the source is $d=2.5$
    kpc.\label{fig:akt}}
\end{figure*}

Similarly, the upper panel of Fig.~\ref{fig:akt} shows how the
spectra change with electron temperature $kT_{\rm e}$ in the
plasma layer in which they are formed. The remaining parameters
of the layer (disk) are the same as those in
Fig.~\ref{fig:ff}. For clarity, the figure shows the radiation
spectra $\nu\,F_{\nu}$ rather than the photon spectra, as in
Figs.~\ref{fig:ff}--\ref{fig:den}.  As the temperature
increases, the spectrum expectedly extends progressively farther
to high energies, but at the same time it also expands to lower
energies, in accordance with the temperature dependence of the
Rayleigh-Jeans spectrum $F_{\nu}\sim kT_{\rm e}$. At the highest
of the temperatures considered, $kT_{\rm e}=150$ keV, saturation
of the spectrum is observed --- the accumulation of photons at
$h\nu\sim 450\ \mbox{\rm keV}\ (= 3\, kT_{\rm e})$ and the
formation of a powerful Wien component.

While discussing the Comptonization of low-frequency photons in
an optically thick cloud ($\tau_0 = \sigma_{\rm T}N_{\rm e}R\ga
1,$ where $R$ is the cloud radius) of hot plasma, Shapiro et
al. (1976) and Sunyaev and Titarchuk (1980) showed that the
formation of the radiation spectrum of an X-ray source is
completely determined by the parameter $y_{\rm C}\simeq
(4kT_{\rm e}/m_{\rm e}c^2)\,\tau_0^2.$ Earlier, a similar
combination $y_{\rm C}\simeq (4kT_{\rm e}/m_{\rm
  e}c^2)\,\sigma_{\rm T} N_{\rm e}\,ct$ was used by Zeldovich
and Sunyaev (1969), Sunyaev and Zeldovich (1970), and Illarionov
and Sunyaev (1975) to investigate the distortions of the cosmic
microwave background spectrum in the early Universe. The larger
this parameter, the greater the modification of the original
spectrum by Comptonization. Although in this paper we consider
the more complex problem of the Comptonization of photons of an
extended bremsstrahlung spectrum, and the direct application of
the parameter $y_{\rm C}=(4kT_{\rm e}/m_{\rm e}c^2)\,\tau_{\rm
  T}^2$ is unjustified, we will specify the range of its
variation corresponding to the spectra presented on the upper
panel in Fig.~\ref{fig:akt}: $y_{\rm C}=0.31-4.70$. The spectra
corresponding to the high temperatures $kT_{\rm e}=150$ and 100
keV in the X-ray range have photon indices $\alpha\simeq 1.14$
and 1.20. In the spectra corresponding to $kT_{\rm e}\la 50$
keV, the exponential cutoff begins to manifest itself at these
energies and their photon index is difficult to estimate. At
energies 20--100 eV $\alpha\simeq 1.03-1.20$.
\begin{figure*}[!t]
  \centering
  \includegraphics[width=0.77\linewidth]{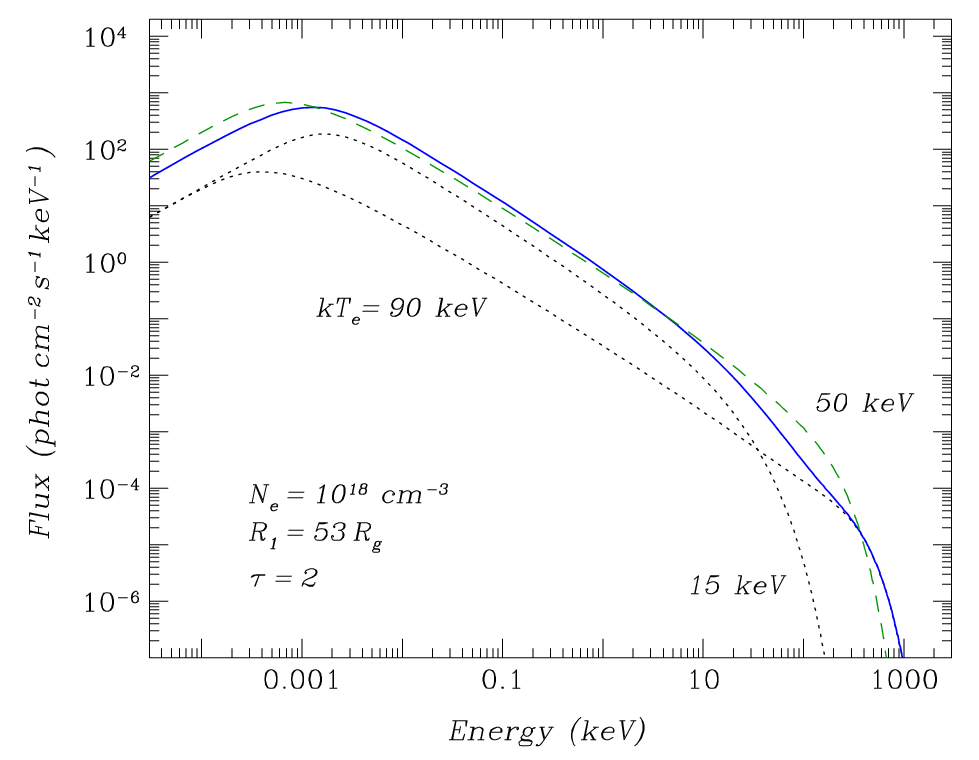}

\caption{\rm The same photon spectrum as that in
  Fig.~\ref{fig:ff}, but assuming that the electron temperature
  in the hot accretion disk zone decreases with radius (blue
  solid line). The other plasma parameters $N_{\rm e}$, $R_1,$
  and $\tau_{\rm T}$ were taken to be the same as those in
  Fig.~\ref{fig:ff}. The dotted lines indicate the spectra of
  the hottest ($kT_{\rm e}=90$ keV) and coldest ($kT_{\rm e}=15$
  keV) rings of the hot disk zone (the components of the overall
  spectrum). For comparison, the radiation spectrum is presented
  for a hot zone of the same area, but with a constant
  temperature, $kT_{\rm e}=50$ keV (green dashed line). In the
  spectrum of the layer with a changing temperature there is a
  power-law segment with a slope $\alpha\sim2.2$ in the range
  20--500 keV.\label{fig:ktevol}}
\end{figure*}

The lower panel of Fig.~\ref{fig:akt} shows the evolution of the
radiation spectra forming in a hot plasma layer with its
transverse optical depth $\tau_{\rm T}$ (at invariable remaining
parameters). The changes in $\tau_{\rm T}$ here correspond to
the changes in $y_{\rm C} = 0.06-9.78$. The photon index of the
X-ray spectra corresponding to $\tau_{\rm T}=5$ and 3 is
$\alpha\simeq1.12$ and 1.25, respectively; for the spectra with
smaller $\tau_{\rm T}$ the photon index defies determination. At
20--100 eV $\alpha\simeq 1.00-1.15$. As in the case of a
variable electron density (Fig.~\ref{fig:den}), the changes in
the spectrum at energies above $h\nu_1$ in this figure are
primarily reduced to the change in its normalization. However,
important changes in the spectral shape occur simultaneously. To
show them, the red dotted and green dashed lines in the figure
indicate the radiation spectrum computed at $\tau_{\rm T}=0.7$
and shifted in normalization to the spectra obtained at
$\tau_{\rm T}=0.4$ and $1.2$, respectively. It can be seen that
as $\tau_{\rm T}$ increases, (1) the transition of the spectrum
to the Rayleigh-Jeans one occurs at higher energies, i.e., the
characteristic energy $h\nu_1$ is shifted rightward, and (2) the
spectrum gradually extends to progressively higher energies. Its
overall curvature also changes, in particular, at the maximum
optical depth of the disk considered $\tau_{\rm T}=5$ evidence
of saturation appears in the spectrum at high energies and the
formation of a separate Wien component begins. Above we have
seen such evidence as the plasma temperature rises to $\sim150$
keV, at which the Compton parameter reaches close (large) values
of $y_{\rm C}\sim 4.7$.
\begin{figure*}[!t]
  \centering
  \includegraphics[width=0.77\linewidth]{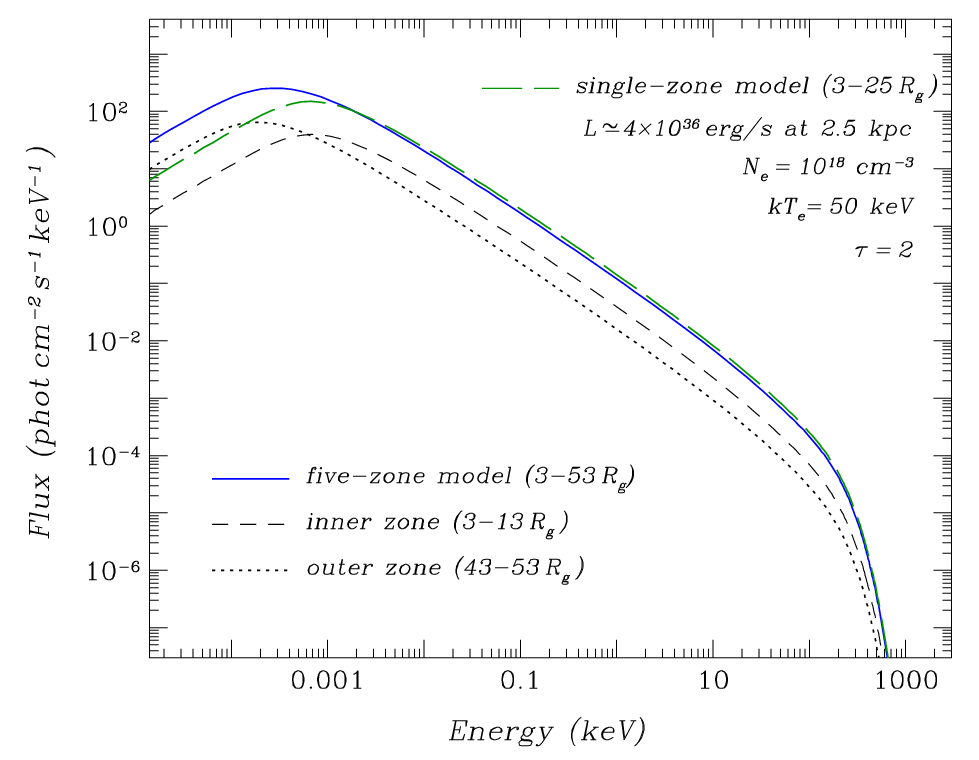}

\caption{\rm The same photon spectrum as that in
  Fig.~\ref{fig:ff}, but assuming that the electron density in
  the hot disk zone decreases with radius, $N_{\rm e}\sim
  R^{-3/2}$ (the blue solid line, see
  Table~\ref{table:neevol}). The other plasma parameters
  $kT_{\rm e}$, $R_1,$ and $\tau_{\rm T}$ were taken to be the
  same as those in Fig.~\ref{fig:ff}. The dashed and dotted
  lines indicate the spectra for the inner and outer rings of
  the hot zone (the components of the overall spectrum). For
  comparison, the green long dashes indicate the radiation
  spectrum of a homogeneous disk (its central hot zone with the
  density $N_{\rm e} = 10^{18}\ \mbox{\rm cm}^{-3}$ bounded by
  the radius $R_1 = 25\, R_{\rm g}$) at the distance $d=2.5$
  kpc.\label{fig:neevol}}
\end{figure*}

\section*{THE MULTICOLOR HOT ZONE}
\noindent
Above, for clarity and simplicity, we assumed that the electron
temperature $kT_{\rm e}$ inside the hot disk zone remains
constant. Although, indeed, Compton scattering inside the zone
leads to certain temperature equalization, it is difficult to
believe that $kT_{\rm e}$ is invariable taking into account a
strong dependence of the energy release in a viscous disk on the
radius, $\sim R^{-3}$ (see Eq.~(\ref{eq:dbbqv}) below).

While discussing Fig.~\ref{fig:cygx1}, we have already noted the
possible changes in the shape of the radiation spectrum for the
hot zone when the temperature gradually decreases with
radius. Consider this effect in more detail. Suppose that the
hot zone in the range of radii from $R_0=3\ R_{\rm g}$ to
$R_1=53\ R_{\rm g}$ is divided into five rings of equal width
$\Delta R=10\ R_{\rm g}$ with a temperature dropping with radius
as $\sim R^{-1}$, i.e. $kT_{\rm e}=90,\ 40,\ 26,\ 19,\ \mbox{\rm
  and}\ 15$ keV within these rings. The farthest zones with a
low electron temperature are characterized by a relatively soft
radiation spectrum, but at the same time have a large area
($=\pi\,[R_{\rm i+1}+R_{\rm i}]\Delta R\sim R$) and contribute
significantly to the integrated spectrum by changing its
slope. This is confirmed by Fig.~\ref{fig:ktevol} that shows the
forming photon spectrum computed for such a hot layer in
comparison with the spectrum of a layer with a constant
temperature, $kT_{\rm e}=50$ keV. It can be seen that the
spectrum of the layer with a variable temperature is generally
steeper; besides, a quasi-power-law segment with a photon index
$\alpha\sim2.2$ appears in the X-ray range 20--500 keV. Note
that the normalization of the spectra at low energies (in the
Rayleigh-Jeans region) also differs noticeably, which is not
surprising, since it is determined by the temperature and area
of the radiating zone. In contrast, the fact that in this region
the fluxes from the two components of the integrated spectrum
(the emissions from the hottest and coldest rings) coincide is a
result of the chosen law of temperature decrease $kT_{\rm e}\sim
R^{-1}.$ Indeed, $F_{\nu}\sim \pi (R_{\rm i}+R_{\rm
  i+1})\,\Delta\,R\ 2h\nu/c^2\,kT_{\rm e}(R_{\rm i}) = const$.

The electron density in the hot disk region does not need to be
constant along the radius either.  Using Eq.~(\ref{eq:Negas})
from the Appendix for a simple radial dependence of the electron
density in the plasma layer maintained in equilibrium in the
vertical direction by the gas pressure and the already tested
division of the hot disk region into five ring zones, it is easy
to construct the integrated spectrum of such a disk --- it is
indicated by the blue solid line in Fig.~\ref{fig:neevol}. The
dashed and dotted lines indicate the spectra of the innermost
(densest) and outermost ring zones. For convenience, the number
density $N_{\rm e}$ for each zone is given in
Table~\ref{table:neevol} together with the luminosity
$\Delta\,L$ of the zone in the entire energy range. The
temperature, $kT_{\rm e}=50$ keV, and the optical depth,
$\tau_{\rm T}=2.0$, were assumed to be the same for all zones.
The luminosity of the entire hot disk region was found to be
$L_{\rm H}=3.8\times 10^{36}\ \mbox{\rm erg s}^{-1}$. Using
Eq.~(\ref{eq:lumin}) based on the distribution of energy release
in the disk, we can estimate its total luminosity (including its
outer $R>R_1$ region) $L_{d}=4.4\times 10^{36}\ \mbox{\rm erg
  s}^{-1}$ and the corresponding accretion rate
$\dot{M}=12\,L_{d}/c^2\simeq 9.2\times
10^{-10}\ M_{\odot}\ \mbox{\rm yr}^{-1}.$ The model luminosity
for each zone estimated using Eq.~(\ref{eq:lumin}) is given in
the right column of Table~\ref{table:neevol}. The comparison of
the calculated and model luminosities shows that the luminosity
at large radii slightly exceeds the rate of energy release,
which is quite explainable, given the fairly rough method used
to estimate the density in the disk (Eq.~(\ref{eq:Negas}) from
the Appendix). In a real accretion disk the temperature and the
optical depth are mutually adjusted so that the disk radiation
matches the energy release.
\begin{table}[th]
 \caption{Parameters of the five-zone model of a disk with
a variable density whose spectrum is presented in
Fig.~\ref{fig:neevol}.\label{table:neevol}}   
\vspace{2mm}

\centering\small

\begin{tabular}{c|@{}c@{}|r@{}c|c|c}\hline\hline
&&&&&\\ [-3mm]  
  Zone&Zone&\multicolumn{2}{c|}{$N_{\rm e},$\bb}
  &$\Delta\,L$\cc&$\Delta\,L_{\rm mod}$\dd\\
  &&&&\multicolumn{2}{c}{}\\ [-8mm]
  &&&&&\\ \cline{5-6}
  &&&&\multicolumn{2}{c}{}\\ [-3mm]
   number&\multicolumn{1}{|@{\,}c@{\,}|}{boundaries\aa}
  &\multicolumn{2}{c|}{$10^{17}\ \mbox{\rm cm}^{-3}$}
  &\multicolumn{2}{c}{$10^{36}\ \mbox{\rm erg s}^{-1}$}\\ \hline
&&&&&\\ [-3mm] 
1& ~3--13&~~~10.56&~&1.23&2.33\\
2& 13--23& 3.12&&0.82&0.77\\
3& 23--33& 1.61&&0.66&0.35\\
4& 33--43& 1.02&&0.57&0.20\\
5& 43--53& 0.72&&0.50&0.13\\ \hline
\multicolumn{6}{c}{}\\ [-2mm]
\multicolumn{6}{l}{\aa\ The inner and outer radii of the zone
  in $R_{\rm g}$.}\\ 
\multicolumn{6}{l}{\bb\ According to the dependence
  (\ref{eq:Negas}) in the Appendix.}\\ 
\multicolumn{6}{l}{\cc\ The luminosity from the computed spectrum.}\\
\multicolumn{6}{l}{\dd\ The luminosity from the energy release
  --- Eq.~(\ref{eq:lumin}).}\\
  \end{tabular}
\end{table} 
 
As can be seen from Fig.~\ref{table:neevol}, the change of the
density in the ring zones has virtually no effect on the shape
of their spectrum, except for its low-frequency part $h\nu\la 1$
eV, but at the same time it changes the normalization of the
integrated spectrum by a factor of $\sim4$. The long green
dashes in the figure indicate the radiation spectrum of a
homogeneous plasma layer with the same temperature and optical
depth as those in the five-zone disk model, but with a constant
density, $N_{\rm e}=10^{18}\ \mbox{\rm cm}^{-3}$. Although this
spectrum closely resembles the integrated photon spectrum of the
five-zone model, to reconcile their normalizations and
luminosities, we had to set the outer radius of the
high-temperature region equal to $R_1=25\ R_{\rm g}$, i.e., the
hot region turns out to be brighter when considering the
homogeneous model with such a density.
\begin{figure*}[!t]
  \centering
  \includegraphics[width=0.77\linewidth]{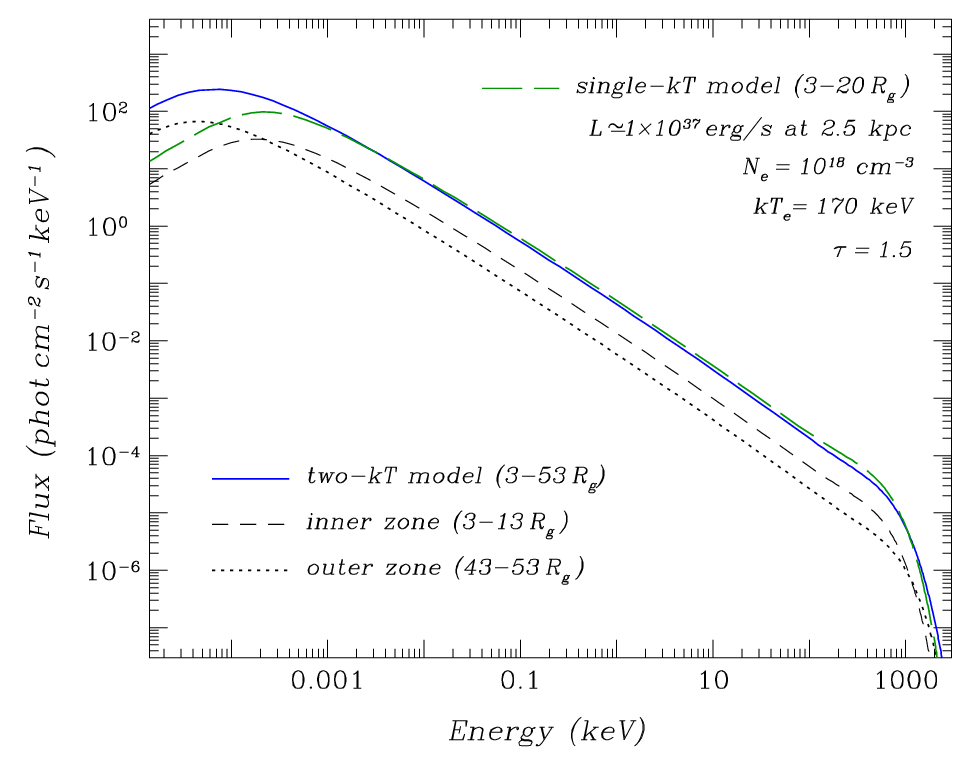}

\caption{\rm The integrated photon spectrum (blue solid line)
  for the hot zone with $R_1=53\,R_{\rm g}$ in the
  two-temperature disk of Shapiro et al. (1976). The accretion
  rate corresponds to the total disk luminosity $L=1\times
  10^{37}\ \mbox{erg s}^{-1}$, the black hole mass is
  $M=10\ M_{\odot}$. The dotted and dashed lines indicate the
  spectra of the hottest ($kT_{\rm e}=231$ keV) outer and
  coldest ($kT_{\rm e}=165$ keV) inner rings of the hot disk
  zone (the components of the overall spectrum). The values of
  $N_{\rm e}$, $kT_{\rm e},$ and $\tau_{\rm T}$ for each ring
  are given in Table~\ref{table:2tmodel}. For comparison, the
  long green dashes indicate the radiation spectrum of a
  single-temperature homogeneous disk (its central hot zone with
  the outer radius $R_1=20\ R_{\rm g}$, the temperature $kT_{\rm
    e}=170$ keV, the density $N_{\rm e}=10^{18}\ \mbox{\rm
    cm}^{-3},$ and the transverse optical depth $\tau_{\rm
    T}=1.5$). \label{fig:2tmodel}}
\end{figure*}

In real accretion disks one might also expect more complex
dependences of the electron temperature and density on the
radius. For example, the two-temperature model of Shapiro et
al. (1976) predicts a weak increase in the electron temperature
outward as $kT_{\rm e}\sim R^{1/4}$, but a dramatic drop in the
proton temperature $kT_{\rm p}\sim R^{-5/4}$, see
Eq.~(\ref{eq:2tmodel-kte}) in the Appendix. This is necessary
for an efficient removal of the gravitational energy being
released in the disk.  The dependence of the electron density on
the radius in this model is given by Eq.~(\ref{eq:2tmodel-ne})
in the Appendix, $N_{\rm e}\sim R^{-9/8}$. Given what has been
said about the imperfection of the previous disk model, it would
be important to compute the radiation field for such a
self-consistent model.

Let us construct the integrated photon spectrum for this model
using the already tested division of the hot disk zone into five
rings. Table~\ref{table:2tmodel} gives the values of $N_{\rm
  e}$, $kT_{\rm e},$ and $\tau_{\rm T}$ calculated from the
above formulas (and other formulas in the Appendix) by assuming
the black hole mass to be $M=10\ M_{\odot}$, the viscosity
parameter to be $\alpha =1,$ and the accretion rate to be
$\dot{M}=2.1\times 10^{-9}\ M_{\odot}\ \mbox{\rm yr}^{-1}$
(corresponding to a total disk luminosity
$L_{d}=\dot{M}c^2/12\simeq 1\times 10^{37}\ \mbox{\rm erg
  s}^{-1}$). The integrated photon spectrum itself is indicated
in Fig.~\ref{fig:2tmodel} by the solid blue line; its two
components corresponding to the innermost and outermost rings
are indicated by the dashed and dotted black lines,
respectively. The luminosity of the computed spectrum is
$L_{H}\simeq 0.96\times 10^{37}\ \mbox{\rm erg s}^{-1},$ i.e.,
slightly higher than should be, given the limited size of the
radiating region $3\,R_{\rm g}\la R\la 53\,R_{\rm g}$
responsible, according to the already mentioned
Eq.~(\ref{eq:lumin}) for 86.7\% of all the disk radiation. This
is probably because the hot disk zone is divided into too coarse
(wide) rings. Note that the purely bremsstrahlung luminosity of
the plasma in the hot zone is $L_{\rm ff}\simeq 1.1\times
10^{36}\ \mbox{\rm erg s}^{-1},$ i.e., Comptonization increases
the plasma luminosity in the zone by a factor $\sim9$.

Despite the differences in plasma parameters, the spectra of the
individual rings are very similar, differing only by the
normalization and the energies of the low-frequency (at energies
$h\nu\la1$ eV) and high-frequency (at energies $h\nu\ga 500$
keV) cutoffs (in accordance with the electron temperature of a
given ring). Comptonization increases the luminosity of the
plasma manyfold and, therefore, even a moderate change in the
temperature and density allows the efficiency of its radiative
cooling to be adjusted to completely remove the gravitational
energy being released in the disk. In this case, as follows from
our computations, the shape of the spectrum in a wide energy
range changes insignificantly.

Again, the long green dashes in Fig.~\ref{fig:2tmodel} indicate
the radiation spectrum of a homogeneous layer with the single
temperature of electrons and protons $kT_{\rm e}=170$ keV, the
density $N_{\rm e}=1\times 10^{18}\ \mbox{\rm cm}^{-3},$ and the
transverse optical depth $\tau_{\rm T}=1.5$. This spectrum
closely resembles the two-temperature disk spectrum, although
the layer radiation is again much more intense --- to reconcile
the spectra and the luminosities, $L\simeq 1\times
10^{37}\ \mbox{erg s}^{-1}$, we had to take the outer radius of
the radiating region to be $R_1=20\, R_{\rm g}$. It seems that
the radiation spectrum of a realistic accretion disk predicted
by the model of Shapiro et al. (1976) or another similar model
can be easily reproduced using a single-temperature disk model
with a uniform density distribution. However, the hot zone of a
realistic self-consistent disk should be much wider in
radius. On the other hand, it is clear that, in any case, the
rings closest to the black hole make a major contribution to the
radiation from the hot zone of an accretion disk without
advection, since the radiation should remove all of the energy,
which is being released at a given radius $\sim R^{-2}$ or, more
precisely, $\sim R^{-2}\left[1-(R_0/R)^{1/2}\right]$, see
Eq.~(\ref{eq:lumin}) below.
\begin{table}[th]
 \caption{Parameters of the two-temperature disk of
Shapiro et al. (1976) whose spectrum is presented in
Fig.~\ref{fig:2tmodel}.\label{table:2tmodel}} 
\vspace{2mm}

\centering\small

\begin{tabular}{c|c|c|c|c}\hline\hline
&&&&\\ [-3mm]
Zone&Zone&$N_{\rm e},$\bb&$\tau_{\rm T}$\cc&$kT_{\rm
  e},$\dd\\ 
number
&\multicolumn{1}{@{\,}c@{\,}}{boundaries\aa}&\multicolumn{1}{|c|}{$10^{17}\ \mbox{\rm 
  cm}^{-3}$}&&keV\\ 
\hline
&&&&\\ [-3mm]
1& ~3--13&6.73& 1.47& 165\\ 
2& 13--23&2.43& 1.29& 188\\ 
3& 23--33&1.43& 1.18& 205\\ 
4& 33--43&1.00& 1.10& 219\\ 
5& 43--53&0.76& 1.05& 231\\ 
\hline
\multicolumn{5}{c}{}\\ [-2mm]
\multicolumn{5}{l}{\aa\ The inner and outer radii of a given zone
  in $R_{\rm  g}$.}\\ 
\multicolumn{5}{l}{\bb\ According to Eq.~(\ref{eq:2tmodel-ne}) in the Appendix.}\\
\multicolumn{5}{l}{\cc\ According to Eq.~(\ref{eq:2tmodel-kte}) in the Appendix.}\\
\multicolumn{5}{l}{\dd\ According to Eq.~(\ref{eq:2tmodel-tau}) in the Appendix.}\\
  \end{tabular}
\end{table} 

The advection-dominated accretion flow (ADAF) models during disk
accretion onto a black hole (Narayan et al. 1998; Yuan and
Narayan 2014; Liu et al. 2025) also predict the appearance of a
two-temperature plasma. As in the model of Shapiro et
al. (1976), in an extended disk region ($R_1\gg 50\, R_{\rm g}$)
the temperature of the electrons being heated by protons remains
very high ($kT_{\rm e}\ga100$ keV). At the same time, the plasma
density and optical depth are smaller than those in Keplerian
disks without advection because of a much higher (transonic)
radial velocity. The advective flows are nearly spherically
symmetric and, therefore, our computations in the approximation
of a plane geometry are inapplicable to them. The question about
the Comptonization efficiency in such models and the validity of
the peculiarities of the formation of the accreting plasma
radiation spectrum noted in the paper remains open.

The advective solutions attract increasingly great
interest because of their stability that the solution of
Shapiro et al. (1976) lacks (Piran 1978). On the other
hand, observations of accreting black holes during
their hard state more likely suggest an instability
of the accretion flow in the high-temperature zone
(chaotic flux fluctuations, shot noise, low-frequency
quasi-periodic oscillations) than its stability.

\section*{THE ANALYTICAL APPROXIMATION}
\noindent
To better understand the generation of radiation from a hot
plasma layer and to check the numerical computations, it is
useful to deduce at least an approximate, but analytical
solution of the problem defined by
Eqs.~(\ref{eq:komp})--(\ref{eq:bcnu}).

While considering a similar problem, but related
to the Comptonization of seed photons in a plasma
cloud, Sunyaev and Titarchuk (1980) proposed to investigate
their spatial diffusion and the motion along the frequency axis
separately\footnote{The possibility of such an approach was
  pointed out long ago in the original paper of Kompaneets
  (1957), in the Appendix.}. The final solution can
be presented in the form of a convolution: 
\begin{equation}\label{eq:uconv}
  F_{\nu}(\nu)=\int^{\infty}_0 G_{\nu}(\nu,u)\,P(u)\,du,
\end{equation}
where $G_{\nu}(\nu,u)$ is the energy distribution of photons
after time $t$ since their emission in which they experience, on
average, $u=\sigma_{\rm T}N_{\rm e}ct$ scatterings, and 
$P(u)\,du$ is the probability for a photon to escape from
the cloud after $u$ scatterings. The photon escape
probability $P(u)$ is determined by the geometry of the
cloud (layer) and the distribution of original photons
in its optical depth (the distribution of the source
of seed photons). If the source is diffuse and its
distribution over the layer as a function of the transverse
Thomson optical depth $\tau$ coincides with the
eigenfunction of the corresponding problem of photon
diffusion in a cloud, then the escape probability takes
a simple form, $P(u)=\beta
\exp{(-\beta u)}$. For a flat layer 
\begin{equation}
  \beta=\pi^2/3 \left(\tau_{\rm T}+4/3\right)^{2}.
\end{equation} 
Having transformed Eq.~(\ref{eq:komp}) in accordance with this
approach and the specified distribution $P(u)$ (assuming that
$\alpha_{\rm T}\gg \alpha_{\rm ff}$), Sunyaev and Titarchuk
(1980) found a general solution for the spectrum of the hard
photons escaping from the plasma layer after their Comptonization:
\begin{equation}\label{eq:spec-st}
  F_{\nu}(x^{\,\prime},x)=F_{\,0}\ (x^{\,\prime})^{\alpha-1}\ x^{-\alpha}
  e^{-x}\times
\end{equation}
$$\times\int^{\infty}_0 t^{\alpha-2}(x+t)^{\alpha+2} e^{-t}\ d\,t$$
 (Eq.~(23) in their paper). The solution was obtained for $x\geq
x^{\,\prime}$, where, as before, 
$x=h\nu/kT_{\rm e}$ and $h\nu^{\,\prime}=x^{\,\prime}\,kT_{\rm e}$
is the energy of the seed photons. The factor 
\begin{equation}\label{eq:f0}
F_{\,0} = \frac{(\alpha-1)(\alpha+2)}{\Gamma (2\alpha+2)\,
  kT_{\rm e}},
\end{equation}
normalizes the total flux of seed photons to unity
and $\Gamma (z)$ is the gamma function (Abramowitz and
Stegun 1964). The photon index of the radiation
being produced is
\begin{equation}
  \alpha=\left(9/4+\gamma\right)^{1/2}-1/2,
\end{equation}
where 
$$\gamma=\beta m_{\rm e}c^2/kT_{\rm e}=\pi^2 m_{\rm e}c^2/3 kT_{\rm e} \left(\tau_{\rm
    T}+4/3\right)^{2}.
$$ 
\begin{figure*}[!t]
\centering
\includegraphics[width=0.73\linewidth]{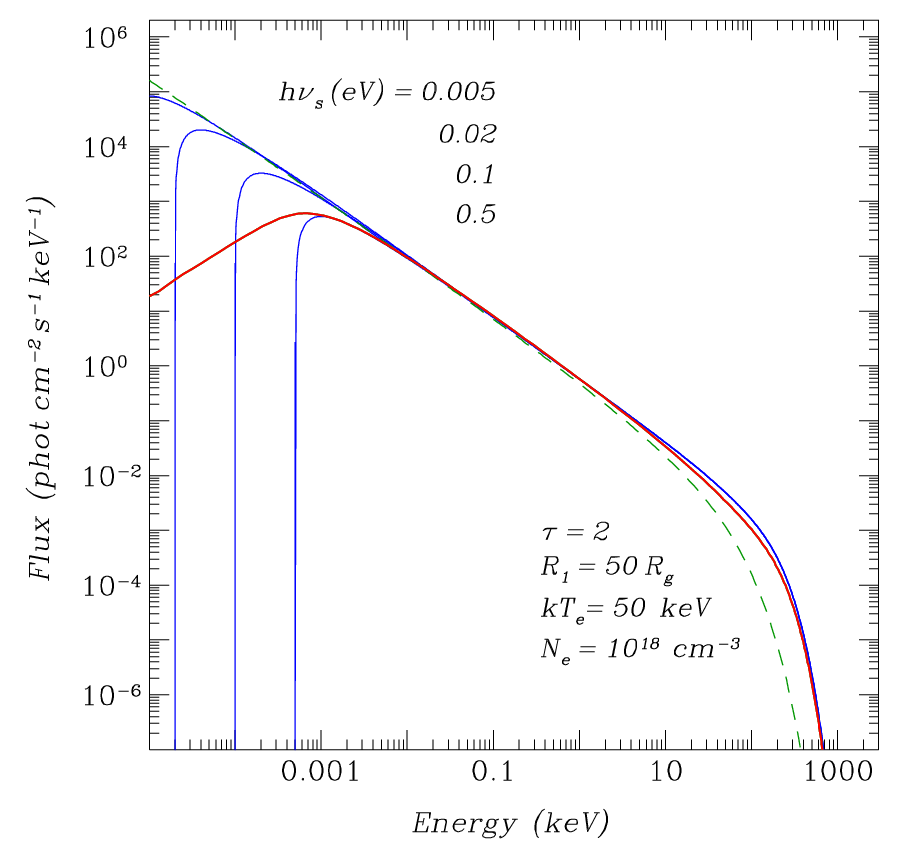}
\caption{\rm The same numerically computed photon spectrum of
  the high-temperature plasma layer as that in Fig.~\ref{fig:ff}
  (thick red line) in comparison with the approximate analytical
  solution under different assumptions about the softest
  bremsstrahlung photons $h\nu_{\rm s}$ involved in
  Comptonization (solid blue thin lines). The dashed line
  indicates the bremsstrahlung spectrum of this plasma
  layer. The distance to the source is $d=2.5$
  kpc.\label{fig:kompan}}
\end{figure*}
\noindent
The spectrum of the radiation being produced in such a layer
during the Comptonization of intrinsic bremsstrahlung photons
can be found by convolving Eq.~(\ref{eq:spec-st}) with the
photon spectrum $B_{\rm ff}$ (Eq.~(\ref{eq:spec-brem})) of the
thermal plasma bremsstrahlung
\begin{equation}\label{eq:spec-an1}
F_{\nu}(x) = \int^x_{x_{\rm s}}  B_{\rm ff}(x^{\,\prime})\, F_{\nu}(x^{\,\prime},x)\ d\,x^{\,\prime}=
\end{equation}
$$=F_{\,1}(x_{\rm s},x)\  x^{-\alpha} e^{-x}
\int^{\infty}_{0}
  t^{\alpha-2}(x+t)^{\alpha+2} e^{-t}\ d\,t.
$$
Here,
\begin{equation}\label{eq:spec-an2}
F_{\,1}(x_{\rm s},x)=F_0\ B_0\left(\frac{S}{d^2}\right)
\int^x_{x_{\rm s}} t^{\alpha-2}\ g(t)\ e^{-t}\ d\,t,
\end{equation}
and $h\nu_{\rm s}=x_{\rm s}kT_{\rm e}$ is the energy of the
softest brems\-strahlung photons that are involved in
Comptonization. The integrals in Eqs.~(\ref{eq:spec-st}) and
(\ref{eq:spec-an2}) can be taken numerically: the first one using
the Gauss-Laguerre quadrature formula and the second one by
more direct methods. The Gaunt factor $g(x)$ can
be taken into account directly using Eq.~(\ref{eq:gaunt}). In the
limit $x_{\rm s}\rightarrow 0$ Eq.~(\ref{eq:spec-an2}) can be expressed via the lower
incomplete gamma function $\gamma(z,x)$ (Abramowitz and
Stegun 1964):
\begin{equation}\label{eq:spec-anf}
  F_{\,1}(x)=B_0\,
  \frac{(\alpha-1)(\alpha+2)\gamma(\alpha-1.13,x)}{\Gamma (2\alpha+2)}
  \left(\frac{S}{d^2}\right).
\end{equation}
For this purpose, Eq.~(\ref{eq:gaunt}) for the Gaunt factor was
approximated, though fairly roughly, by a power law, $g(x)=1.7\,
x^{-0.13}$.

The radiation spectra computed from the quasi-analytical
formulas~(\ref{eq:spec-an1}) and (\ref{eq:spec-an2}) are
indicated in Fig.~\ref{fig:kompan} by the solid blue lines. Our
computations were performed for the same parameters of the
plasma layer as those used to construct Fig.~\ref{fig:ff} under
different assumptions about the minimum energy $h\nu_{\rm s}$
of the bremsstrahlung photons involved in Comptonization. The
limitation on $h\nu_{\rm s}$ is important, since, as has been
shown above, the bremsstrahlung photons at energies $h\nu\la 1$
eV are efficiently absorbed and the forming radiation spectrum
has a decaying Rayleigh-Jeans form. Figure~\ref{fig:kompan}
confirms the strong dependence of the shape and slope of the
radiation spectrum forming due to Comptonization on $h\nu_{\rm
  s}$, but only at sufficiently low energies $h\nu\la 100$
eV. In the X-ray range the forming spectrum has a single shape
independent of $h\nu_{\rm s}$. For comparison, the red thick
line in the figure indicates the numerically computed radiation
spectrum (the one shown in Fig.~\ref{fig:ff}). The dashed green
line indicates the plasma bremsstrahlung spectrum assuming that
the plasma layer is optically thin.

It can be seen that although the analytical and numerically
computed spectra have much in common and both suggest a great
hardening of the radiation from the plasma layer compared to its
bremsstrahlung, they have noticeable differences. In particular,
the analytical spectrum extends as a power law to higher
energies than does the numerically computed one, i.e., it
predicts a harder radiation spectrum. This is probably because
of the noticeably higher assumed concentration of seed photons
to the central disk plane in our analytical consideration
compared to their homogeneous initial distribution in our
numerical computations. Accordingly, the effective Thomson
optical depth of the plasma layer for the seed photons in our
approximate consideration is larger than that in our accurate
numerical computations. Therefore, the Comptonization process
itself is also more efficient.

However, we repeat that, on the whole, our analytical
consideration confirms the conclusions drawn from our numerical computations.
\begin{figure*}[!t]
 \centering
 \includegraphics[width=0.69\linewidth]{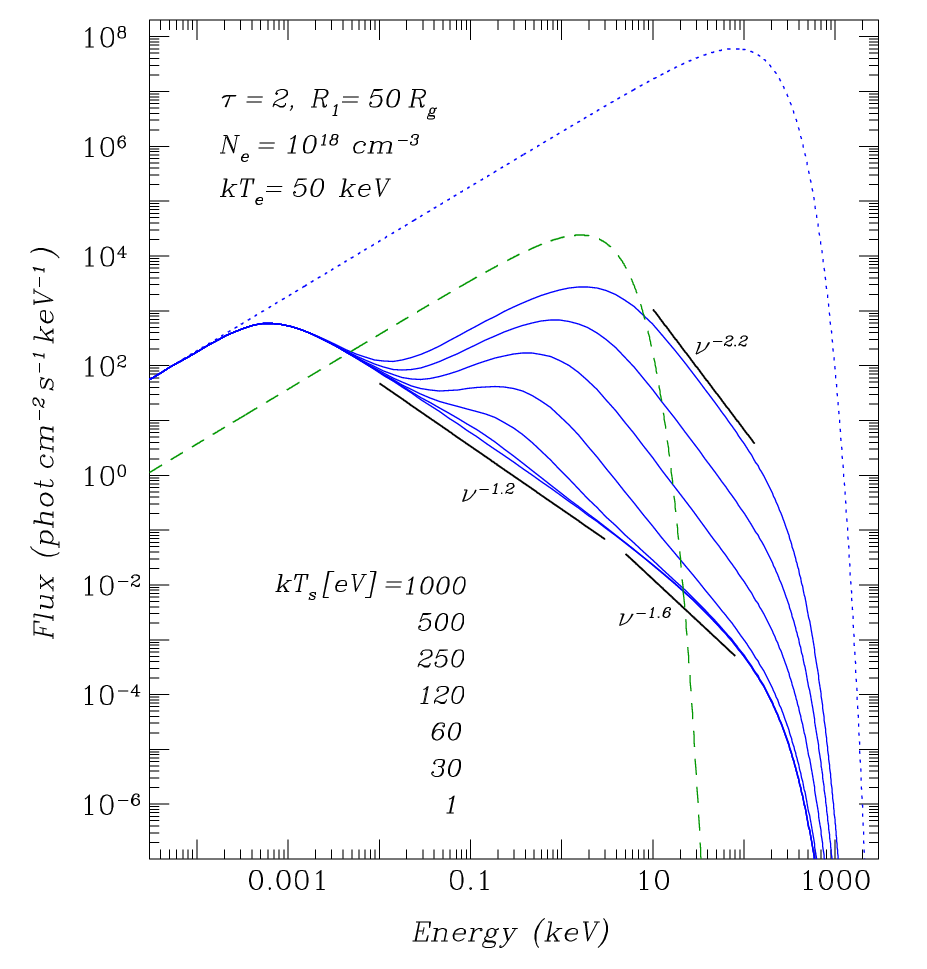}

\caption{\rm The same photon spectra as those in
  Fig.~\ref{fig:ff}, but in the ``corona'' model (assuming that
  the hot plasma layer with the same parameters as those in
  Fig.~\ref{fig:ff} is irradiated from one side by blackbody
  radiation from a cold underlying surface with a temperature
  $T_{\rm s}$). The different spectra indicated by the blue
  solid lines correspond to the different values of $kT_{\rm s}$
  specified in the figure. The dotted blue line indicates the
  Planck spectrum corresponding to the corona temperature
  $kT_{\rm e}=50$ keV; the dashed green line indicates the
  Planck spectrum corresponding to the temperature of the
  underlying surface $kT_{\rm s}=1$ keV. The black segments
  (parts of the power-law spectrum) characterize the slope of
  the computed spectrum closest to them in the corresponding
  energy range.\label{fig:soft2.0}}
\end{figure*}

\section*{THE CONTRIBUTION OF EXTERNAL PHOTONS}
\noindent
There were the various possibilities of the penetration of
external photons into the hot region of the accretion disk
discussed. These can be:\\ 
\indent
(1) the ultraviolet and soft X-ray photons emitted
by its outer cold region (primarily near its inner edge
adjacent to the boundary of the hot zone);\\ 
\indent
(2) the photons emitted by the cold dense disk
above which a high-temperature rarefied ``corona'' is
formed (Galeev et al. 1979; Haardt and Maraschi
1993);\\  
\indent
(3) the synchrotron photons emitted by high-temperature thermal
or relativistic nonthermal electrons in a magnetized hot plasma
(if nonthermal electrons are present there).\\

The first option looks most natural, since some number of
photons from the periphery of the disk should inevitably enter
its bloated hot central zone. Concurrently, some number of hard
X-ray photons from the central zone should irradiate the surface
of the cold outer disk region, heat it, and, being partially
reflected, form some features in the overall radiation spectrum
of the source --- a component of the reflected continuum
radiation (Compton bump), a photoabsorption edge, and emission
lines of highly ionized iron ions (Garcia and Kallman 2010;
Gilfanov 2010). However, in the approach implemented in this
paper, it is most convenient to ensure the penetration of
external photons into the hot zone in the ``corona'' model, in
which an opaque underlying surface (cold disk) with a
temperature $T_{\rm s}\ll T_{\rm e}$ and corresponding radiation
with a blackbody spectrum is adjacent to the hot plasma layer
from its one side. In this case, the boundary condition for
Eq.~(\ref{eq:komp}) at the lower boundary (Eq.~(\ref{eq:bclow}))
is replaced by
\begin{equation}
\left.
\frac{1}{2}U_{\nu} \right|_{\tau=\tau_{\rm T}}=\pi B_{\nu}(T_{\rm s}).
\end{equation}
Note that the total transverse optical depth of this layer is
twice as large as $\tau_{\rm T}$ that characterizes the optical
depth of the hot zone above the underlying surface (from its one
side). Nevertheless, while presenting the results in this
section, we will use $\tau_{\rm T}$ for the convenience of a
comparison with the hot layer that has no underlying surface.

The solid blue lines in Fig.~\ref{fig:soft2.0} indicate the
computed spectra of the radiation produced in such a corona at
the same high-temperature plasma parameters as those in
Fig.~\ref{fig:ff}, but at different temperatures of the
underlying surface $kT_{\rm s}$. It can be seen that the
additional photons emitted by an underlying surface with
$kT_{\rm s} \ga 30$ eV changes radically the shape of the
produced radiation spectrum. The spectral slope (photon index)
at $h\nu\ga 1$ keV reaches $\alpha\simeq 2.2,$ whereas the slope
of the radiation spectrum for a corona without an underlying
surface (or at a temperature of the underlying surface $kT_{\rm
  s} \la 1$ eV) remains close to its standard value
$\alpha\simeq 1.6$ typical for the observed sources --- black
hole candidates in their canonical hard state. In the range 10
eV -- 1 keV the radiation spectrum for the corona without an
underlying surface is even flatter, $\alpha\simeq 1.2.$
\begin{figure*}[!t]
 \centering
 \includegraphics[width=0.69\linewidth]{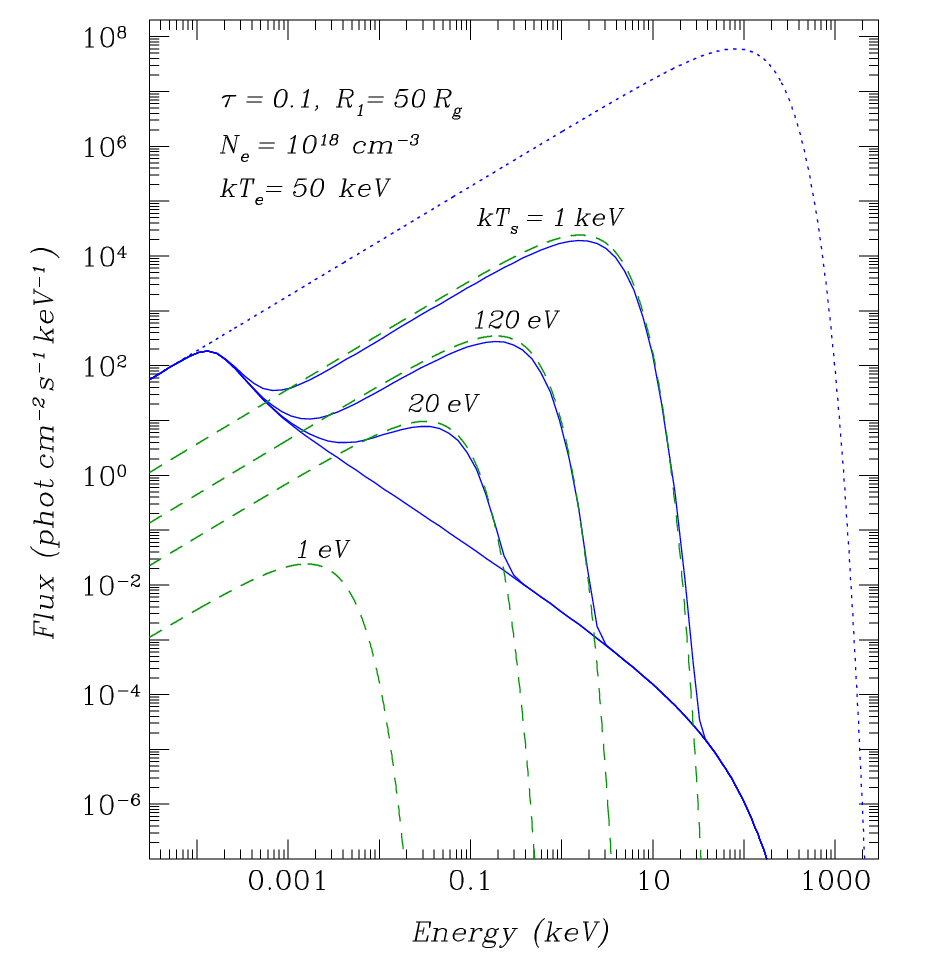}

\caption{\rm The same photon spectra as those in
  Fig.~\ref{fig:soft2.0}, but for the optical depth of the hot
  corona $\tau_{\rm T}=0.1$. The computations are presented only
  for the four temperatures of the underlying surface $kT_{\rm
    s}$ specified in the figure. The dashed green lines indicate
  the corresponding Planck spectra. It can be seen that at the
  lowest temperature, $kT_{\rm s}=1$ eV, the underlying-surface
  photons do not affect the radiation spectrum of the
  corona. The dotted blue line indicates the Planck spectrum
  corresponding to the corona temperature $kT_{\rm e}=50$
  keV. \label{fig:soft0.1}}
\end{figure*}

At lower energies, $h\nu\la 10$ eV, the radiation spectrum of
the corona has a single shape independent of the presence or the
temperature of external photons. Such a shape is typical for a
high-temperature plasma; its formation was explained when
examining Fig.~\ref{fig:ff}. Note that the range $h\nu\la 10$ eV
includes the infrared, optical, and near-ultraviolet ranges,
i.e., the external photons entering the inner disk zone should
have no effect on the observations of its radiation spectrum in
these ranges.

In contrast, at energies $10\ \mbox{eV}\ \la h\nu\la 1$ keV the
presence of external photons affects critically the shape of the
spectrum; even its type changes --- instead of the decreasing
one it becomes an increasing one, approaching the Rayleigh-Jeans
spectrum in slope, but only in slope --- the dashed green line
in Fig.~\ref{fig:soft0.1} indicates the real Planck radiation
spectrum corresponding to the underlying-surface temperature
$kT_{\rm s}=1$ keV. It is at this temperature that the greatest
changes are observed in the spectrum of the corona. It can be
seen that the ``hump'' in the spectrum of the corona is greatly
shifted toward higher energies relative to the Planck spectrum
and has a lower amplitude. This is a result of the Compton
scattering of Planck photons from the underlying surface by
high-temperature electrons of the corona.

To make sure that this is the case, we computed the radiation
spectra of an optically thin corona ($\tau_{\rm
  T}=0.1$)\footnote{The used equations were obtained formally in
  the approximation of a large optical depth for scattering
  ($\tau_{\rm T}\ga 1$). However, they lead to reasonable
  results in the limit of a small optical depth as well. In
  particular, this was shown when computing the distortions of
  the cosmic microwave background spectrum in the hot gas of
  galaxy clusters (Sunyaev and Zeldovich 1980; Sunyaev 1980;
  Zeldovich and Sunyaev 1982).} for four temperatures: $kT_{\rm
  s}=1000,\ 120,\ 20,\ \mbox{\rm and}\ 1$ eV. They are presented
in Fig.~\ref{fig:soft0.1} in comparison with the original Planck
spectra of the underlying surface.  As expected, in this case
the spectra of the corona in their central part differ little
from the corresponding Planck spectra, but at low and high
energies they are still transformed into the spectrum of the
intrinsic high-temperature plasma radiation. The spectrum
corresponding to the lowest temperature of the underlying
surface considered, $kT_{\rm s}=1$ eV, is an exception. In this
case, the photons emitted by it do not affect the radiation
spectrum of the corona.
\begin{figure*}[!t]
 \centering
 \includegraphics[width=0.69\linewidth]{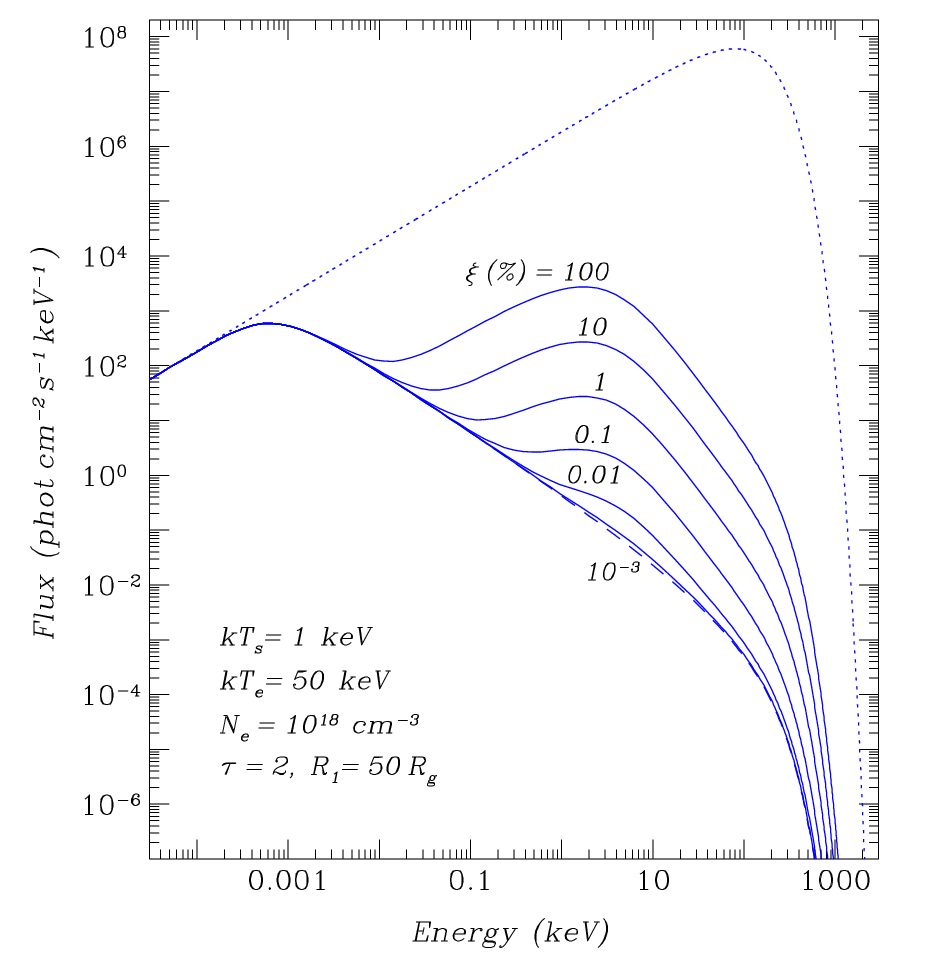}

\caption{\rm The same photon spectra as those in
  Fig.~\ref{fig:soft2.0} at the temperature of the underlying
  surface $kT_{\rm s}=1$ keV, but assuming that only some
  fraction $\xi$ of the photons emitted by it enter the hot
  layer (corona). The values of $\xi$ (in \%) are specified near
  the spectra. The dashed line indicates the spectrum of the
  corona without an underlying surface; the dotted line
  indicates the Planck spectrum corresponding to the corona
  temperature $kT_{\rm e}=50$ keV. The influence of the
  underlying surface disappears at
  $\xi\la10^{-5}$.\label{fig:soft-num}}
\end{figure*}

Similarly, in the case of a significant optical depth,
$\tau_{\rm T}=2$, illustrated in Fig.~\ref{fig:soft2.0}, as the
temperature of the underlying surface decreases below $kT_{\rm
  s}\sim 20-30$ eV, the external photons also cease to affect
tangibly the shape and intensity of the corona radiation
spectrum.  As in the previous case, this is initially associated
with the decrease in the number of blackbody photons as the
temperature decreases ($N_{\rm s}\sim T_{\rm s}^3$), but
subsequently, on reaching $kT_{1}\simeq (1/3)\,h\nu_1\la 0.1$
eV, the cause becomes different --- the external photons of such
low energies begin to be efficiently absorbed in the
bremsstrahlung processes in the hot plasma.

This is true for any external photons of such energies,
irrespective of their origin. In particular, for this reason,
the cyclotron photons produced in the inner disk zone should
have no influence on the forming radiation spectrum. The
magnetic induction in the accretion disk does not exceed $B\la
10^7$~G (Shakura and Sunyaev 1973) and, therefore, the energy of
the cyclotron photons is $h\nu_{\rm B}=h (eB/m_{\rm e}c)/2\pi
\la 0.12$ eV and they should be efficiently absorbed. This was
noted long ago by Shapiro et al. (1976). The aforesaid is partly
true for the synchrotron radiation as well, whose power maximum
occurs at an energy $h\nu_{\rm c}\simeq (3/2) h\nu_{\rm
  B}\,\gamma^2,$ where $\gamma$ is the Lorentz factor of
nonthermal electrons (Lang 1974). The overwhelming number of
them in the hot zone should have $\gamma\sim 3$. Therefore, even
at a magnetic induction $B\sim 10^7\ \mbox{\rm G}$ the energy of
the synchrotron photons is $h\nu_{\rm c}\la 1.5$ eV and they are
also actively absorbed. The possibility of the production of
synchrotron photons in accretion disks and their involvement in
Comptonization were studied in more detail by Wardzinski and
Zdziarski (2000), Veledina et al. (2011), and Dexter et
al. (2021).
\begin{figure*}[!t]
 \centering
 \includegraphics[width=0.69\linewidth]{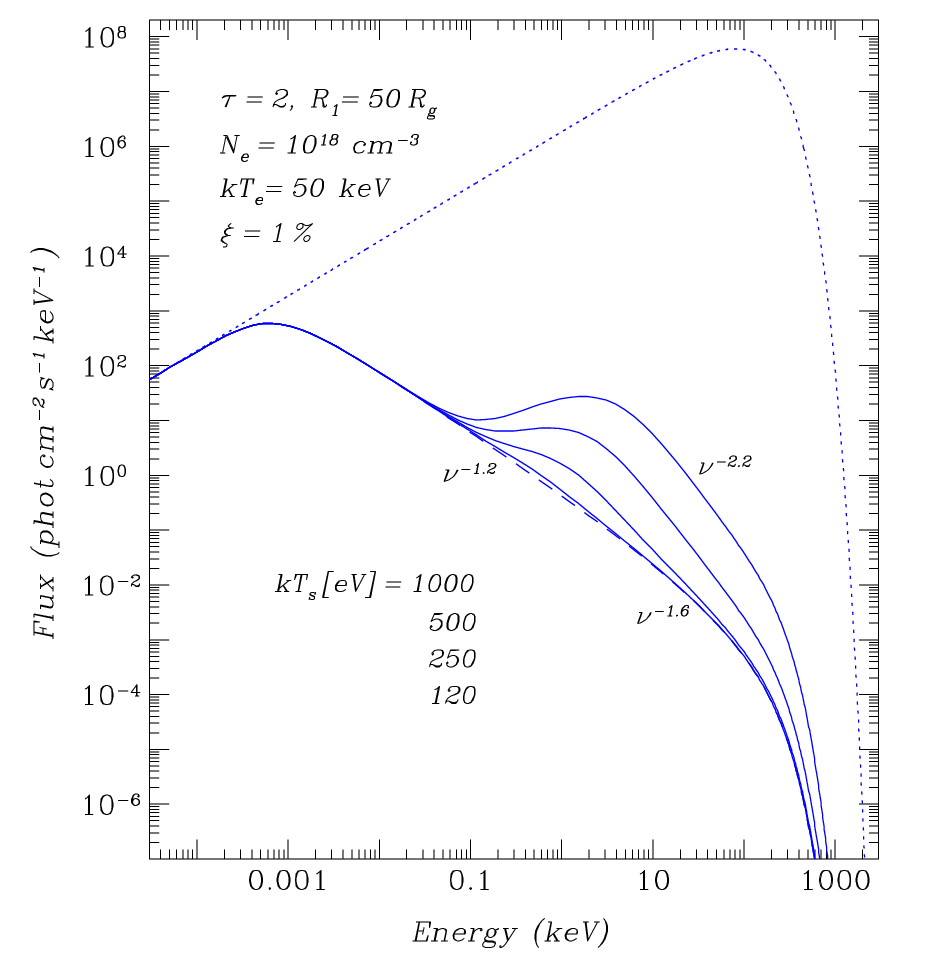}

\caption{\rm The same photon spectra as those in
  Fig.~\ref{fig:soft2.0}, but assuming that only 1\% of the
  photons emitted by the underlying surface enter the hot layer
  (corona). Various temperatures of the underlying surface
  $kT_{\rm s}$ (specified in the figure) are considered.  The
  dashed line indicates the spectrum of the corona without an
  underlying surface; the dotted line indicates the Planck
  spectrum corresponding to the corona temperature $kT_{\rm
    e}=50$ keV. The influence of the underlying surface
  disappears at $kT_{\rm s}\la 0.1$ keV.\label{fig:soft-1pc}}
\end{figure*}
\begin{figure*}[!t]
 \centering
 \includegraphics[width=0.70\linewidth]{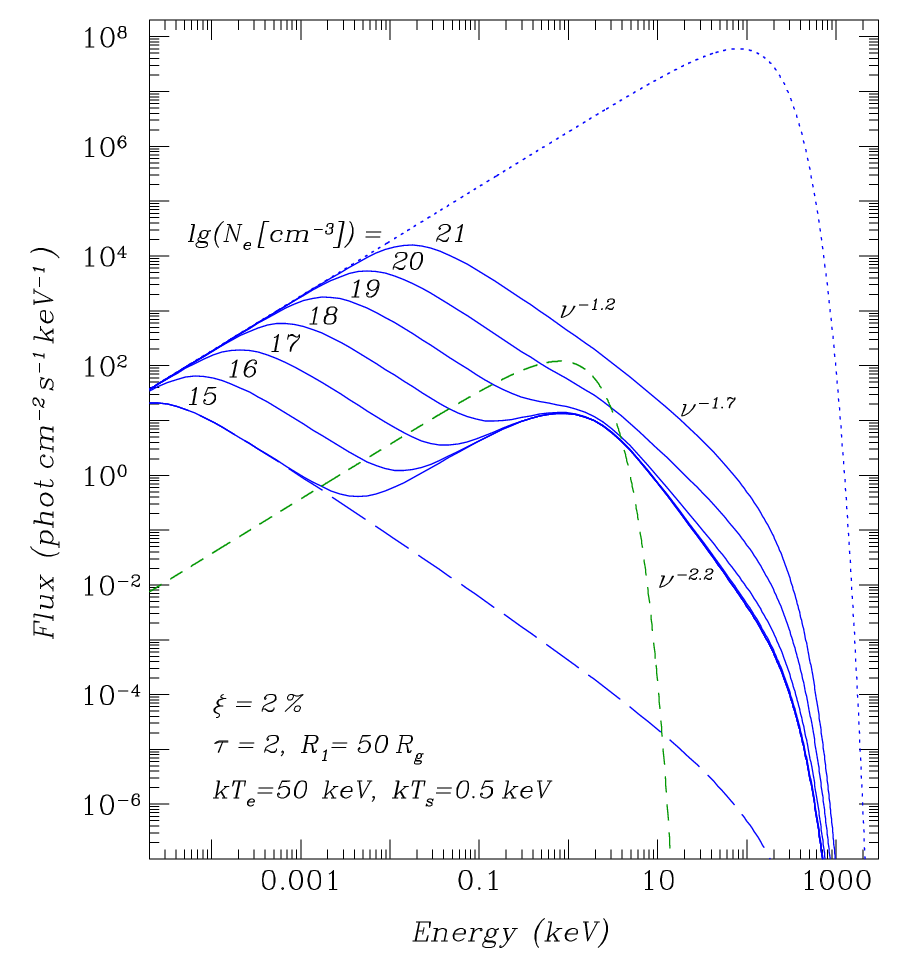}

\caption{\rm The same photon spectra as those in
  Fig.~\ref{fig:soft-1pc} forming in a hot corona with $kT_{\rm
    e}=50$ keV, $\tau_{\rm T}=2.0$, the underlying-surface
  temperature $T_{\rm s}=0.5$ keV, the fraction of the photons
  emitted by it that enter the hot layer $\xi=2$\%, and
  different electron densities $N_{\rm e}$ (solid blue
  lines). The short dashes indicate the spectrum of the
  underlying surface, the long dashes indicate the spectrum of
  the corona without an underlying surface for $N_{\rm
    e}=10^{15}\ \mbox{\rm cm}^{-3}$, and the dotted line
  indicates the Planck spectrum corresponding to the corona
  temperature. The underlying-surface photons subjected to
  Comptonization determine the X-ray spectrum for $N_{\rm
    e}\la 10^{19}\ \mbox{\rm cm}^{-3}.$ \label{fig:soft-den}}
\end{figure*}

In the picture of the corona considered the number of low-energy
photons penetrating into the hot layer is completely determined
by the underlying-surface temperature $kT_{\rm s}$. The area of
the surface through which they penetrate remains invariable and
maximally possible (equal to the area $S$ of the layer itself).
In real conditions, for example, when the photons enter the hot
disk zone from the outer cold region, their number can be
noticeably smaller. This possibility is investigated in
Fig.~\ref{fig:soft-num}. It is assumed that the temperature of
the underlying surface is $kT_{\rm s}=1$ keV, but only a
fraction $\xi\la 1$ of the photons emitted by it penetrate into
the hot layer. Each of the solid curves corresponds to the
computation for a specific value of $\xi$. The figure shows that
the photons of such a (fairly hot) underlying surface cease to
influence the shape of the spectrum for the high-temperature
layer only at $\xi\la 10^{-5}$.

To demonstrate how small this fraction is, suppose that the
photons from the cold disk region enter the hot zone through its
outer boundary (its end with an area $2\pi R_1\times 2H$); the
disk here should then have been unrealistically thin, $2H\simeq
\xi (\pi R_1^2)/(2\pi R_1)\la 7.5\ (\xi/10^{-5}) (R_1/50\,R_{\rm
  g})$~m. However, the photons entering the hot zone in this way
rapidly leave it and are able to change the forming radiation
spectrum only locally, near the outer boundary. The picture
where the photons from the cold disk regions enter the central
zone through a more or less uniform illumination of its surface
looks more realistic.
\begin{figure*}[!t]
 \centering
 \includegraphics[width=0.75\linewidth]{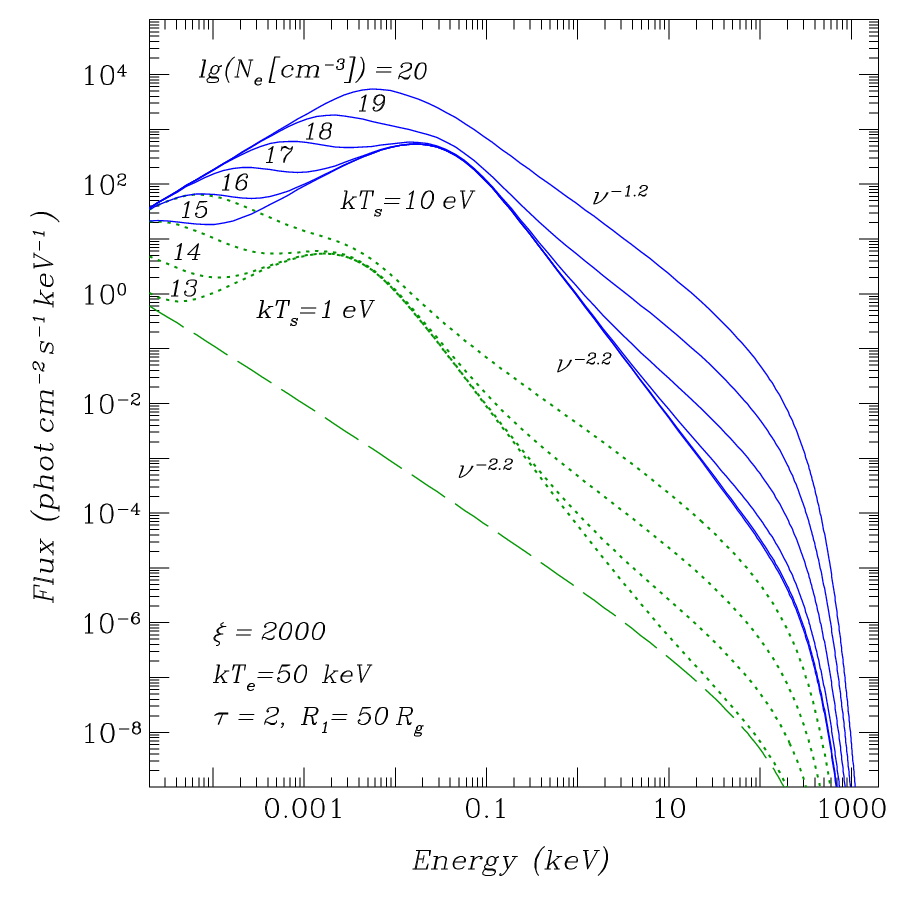}

\caption{\rm The same photon spectra for various electron
  densities $N_{\rm e}$ as those in Fig.~\ref{fig:soft-den}, but
  assuming the temperature of the underlying surface to be
  $kT_{\rm s}=10$ eV (blue solid lines) or $1$ eV (green dotted
  lines); in both cases, the number of photons emitted by it
  that enter the hot layer is a factor of 2000 larger than that
  in the corresponding blackbody spectrum ($\xi=2000$). The long
  dashes indicate the spectrum of the corona without an
  underlying surface (for $N_{\rm e}=10^{13}\ \mbox{\rm
    cm}^{-3}$).\label{fig:soft-den2}}
\end{figure*}

Figure~\ref{fig:soft-1pc}, like Fig.~\ref{fig:soft2.0},
investigates the dependence of the shape of the hot-layer
spectrum on the temperature of the underlying surface $T_{\rm
  s}$, but provided that only $\xi=1$\% of its blackbody photons
enter the zone.  The influence of additional photons,
irrespective of $T_{\rm s}$, manifests itself only at energies
$h\nu\ga30$ eV. At lower energies the hot-plasma radiation
spectrum is insensitive to such photons, while at $kT_{\rm s}\la
0.1$ keV the influence of external photons on the spectrum is
insignificant even at high energies. To some extent this is
because the number of photons in the blackbody spectrum
decreases rapidly as its temperature decreases. Note that the
number of blackbody photons with a specific temperature emitted
by the surface of a viscous accretion disk drops with $T_{\rm
  s}$ more slowly than is assumed in the figure. The reason is
that the disk surface temperature decreases with radius as
$T_0\,(R/R_0)^{-3/4}$ (see below), but at the same time the area
of its local surface increases as $2\pi R\,\Delta
R$. Accordingly, the number of photons that are emitted by the
outer disk ring with the temperature $T_{\rm s}$ and then enter
the hot zone is $N_{\rm s}(T_{\rm s})\sim T_{\rm s}^3 R\Delta R
\sim R^{-5/4}\sim T_{\rm s}^{5/3}$. For blackbody radiation from
a surface with an invariable area and a constant temperature, as
in Fig.~\ref{fig:soft-1pc}, the number of photons is $N_{\rm
  s}(T_{\rm s})\sim T_{\rm s}^3$.

From all of what has been said (and shown) it seems that the
X-ray spectrum of a high-temperature plasma can have a slope
typical for X-ray novae and other accreting black holes in their
hard state only in the absence of additional soft photons
entering the plasma layer from outside. Is this actually the
case?

First of all, let us examine what happens to the spectrum as the
plasma density decreases (but its optical depth is
retained). Obviously, the purely bremsstrahlung flux from the
plasma should decrease rapidly and the flux of all its intrinsic
radiation should decrease accordingly. Figure~\ref{fig:soft-den}
shows the photon spectra from a plasma layer with $kT_{\rm
  e}=50$ keV, $\tau_{\rm T}=2.0$, and different electron
densities $N_{\rm e}$. The layer is irradiated from one side by
the underlying surface with $kT_{\rm s}=0.5$ keV with the
efficiency (the fraction of the blackbody photons entering the
hot layer) $\xi=2$\%. It can be seen that for $N_{\rm e}\leq
10^{18}\ \mbox{\rm cm}^{-3}$ the underlying-surface photons
subjected to Comptonization in the hot plasma layer completely
determine the shape and normalization of the X-ray and soft
gamma-ray plasma spectrum. However, the photon index of the
power-law part of the spectrum, $\alpha\simeq 2.2$, exceeds
noticeably $\alpha\sim 1.6$ typical for the hard state of
accreting black holes. At energies smaller than $\la 10-100$ eV
we observe an intrinsic hot-plasma radiation spectrum with a
slope $\alpha\simeq 1.2$, whose normalization depends strongly
on the electron density. Finally, in the case of a high electron
density $N_{\rm e}\ga 10^{19}\ \mbox{\rm cm}^{-3}$ the external
photons become unimportant initially in the hard X-ray range and
subsequently in the entire energy range --- the intrinsic
spectrum of the hot plasma layer is observed everywhere. As has
already been said, this spectrum agrees most closely with the
canonical hard X-ray spectrum of accreting black holes. To
summarize, it can be said that the radiation spectrum of the
plasma layer depends strongly on the electron density and is
formed under conditions of complex interference between the
intrinsic radiation of the layer and the radiation of the seed
photons subjected to Comptonization that enter it from outside.

Figure~\ref{fig:soft-den2} gives more examples of the dependence
of the photon spectrum forming in the hot layer on the electron
density. Here the temperature of the underlying surface is taken
to be $kT_{\rm s}=1$ eV (the corresponding spectra are indicated
by the green dotted lines) and 10 eV (the spectra are indicated
by the blue solid lines). In both cases, the number of photons
in the corresponding Planck spectrum was increased by a factor
of 2000 ($\xi=2000$). The figure confirms the features in the
behavior of the radiation spectrum noted in
Fig.~\ref{fig:soft-den}, but for $kT_{\rm s}=10$ eV the
contribution of external photons in the X-ray and soft gamma-ray
range ceases to be felt much earlier --- even at an electron
density $N_{\rm e}\la 10^{16}\ \mbox{\rm cm}^{-3}$ (and at
$N_{\rm e}\la 10^{13}\ \mbox{\rm cm}^{-3}$ for $kT_{\rm s}=1$
eV). For such a density the dashed green line in the figure
indicates the radiation spectrum of the corona without an
underlying surface; it can be seen that the green dotted line
corresponding to this density merges with the spectrum indicated
by the dashed line at energies $h\nu\ga20-30$ keV.

It should be emphasized that the number of underlying-surface
photons in these examples was increased manyfold compared to
their number in the blackbody spectrum and probably exceeds
noticeably the number of soft photons capable of penetrating
into the central hot disk zone in any reasonable scenario.
Thus, Fig.~\ref{fig:soft-den2} confirms the above conclusion that
it is impossible to change the intrinsic plasma radiation
spectrum using external seed photons with energies $h\nu\la10$
eV and their Comptonization in a sufficiently dense, $N_{\rm
  e}\la 10^{17}\ \mbox{\rm cm}^{-3}$, plasma.

This is important, since it is difficult to completely avoid the
penetration of photons from the outer disk regions into the
central hot zone. At the same time, it follows from what has
been said above that if these are photons of moderately high
energies $h\nu\la 100$ eV, then they will not change noticeably
the forming spectrum. Thus, the surface temperature of the cold
outer disk in the immediate vicinity of its boundary with the
hot zone (at radii $R\ga R_1$), where these photons are actually
emitted, turns out to be one of the most important factors
controlling the transition between the hard and soft states of
accreting black holes (along with the size $R_1$ of the hot zone
and the radius $R_1^*<R_1$ of the fairly dense part of the hot
zone that determines the normalization of its hard flux).
\begin{figure*}[!t]
\centering
\includegraphics[width=0.76\linewidth]{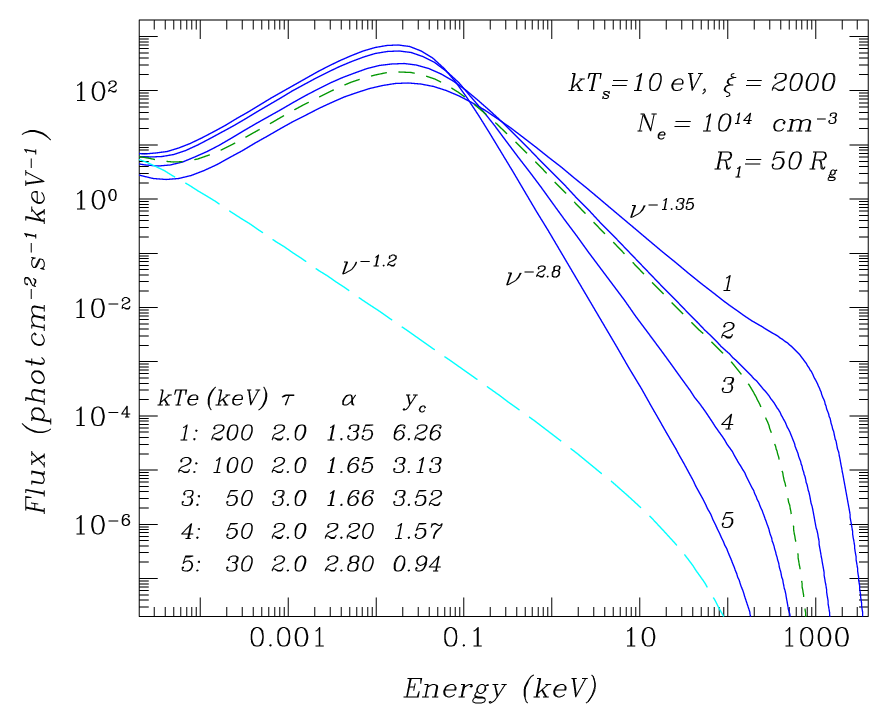}

\caption{\rm The photon spectra of a hot corona with the
  underlying-surface temperature $kT_{\rm s}=10$ eV and the
  electron density $N_{\rm e}=10^{14}\ \mbox{\rm cm}^{-3}$ that
  demonstrate the dependence of the shape and slope on the
  electron temperature $kT_{\rm e}$ (blue solid lines) and the
  optical depth $\tau$ (green dashes). This resembles that shown
  in Fig.~\ref{fig:akt}, but attention is drawn to the
  Comptonization of external (seed) underlying-surface
  photons. The number of external photons entering the hot layer
  is a factor of 2000 larger than that in the blackbody spectrum
  corresponding to the underlying-surface temperature
  ($\xi=2000$). The long dashes indicate the spectrum of a
  corona with $kT_{\rm e}=30$ keV without an underlying
  surface. The normalization is determined by the radius
  $R_1=50\ R_{\rm g}$ and the distance $d=2.5$
  kpc.\label{fig:soft-alpha}}
\end{figure*}
\section*{THE PHOTON INDEX OF THE SPECTRA}
\noindent
Figures \ref{fig:soft2.0}--\ref{fig:soft-den2} demonstrate that
when the soft seed photons entering the hot plasma layer from
outside are Comptonized, an additional hard radiation component
with a photon index $\alpha\sim 2.2$ independent of the
temperature of the soft photons and their flux is formed. We get
the impression that such a spectral slope is some general unique
property of a high-temperature and fairly dense plasma. Of
course, this is not the case. The value of $\alpha$ is
determined by the parameters of the plasma layer $kT_{\rm e}$
and $\tau_{\rm T}$ in much the same way as the photon index of
the intrinsic radiation from the plasma layer is determined by
them (as shown in Fig.~\ref{fig:akt}).

Figure~\ref{fig:soft-alpha} illustrates the dependence of the
corona spectrum on these parameters. The electron density is
assumed to be low, $N_{\rm e}=10^{14}\ \mbox{\rm cm}^{-3}$, to
minimize the intrinsic radiation of the plasma layer. Actually,
the figure shows the dependence of the slope of the spectrum
forming during the Comptonization of low-energy seed photons in
the hot plasma layer that was studied in detail by Sunyaev and
Titarchuk (1980). Spectra 1--5 in Fig.~\ref{fig:soft-alpha} are
characterized by photon indices in the range $\alpha=1.35-2.80$.
The values of the photon index are specified for each presented
spectrum, along with the values of the Compton parameter $y_{\rm
  C}=(4kT_{\rm e}/m_{\rm e}c^2)\,\tau_{\rm T}^2$. It can be seen
that $y_{\rm C}$ changes in the range from 0.94 to 6.26. A
radiation spectrum with a photon index $\alpha\simeq1.6-1.7$
close to the canonical one is formed at a temperature $kT_{\rm
  e}\simeq 100$ keV and the optical depth $\tau_{\rm T}=2.0$
(spectrum 2) or at $kT_{\rm e}\simeq 50$ keV and $\tau_{\rm
  T}=3.0$ (spectrum 3). Naturally, different temperatures lead
to different characteristic energies of the high-energy cutoff
in the spectrum, but the power-law parts of the spectra
approximately coincide. The long dashes in the figure indicate
the intrinsic (formed without external photons) radiation
spectrum of the hot layer with $kT_{\rm e}=30$ keV and
$\tau_{\rm T}=2.0$, the same as those for spectrum 5. The photon
index of this spectrum is $\alpha=1.2$. Note that the difference
in spectral normalization in this figure at very low energies
$h\nu\la0.1$ eV is related to the weak $\sim T_{\rm e}^{-1/2}$,
but present temperature dependence of the bremsstrahlung flux
from the plasma layer.

Figure~\ref{fig:soft-alpha} confirms that a radiation spectrum
with a slope (photon index) $\alpha\simeq1.6-1.7$ and an
exponential cutoff energy $E_{\rm c}\sim 50-100$ keV typical for
the hard canonical state of accreting black holes can be
produced through the penetration of a large number of seed
photons into the hot plasma layer, although this will require a
fairly fine tuning of plasma parameters. It is even more
difficult to produce such a canonical spectrum from external
photons against the background of intense intrinsic plasma
radiation. This is because the shapes of the intrinsic and
induced radiation spectra are different and, most importantly,
because their dependences on plasma parameters are different.
\begin{figure*}[!t]
\centering
\includegraphics[width=0.78\linewidth]{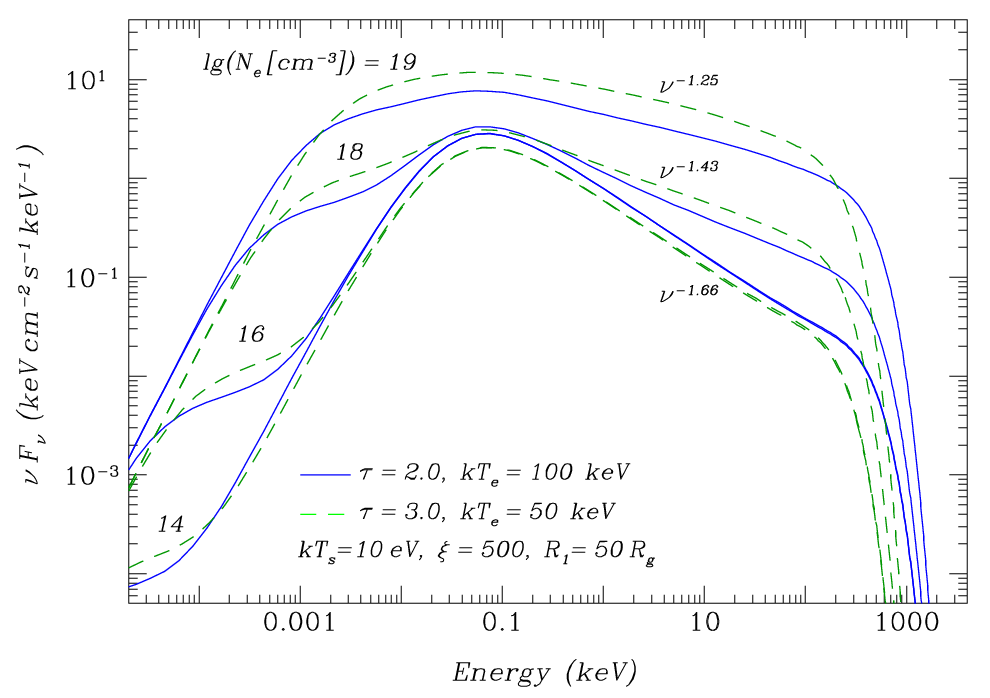}

\caption{\rm The $N_{\rm e}$ dependence of the radiation
  spectrum for a hot corona with the optical depth $\tau_{\rm
    T}$ and temperature $kT_{\rm e}$ at which the best agreement
  with the canonical spectrum of black holes was achieved in the
  case of a low electron density ($N_{\rm e}=10^{14}\ \mbox{\rm
    cm}^{-3}$), i.e., predominant Comptonization of seed photons
  from the underlying surface (see Fig.~\ref{fig:soft-alpha}).
  The temperature of the underlying surface is $kT_{\rm s} = 10$
  eV; the number of photons entering the hot layer is a factor
  of 500 larger than that in the blackbody spectrum
  corresponding to its temperature ($\xi=500$). As $N_{\rm e}$
  increases to realistic values 
  ($10^{16},\ 10^{18},\ \mbox{\rm and}\ 10^{19}\ \mbox{\rm
    cm}^{-3}$), the spectral slope changes noticeably. Although
  the spectra themselves are presented in the form
  $\nu\,F_{\nu}$, the photon indices are specified near them to
  avoid confusion. The normalization of the spectra is
  determined by the radius $R_1=50\ R_{\rm g}$ and the distance
  to the source $d=2.5$ kpc.\label{fig:soft-alphac}}
\end{figure*}

Figure~\ref{fig:soft-alphac} shows how the shape and slope of
two radiation spectra, which corresponded most closely to the
canonical X-ray spectrum of accreting black holes in their hard
state under conditions of predominant Comptonization of seed
photons (at a very low density, $N_{\rm e}=10^{14}\ \mbox{\rm
  cm}^{-3}$, see Fig.~\ref{fig:soft-alpha}), change as the
electron density increases to realistic values. For clarity, the
radiation spectra $\nu\,F_{\nu}$ rather than the photon spectra
$F_{\nu}$, as in previous figures, are presented.  The spectrum
computed for the optical depth $\tau_{\rm T}=2.0$ and the
temperature $kT_{\rm e}=100$ keV indicated in the figure by the
blue solid lines becomes a power-law one with a photon index
$\alpha\simeq 1.43$ (in the range 0.1--200 keV) as the density
rises to $N_{\rm e}=10^{18}\ \mbox{\rm cm}^{-3}$; as the density
rises further to $N_{\rm e}=10^{19}\ \mbox{\rm cm}^{-3}$, its
slope decreases to $\alpha\simeq 1.25.$ This is accompanied by a
noticeable (by two orders of magnitude) change in the hard X-ray
flux. Even more interesting transformations occur at $h\nu\la10$
eV, where the Rayleigh-Jeans spectrum turns into a gently
sloping radiation spectrum with a photon index $\alpha\sim 0.75$
as the density rises. At $N_{\rm e}\ga 10^{19}\ \mbox{\rm
  cm}^{-3}$ the Comptonization of the seed photons entering the
plasma layer from outside ceases to affect noticeably the shape
and normalization of the spectrum (at least at the (far from
small) number of these photons adopted here).

The radiation spectrum computed for $\tau_{\rm T}=3.0$ and
$kT_{\rm e}=50$ keV and indicated in Fig.~\ref{fig:soft-alphac}
by the green dashed lines changes in a similar way. The
radiation flux in this spectrum increases even more rapidly and
dramatically as the density rises.

The main conclusion that follows from this figure is that at
realistic densities of the plasma layer, $N_{\rm
  e}=10^{16}-10^{19}\ \mbox{\rm cm}^{-3}$, the radiation
spectrum forming in it due to Comptonization (its normalization,
the slope of the power-law part, the shape of the exponential
cutoff, and the shape of the low-frequency radiation spectrum)
depends not only on the Thomson optical depth of the layer and
its temperature, as is commonly assumed, but also on its
density.

At realistic fluxes of the additional photons that enter the
high-temperature plasma layer (the central zone of the accretion
disk), for example, from its outer cold zone, and are then
involved in Comptonization, their influence on the forming
radiation spectrum ceases to be felt (against the background of
Comptonized intrinsic plasma bremsstrahlung photons) already at
electron densities $N_{\rm e}\ga 10^{19}\ \mbox{\rm
  cm}^{-3}$. For the seed photons to exert some influence on the
X-ray spectrum even at lower densities, their number entering
the plasma should be very large. Previously, we have already
noted that, for the same reason, these should not be too
low-frequency photons.

On this basis, it seems that a hard radiation spectrum of X-ray
novae and other accreting black holes with a slope typical for
their canonical hard state can be realized most easily and
naturally under conditions when the intrinsic Comptonized
radiation of the hot plasma in the central disk zone undistorted
by external photons is observed. Of course, it is necessary that
this radiation be also able to ensure an efficient removal of
the heat being released in the disk during the viscous
dissipation of gravitational energy of accreting matter.
\begin{figure}[t]
  \centering
  \includegraphics[width=1.02\linewidth]{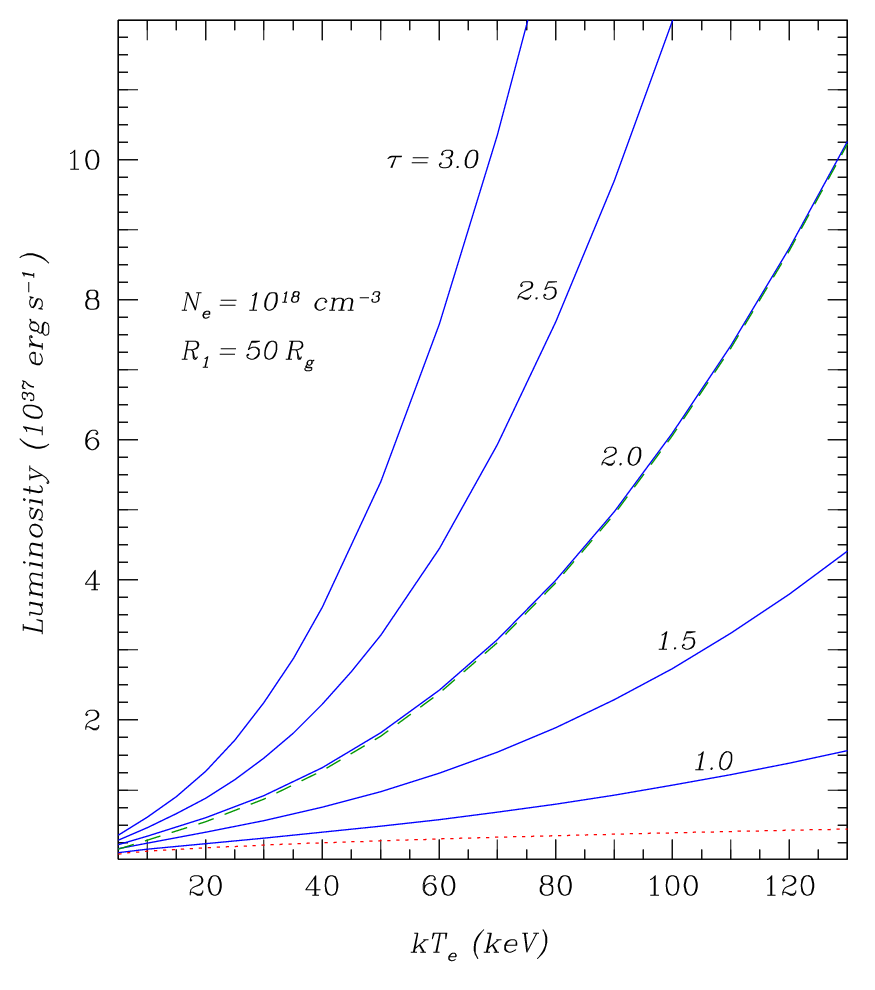}
  \caption{\rm The luminosity of the hot accretion disk zone as
    a function of the electron temperature $kT_{\rm e}$ and
    transverse Thomson optical depth $\tau_{\rm T}$. The solid
    lines indicate the luminosity in the entire energy range;
    the dashed line (for $\tau_{\rm T}=2.0$) indicates the
    luminosity at energies $h\nu > 1$ keV. Other disk
    parameters, $N_{\rm e}$ and $R_1$, were taken to be the same
    as those in Fig.~\ref{fig:ff}. The dotted line indicates the
    purely bremsstrahlung luminosity (for a layer with
    $\tau_{\rm T}=1.0$). \label{fig:alum}}
\end{figure}

\section*{THE LUMINOSITY OF THE HOT ZONE}
\noindent
Although the shape of the radiation spectrum for the hot disk
zone is determined by its physical parameters, the electron
temperature $kT_{\rm e}$, the density $N_{\rm e},$ and the
optical depth $\tau_{\rm T}$, the plasma properties in this zone
and the characteristics of its radiation spectrum are actually
specified by the accretion rate or (as long as there is no
advection) its total luminosity. The temperature $kT_{\rm e},$
the optical depth $\tau_{\rm T},$ and the radius $R_1$ (and the
fraction $\xi$ of the soft photons penetrating into the inner
zone related to it) are adjusted so that the X-ray luminosity of
this zone $L_{\rm X}$ corresponds to the release of
gravitational energy by the accreting matter in it.

The solid lines in Fig.~\ref{fig:alum} indicate the luminosity
of the hot zone as a function of $kT_{\rm e}$ and $\tau_{\rm
  T}$. The electron density, $N_{\rm e}= 10^{18}\ \mbox{\rm
  cm}^{-3}$, and the outer radius of the zone, $R_1=50\ R_{\rm
  g},$ were fixed. These parameters determine the absolute
luminosity
\begin{equation}\label{eq:lum}
L_{\rm H}=2\pi\,S(R_1)\int^{\infty}_0 h\nu\,F_{\nu}(\nu)\,d\,\nu,
\end{equation}
$R_1$ enters into the normalization explicitly (via the area
$S=\pi\,[R_1^2-R_0^2])$, while $N_{\rm e}$ enters into it
approximately ($\sim N_{\rm e}$, see the discussion above, where
it was shown that the amplitude, but not the shape of the hard
part of the hot-disk radiation spectrum, which actually
determines the luminosity, depends on the density). In the
formula for the luminosity the coefficient 2 on the right-hand
side takes into account the radiation from the upper and lower
disk surfaces, while $\pi$ takes into account the integration
over the radiation escape angle. On this basis, it is easy to
recalculate the luminosity presented in Fig.~\ref{fig:alum} for
the hot zone of the needed size and density. However, it should
be kept in mind that although, to a first approximation, the
outer radius of the hot zone, $R_1$ determines only its surface
area, in fact, it is implicitly also responsible for the
fraction and temperature $kT_{\rm s}$ of the external photons
penetrating into the hot zone, and these photons, as shown
above, are able to radically change its radiation spectrum and,
according, luminosity.

The luminosity of the hot accretion disk zone for the
temperature, $kT_{\rm e}=50$ keV, the optical depth, $\tau_{\rm
  T}=2$, and the outer radius, $R_1=50\ R_{\rm g}$, adopted by
us as typical ones, approaches $L_{\rm H}\simeq (1.5-2)\times
10^{37}\ \mbox{\rm erg s}^{-1}$ in good agreement with the X-ray
luminosities of the sources Cyg~X-1, 1E~1740.7-2942, and
GRS~1758-258 in their hard state considered above.

It should be emphasized that although Fig.~\ref{fig:alum}
presents the luminosity $L_{\rm H}$ in the entire energy range,
it differs little from the X-ray luminosity $L_{\rm X}$. This is
illustrated for $\tau_{\rm T}=2$ by the dashed (green) curve
that indicates the luminosity at $h\nu\geq 1$ keV. The dotted
(red) line in the figure indicates for $\tau_{\rm T}=1$ the
purely bremsstrahlung luminosity of a layer with a given
temperature and density (Lang 1974):
\begin{equation}\label{eq:lumff}
  L_{\rm ff}\simeq1.43\times10^{-27}\, N_{\rm e}\,T_{\rm
  e}^{1/2}\,(\tau_{\rm T}/\sigma_{\rm T})\,S\ \mbox{erg
    s}^{-1}.
\end{equation}
For larger $\tau_{\rm T}$ the luminosity $L_{\rm ff}$ increases
proportionally. It can be seen that the Compton processes 
increase the luminosity of the layer manyfold; the
higher its temperature and the larger its optical depth,
the more dramatic this increase.
\begin{figure}[t]
  \centering
  \includegraphics[width=1.02\linewidth]{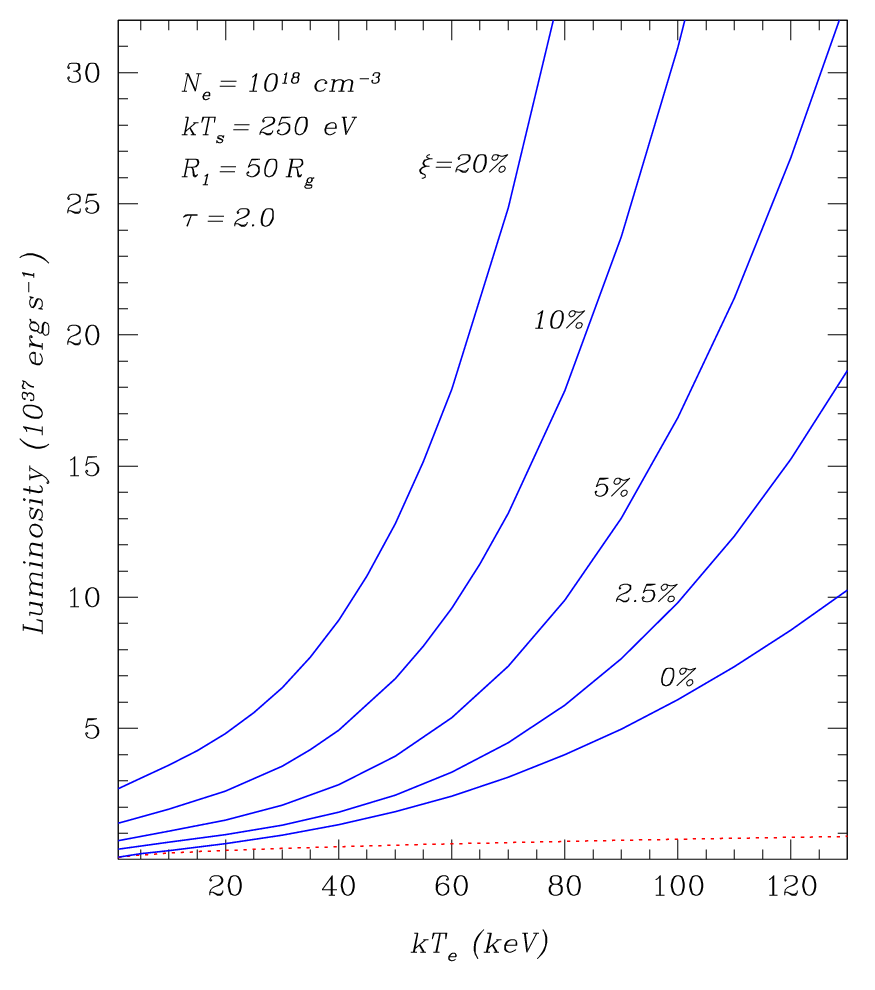}
  \caption{\rm The luminosity of the hot accretion disk zone as
    a function of the electron temperature $kT_{\rm e}$ in the
    corona model.  Other parameters, $\tau_{\rm T}$, $N_{\rm
      e}$, and $R_1$, were taken to be the same as those in
    Fig.~\ref{fig:ff}. The temperature of the underlying surface
    is $kT_{\rm s}=0.25$ keV. The solid lines indicate the
    luminosity under various assumptions about the fraction
    $\xi$ of the underlying-surface photons entering the hot
    layer; the dotted line indicates the purely bremsstrahlung
    luminosity (for a layer with $\tau_{\rm
      T}=2.0$).\label{fig:alums}}
\end{figure}

The luminosity of the hot plasma layer increases even more
dramatically if additional photons of moderately high energies
being involved in Comptonization enter it from
outside. Figure~\ref{fig:alums} presents the computed luminosity
of such a layer in the corona model as a function of the
electron temperature $kT_{\rm e}$ (as in Fig.~\ref{fig:alum},
but at a fixed transverse optical depth, $\tau_{\rm
  T}=2.0$). The temperature of the underlying surface was also
fixed at $kT_{\rm s}=0.25$ keV, but the fraction $\xi$ of the
underlying-surface photons entering the hot zone (in percent of
their number in the Planck spectrum) varied. It can be seen that
even an apparently small ($\xi\sim 2-5$\%) number of such
photons increases the luminosity of the plasma layer manyfold
compared to its intrinsic luminosity (let alone its purely
bremsstrahlung luminosity). The hotter the plasma in this layer,
the more efficient the Comptonization. Note that the external
photons entering the hot layer bring some initial luminosity
$L_{\rm s}$. Within the corona model this luminosity is $L_{\rm
  s}=2\xi\,S(R_1)\ \sigma T_{\rm s}^4,$ where $\sigma\simeq
5.67\times 10^{-5}\ \mbox{\rm erg cm}^{-2} \mbox{\rm s}^{-1}
\mbox{\rm K}^{-4}$
is the Stefan-Boltzmann constant. The contribution of $L_{\rm
  s}$ to the total luminosity of the plasma layer $L_{\rm H}$
can be seen in the left part of the figure at low temperatures
$kT_{\rm e}\la 10$ keV, where the Comptonization is not yet very
efficient. Obviously, the luminosity of the external photons in
the cases considered exceeds the plasma bremsstrahlung
luminosity several-fold.

Knowing the luminosity of the central hot zone
and considering a viscous accretion disk without advection, we
can estimate the total accretion rate $\dot{M}$ and the
luminosity $L_{\rm O}$ of the outer cold, optically
opaque disk regions. The radiation flux $Q_{\mathrm vis}(R)$ from
unit disk surface related to the viscous energy dissipation in
its interiors is given by the expression 
(Shakura and Sunyaev 1973)
\begin{equation}\label{eq:dbbqv}
  Q_{\mathrm vis}=\frac{3}{8\,\pi}\,\frac{GM\dot{M}}{R^{3}}\
  \phi(R/R_0)\simeq\frac{\dot{M}c^2}{16\pi\,R^2}
  \left(\frac{R_0}{R}\right),
\end{equation}
where $\phi(r)=1-r^{-1/2}$, $R_0=3\,R_{\rm g}=6GM/c^2$, and
$r=R/R_0$ (for simplicity, the black hole is assumed to be a
Schwarzschild one). The last part of this expression corresponds
to the limit $R\gg R_0$. Integrating the middle part of
Eq.~(\ref{eq:dbbqv}) over the area of the upper and lower disk
surfaces within the range from $R_0$ to $R_1$, we find the
luminosity of its hot inner zone
\begin{equation}\label{eq:lumin}
L_{\rm H} =
\frac{\dot{M}c^2}{12}\left[1-3\left(\frac{R_0}{R_1}\right)+
2\left(\frac{R_0}{R_1}\right)^{3/2}\right].
\end{equation}
Obviously, the luminosity of the entire disk is $L_{\rm
  d}=\dot{M}c^2/12$. For a black hole with $M=10\ M_{\odot}$ the
inner zone with the outer radius $R_1=50\ R_{\rm g}$ has a
luminosity $L_{\rm H}\simeq L_{\rm X}=1\times 10^{37}\ \mbox{\rm
  erg s}^{-1}$ if the accretion rate is $\dot{M}\simeq
1.6\times10^{17}\ \mbox{g s}^{-1}\simeq 2.5\times
10^{-9}\ M_{\odot}\ \mbox{\rm yr}^{-1}.$ At the same time, the
luminosity of the outer disk region is an order of magnitude
lower, $L_{\rm O}\simeq 1.8\times 10^{36}\ \mbox{\rm erg
  s}^{-1}\simeq 460\ L_{\odot}$.
\section*{OPTICAL OBSERVATIONS}
\noindent
One of the consequences of the results of our computations
presented in Figs.~\ref{fig:ff} and \ref{fig:cygx1} is the
conclusion about a natural extension of the hard X-ray spectrum
observed from accreting black holes that is formed in the
high-temperature plasma of the central zone of their accretion
disk to the near-infrared, optical, and long-wavelength
ultraviolet ranges. The spectrum extends as a power law; during
the hard state of the sources its shape changes only at energies
below $h\nu\la 0.3$ eV, where it transforms into the
Rayleigh-Jeans spectrum corresponding to the plasma temperature
in this zone. A single power-law spectrum could be observed from
the sources in this state under favorable conditions in both
ultraviolet and soft X-ray ranges if it were not for the
distortions related to strong interstellar absorption typical
for these ranges.

During the soft or two-component states of the sources the
intrinsic radiation spectrum of the high-temperature plasma in
the central disk zone also reaches the optical and infrared
ranges. However, (1) its slope will no longer coincide with the
slope of the X-ray spectrum forming under the Comptonization of
not only the intrinsic bremsstrahlung photons, but also the
additional photons entering this zone from the cold outer disk
regions (or from the cold underlying surface appearing under the
hot plasma layer) and (2) this radiation, as a rule, will be
noticeably weaker than the radiation from the outer disk regions
because of the dramatic decrease in the size of the central zone
(see the next section).

The conclusion about the production of optical and infrared
(OIR) radiation from accreting black holes in the hot central
disk zone partly contradicts the existing view (Shakura and
Sunyaev 1973; Lyuty and Sunyaev 1976) that this radiation is
associated with the outer disk regions (or the irradiation of
the surface of the optical companion star by X-rays from the
central zone). Nevertheless, Grebenev et al. (2013, 2014, 2016,
2020) have previously shown that the observations of some X-ray
novae in their hard state directly point to a common origin of
their OIR and X-ray radiations. Moreover, Grebenev et al. (2016)
demonstrated that even during the high (two-component) state of
the X-ray nova MAXI~J1828-249 its OIR radiation was described by
the extension of the hard X-ray spectrum.  Although the
radiation from the outer disk regions was observed in soft
X-rays, it did not contribute to the OIR spectrum even with
taking into account the possible irradiation by X-rays from the
center. The outbursts of X-ray novae are revealing because,
first, they attract widespread attention of observers and are
often accompanied by simultaneous observations in various
spectral ranges and, second, the X-ray novae are low-mass binary
systems and, accordingly, the OIR observations of their
accretion disk, as a rule, are not made difficult by the
contribution of the optical radiation from the companion star.
\begin{figure*}[!t]
  \centering
  \includegraphics[scale=0.69]{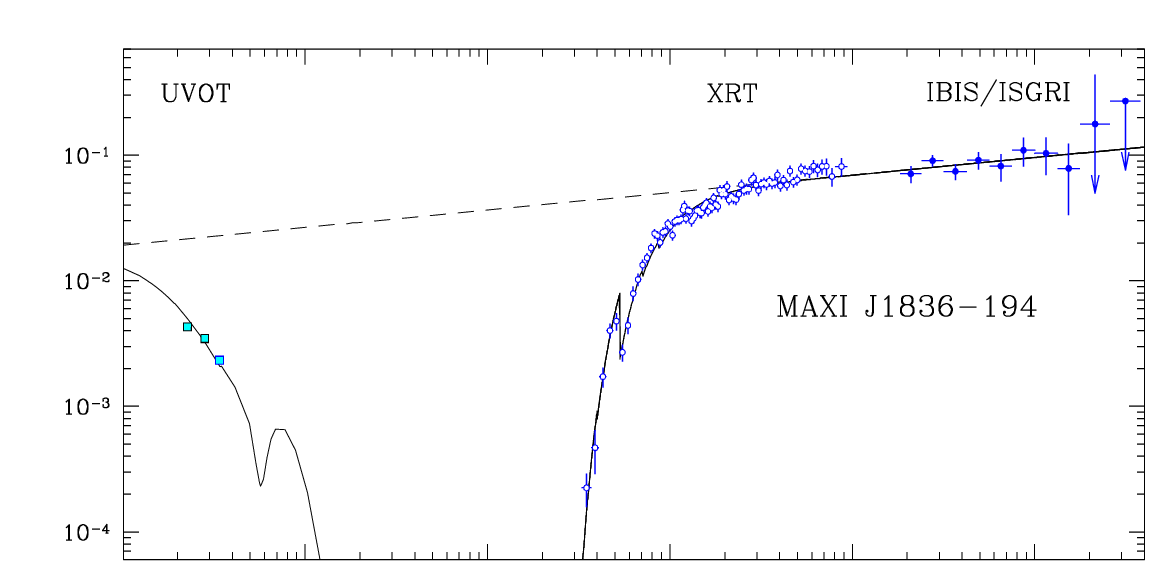}
  \includegraphics[scale=0.69]{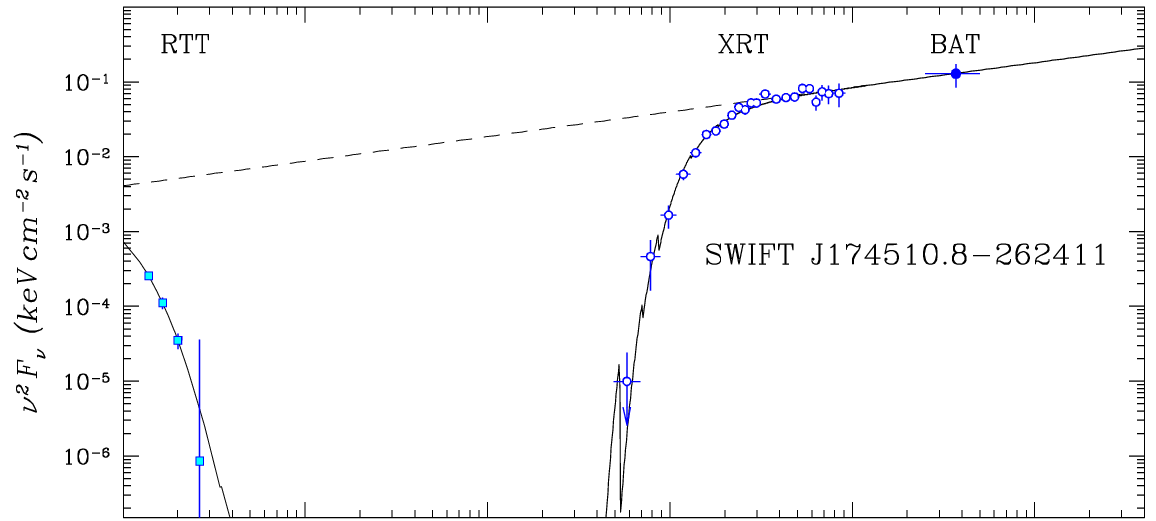}
  \includegraphics[scale=0.69]{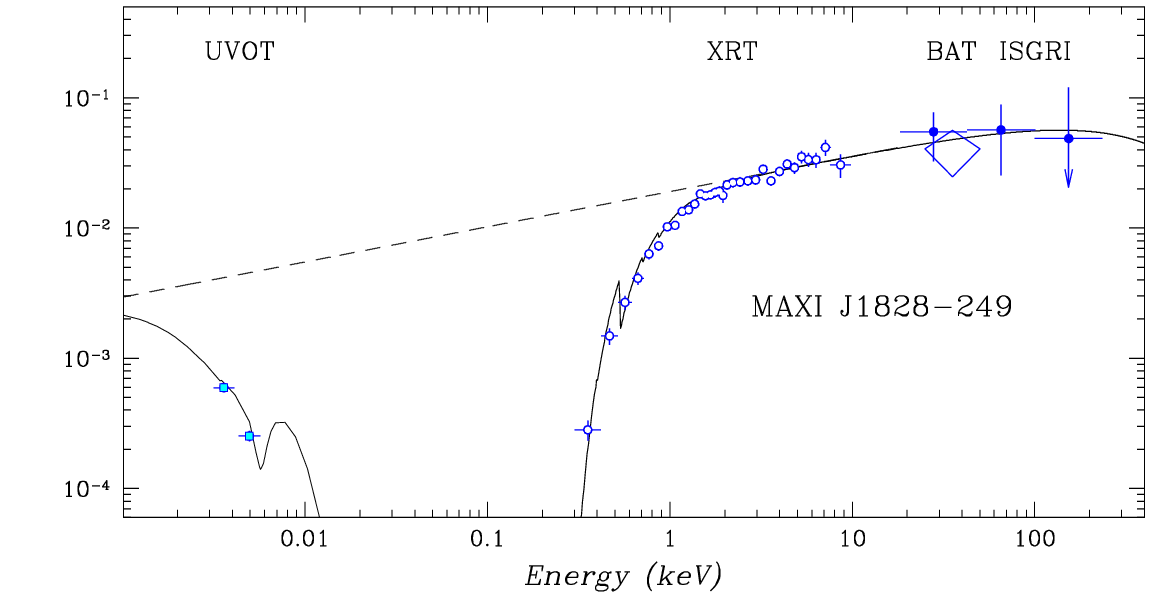}

\caption{\rm Broadband radiation spectra for the X-ray novae
  MAXI J1836-194, SWIFT J174510.8-262411, and MAXI J1828-249
  during their hard spectral state. The measurements were
  performed by the SWIFT (BAT, XRT, and UVOT) and INTEGRAL
  (IBIS/ISGRI) observatories as well as by the optical RTT-150
  telescope during their 2011--2012 (Grebenev et al. 2013),
  2012--2013 (Grebenev et al. 2014), and 2013--2014 (Grebenev et
  al. 2016) outbursts. It can be seen that after their
  correction for absorption, all of the observational data from
  the infrared and optical to hard X-ray ranges can be fitted by
  a single power law.\label{fig:oir-xray}}
\vspace{-0mm}
\end{figure*}

Figure~\ref{fig:oir-xray} presents the broadband radiation
spectra for three X-ray novae, MAXI J1836-194, SWIFT
J174510.8-262411, and MAXI J1828-249, measured during their hard
state at the decaying phase of their 2011--2012, 2012--2013, and
2013--2014 outbursts, respectively. We used data from the BAT,
XRT, and UVOT telescopes onboard the SWIFT orbital observatory,
the IBIS/ISGRI telescope onboard the INTEGRAL observatory, and
the ground-based optical RTT-150 telescope. It can be seen that
the spectra of all these sources are severely distorted by
interstellar absorption; nevertheless, they can be successfully
fitted by a simple power law in the entire energy range
considered (from 1 eV to 400 keV). The best-fit parameters for
these spectra are given in Table~\ref{table:oir-xray}.
\begin{table}[t]
 \caption{Parameters of the radiation spectra for the three
   X-ray novae in the hard state shown in
   Fig.~\ref{fig:oir-xray}.\label{table:oir-xray}}

 \vspace{-2mm}

\begin{center} \small
  \begin{tabular}{l|c|c|c}\hline\hline
  Source&$\alpha$\aa&$N_{\rm H}$\bb& Reference\cc\\
  & & &  \\ [-3mm]
  \hline
  & & &  \\ [-2mm] 
MAXI J1836-194        &1.86&0.29&2013\\
SWIFT J174510.8-262411&1.67&1.21&2014\\ 
MAXI J1828-249        &1.73&0.23&2016\\ 
\hline
\multicolumn{4}{l}{}\\ [-1mm]
\multicolumn{4}{l}{\aa\ The photon index.}\\
\multicolumn{4}{l}{\bb\ The hydrogen column density,
  $10^{22}\ \mbox{\rm cm}^{-2}$.}\\ 
\multicolumn{4}{l}{\cc\ Grebenev et al. (2013, 2014, 2016).}\\
\end{tabular}
\end{center}
\vspace{-5mm} 
\end{table}
\begin{figure*}[!t]
  \centering
  \includegraphics[scale=0.92]{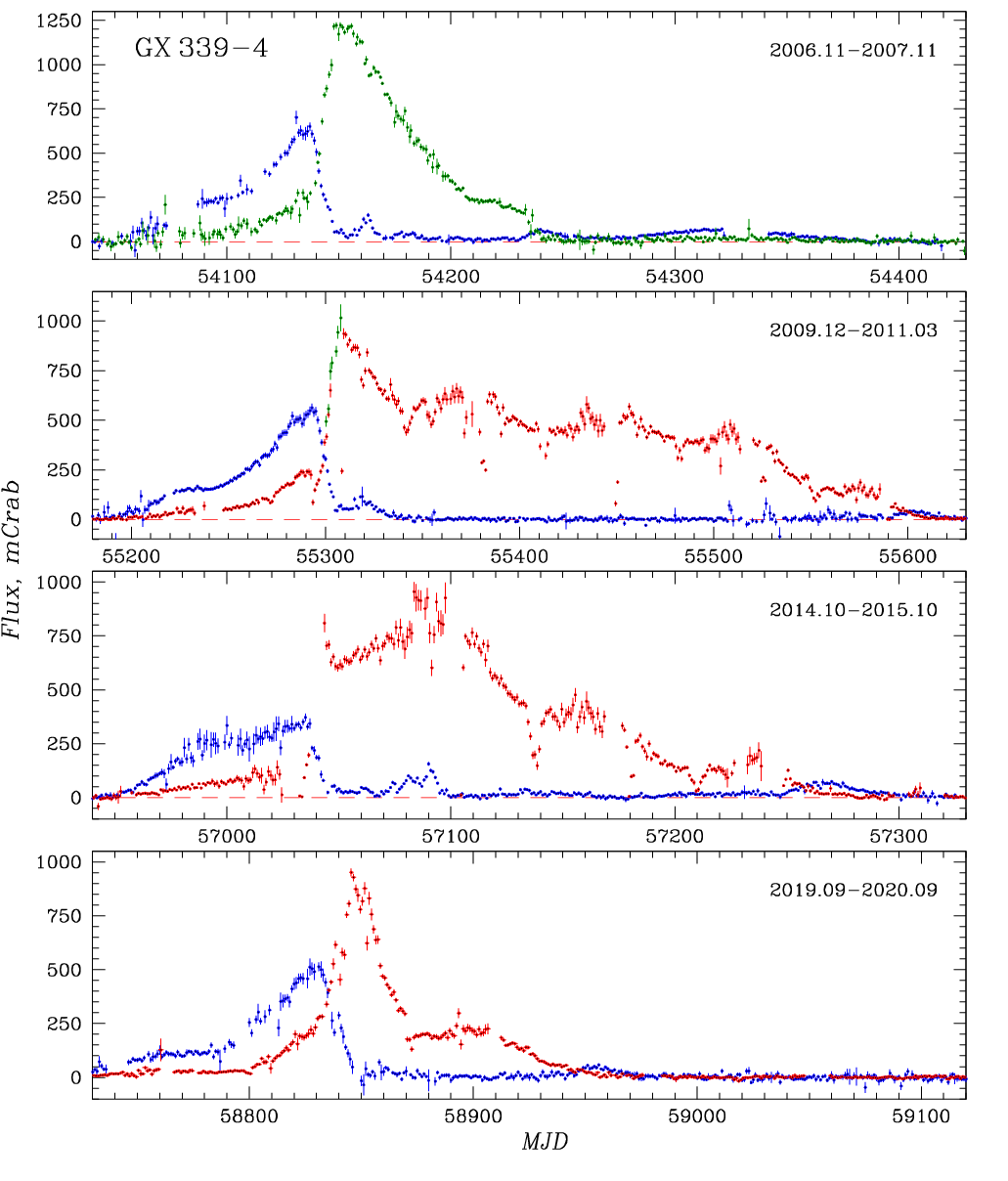}

\caption{\rm Light curves for four powerful outbursts of the
  X-ray transient GX 339-4 in which nonstationary accretion onto
  a black hole occurs (see Grebenev et al. 2020). The
  measurements were performed by SWIFT/BAT in the hard X-ray
  range 15--50 keV (blue dots) and by RXTE/ASM (green dots) and
  MAXI/GSC (red dots) in the soft X-ray range 1.3--3 or 2--4
  keV, respectively. At the initial outburst phase, during the
  hard state of GX 339-4, the hard and soft X-ray fluxes
  correlated, suggesting a common radiation
  mechanism.\label{fig:flares}}
\vspace{-0mm}
\end{figure*}

The distortions of the spectra in the soft X-ray range were
described by the fit from Morrison and McCammon (1981) for the
photoabsorption cross section in an interstellar medium with
normal solar abundances of elements and a hydrogen column
density toward the source $N_{\rm H}$ (we used the {\sc wabs\/}
code in the XSPEC\footnote{heasarc.gsfc.nasa.gov/docs/software/xspec/} NASA 
HEASARC); the distortions in the OIR and near-ultraviolet ranges
were described by the fit of the interstellar extinction in this
direction from Cardelli et al. (1989) and Draine (2003) (the
{\sc redden\/} code in the same package, the color correction
E(B-V), which is a parameter of the code, was set equal to
E(B-V) $\simeq 1.79\times 10^{-22}\ N_{\rm
  H}\ \mbox{cm}^{2}$). The fits to the spectra are indicated in
the figure by the solid lines; the undistorted power-law spectra
are indicated by the dashed lines.

No evidence for the presence of other radiation components in
the spectra presented in the figures is seen. However, given the
strong absorption observed in the range 10--300 eV, this
possibility cannot be ruled out completely. Note that serious
constraints on the contribution of the outer disk regions to the
soft X-ray spectrum follow even from the observations of the
light curves for X-ray nova outbursts. In particular, at the
initial phase of the outbursts of GX 339-4, when it is in the
hard state, its light curve in the soft X-ray range \mbox{1--4} keV
always completely traces the changes in the hard X-ray flux
(Fig.~\ref{fig:flares}, for more details, see Grebenev et
al. 2020). This directly points to a common origin of the soft
and hard X-ray radiations from the source at this outburst
phase.  Note that the hard spectral state lasts at this rising
outburst phase for a very long time; the total X-ray flux reaches
a much higher level than that at the time of the soft-to-hard
state transition at the decaying outburst phase. Obviously, the
transition between the states is regulated not only by the
instantaneous accretion rate, but also by the prehistory of its
change.

\section*{THE SPECTRUM OF THE OUTER ACCRETION DISK REGIONS}
\noindent
If the contribution of the cold outer accretion disk regions to
the total radiation spectrum of a source cannot be completely
measured due to the interstellar absorption in the ultraviolet
and soft X-ray ranges, then we can attempt to estimate it
theoretically. To a first approximation, the photon spectrum of
the disk region located between radii $R_1$ and $R_2$ can be
described as a superposition
\begin{equation}\label{eq:dbbfl}
F_{\nu}=\frac{2\pi\cos{i}}{d^2\,h\nu}\,\int_{R_1}^{R_2}{B_{\nu}(T)RdR}
\end{equation}
of the Planck spectra $B_{\nu}(T)$ corresponding to rings with
different surface temperatures $T(R)=(Q_{\mathrm
  vis}/\sigma)^{1/4}$ (the so-called ``multicolor'' accretion
disk, Shakura and Sunyaev 1973; Mitsuda et al. 1984). Here, $i$
is the disk inclination, $\sigma$
is the Stefan-Boltzmann constant, and $Q_{\mathrm vis}(R)$ is
the local energy flux emitted by the disk surface at a given
radius $R$ (in accordance with the dissipation of gravitational
energy in the disk, as defined by Eq.~(\ref{eq:dbbqv})). The
disk surface temperature in the limit $R\gg R_0$ depends on the
radius as $T(r) \simeq T_0\,r^{-3/4},$ where
$T_0=(\dot{M}c^2/16\pi R_0^2 \sigma)^{1/4}$. Recall that we
assume the black hole to be a Schwarzschild one.  Therefore,
$R_0=3\,R_{\rm g}=6GM/c^2$ and $r=R/R_0$. Equation
(\ref{eq:dbbfl}) takes the form
$$F_{\nu}\simeq \frac{8 \pi}{3} \left(\frac{R_0^2}{d^2}\right)
\frac{\cos{i}}{h\nu} \int_{T_2}^{T_1} 
{B_{\nu}(T)\left(\frac{T}{T_0}\right)^{\!-11/3}\frac{dT}{T_0}}=$$
\begin{equation}\label{eq:dbbflf}
=\frac{30}{\pi^5} \left(\frac{L_d}{d^2}\right)\frac{\cos{i}}{h\nu_0^2}
\left(\frac{\nu}{\nu_0}\right)^{-2/3} 
\int_{x_1}^{x_2}{\frac{x^{5/3}\,dx}{e^x-1}}.
\end{equation}
In the derivation we took into account the fact that the
Stefan-Boltzmann constant is expressed via the physical
constants as $\sigma=(2/15)\,\pi^5k^4/h^3c^2$. We introduced the
notations $\nu_0=kT_0/h$, $x_1=h\nu/kT_1$, and $x_2=h\nu/kT_2$,
where $T_1$ and $T_2$ are the surface temperatures at the inner
and outer boundaries of the cold disk region. As has already
been mentioned, $L_{d}$ is the total disk luminosity;
$L_{d}=\dot{M} c^2/12$ in the case of a Schwarzschild black
hole. Note the powerlaw dependence of the radiation flux with
the exponent (photon index) $\alpha=2/3$. The frequency range in
which this dependence holds is determined by the radii $R_1$ and
$R_2$. In Appendix~II we show that the above formula can be
simplified even more under some conditions.

\begin{sidewaysfigure*}
  \centering
  \includegraphics[width=0.9\textwidth]{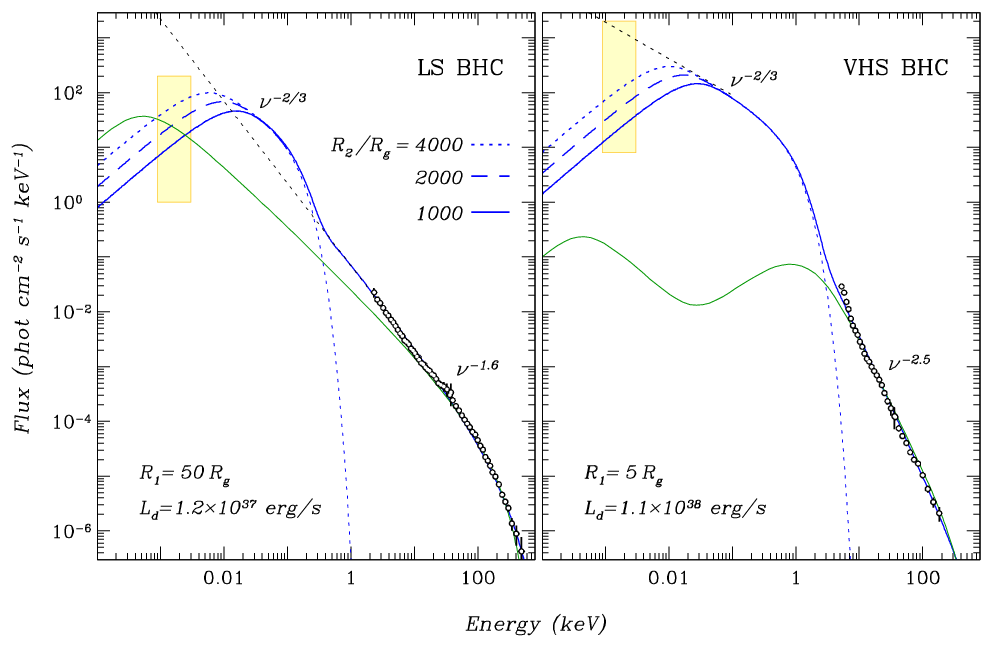}
  
\caption{\rm Broadband photon spectra for an accreting black
  hole in the hard (left) and soft(right) states according to
  the universally accepted piture of their formation (the
  truncated disk model, see the text). The dotted, dashed, and
  solid lines indicate the spectra under different assumptions
  about the radius of the outer boundary of a blackbody disk
  $R_2$; the radius of its inner boundary is $R_1=50\ R_{\rm g}$
  (left) and $5\ R_{\rm g}$ (right). The hot zone responsible
  for the hard X-ray radiation lies between the radii
  $R_0=3\ R_{\rm g}$ and $R_1$. Its radiation is indicated by
  the green solid line. For the soft state this radiation was
  computed by taking into account the multiple Compton
  scattering of photons from the outer disk region. In the
  optical and infrared ranges (light-yellow sector) the flux
  from the hot zone during the hard state of the source can
  exceed the flux from the vast outer region. The dotted
  straight line indicate the power-law extension of the X-ray
  spectrum on the left and the asymptotics of the law
  $\nu^{-2/3}$, according to Eq.~(\ref{eq:dbbflr}) in the
  Appendix, on the right.\label{fig:dbb-spec}}
\end{sidewaysfigure*}

Equation~(\ref{eq:dbbflf}) was used to numerically compute the
contribution of the outer accretion disk region to the broadband
(from optical to hard X-ray) spectrum of its radiation. The blue
solid lines in Fig.~\ref{fig:dbb-spec} represent the computed
photon spectra forming during two typical states of accreting
black holes (in accordance with their interpretation in the
truncated disk model; see, e.g., Grebenev et al. 1997a;
Zdziarski and Gierlinski 2004; Done et al. 2007; Gilfanov
2010). The source was assumed to be at the distance $d=10$ kpc.

  
The first model case (Fig.~\ref{fig:dbb-spec}, left panel) suggests
that the rate of accretion onto the black hole is
$\dot{M}=2.5\times 10^{-9}\ M_{\odot}\ \mbox{\rm yr}^{-1}$,
providing its total luminosity $L_d\simeq 1.18\times10^{37}\,
\mbox{erg s}^{-1}$. For a black hole with $M=10\ M_{\odot}$ it
corresponds to $\sim 0.9$\% of the critical Eddington luminosity
$L_{\rm c}= 4\pi m_p c\, GM/\sigma_{\rm T}\simeq 1.38\times
10^{38}\ (M/M_{\odot})\ \mbox{erg s}^{-1}$. At such a luminosity
the source is usually in its hard spectral state.  The outer
radius of the hot disk zone is taken to be $R_1=50\, R_{\rm
  g}$. According to Eq.~(\ref{eq:lumin}), the luminosity of this
zone accounts for 85\% of the total disk luminosity and is
$L_{\rm H}\simeq 1.0\times 10^{37}\ \mbox{erg s}^{-1}$.  Its
spectrum (green line) was obtained by assuming that the optical
depth of the central hot zone is $\tau_{\rm T}=1.5$, the
electron temperature is $kT_{\rm e}=60$ keV, and the density is
$N_{\rm e}=10^{18}\ \mbox{\rm cm}^{-3}$. Such plasma parameters
provide an approximate correspondence of the luminosity of this
zone to its expected luminosity ($L_{\rm H}$). Here, we did not
try to achieve close agreement, since the figure is intended to
show an approximate shape of the radiation spectrum for the
source in its hard state.  For comparison, we present the
measured GRANAT spectrum of the source Cyg~\mbox{X-1} in this state
(the same as that on the upper panel of Fig.~\ref{fig:cygx1}),
that was properly normalized.

Note that the radiation spectrum of the outer ($R\gg R_1$) disk
region (dotted blue line) manifests itself only at energies
$h\nu\ll 1$ keV, in a range that, as a rule, is already
inaccessible to observations. In contrast, the contribution of
this disk region to the optical and infrared radiation depends
critically on the outer radius $R_2$ of the entire
disk. Figure~\ref{fig:dbb-spec} presents the spectra for three
values of this radius, $R_2=4000,\ 2000,\ \mbox{\rm
  and}\ 1000\ R_{\rm g}$ (the dotted, dashed, and solid blue
lines, respectively). At all these values the photon spectrum of
the outer disk region in the optical and infrared ranges has a
Rayleigh-Jeans form $\sim\nu^{1}$; it has nothing in common with
the spectrum of a ``multicolor'' accretion disk $\sim
\nu^{-2/3}.$ Let us show that the chosen values of $R_2$ are
realistic.

As a rule, the orbital periods $P_{b}$ of accreting black
holes (and primarily X-ray novae) are several hours
(Cherepashchuk 2013). This ensures that the normal star fills its
Roche lobe and its mass can be efficiently transferred through
the inner libration point (L1). According to Paczynski (1971),
the size of the Roche lobe $R_L$ for a black hole of mass 
$M$ can be determined with a reasonable accuracy, provided that
the mass of its optical companion $M_{\rm O}$ lies in the range
$0.05\ M < M_{\rm O} < 3.3\ M,$ from the formula
\begin{equation}\label{eq:roche}
  R_{L}=A \left[0.38+0.2\, \lg\left(\frac{M}{M_{\rm
        O}}\right)\right]. 
\end{equation}
Here, $A$ is the separation between the binary components that
can be found from the Kepler law
\begin{equation}
A=\left[\left(\frac{P_{b}}{2\pi}\right)^2 G (M + M_{\rm O})\right]^{1/3}. 
\end{equation}
For $M = 10 M_{\odot}$ and a low-mass ($M_{\rm O}\sim
0.5-1.0\ M_{\odot}$) companion we obtain $A\simeq 5.3\times
10^3\ P_5^{2/3}\ R_{\rm g}$ and the radius of the Roche lobe is
$R_L\sim (3.1-3.4)\times 10^3\ P_5^{2/3}\ R_{\rm g}$. Here, the
orbital period is presented in the form
$P_b=5\ P_5\ \mbox{h}$. For a massive companion, for example,
$M_{\rm O}=20\ M_{\odot}$, we have $A\simeq 7.5\times
10^3\ P_5^{2/3}\ R_{\rm g}$ and the radius of the Roche lobe for
the black hole is $R_L\simeq 2.4\times 10^3\ P_5^{2/3}\ R_{\rm
  g}$. Obviously, the accretion disk radius should be slightly
smaller than $R_L$, so that the values of $R_2$ being used are
quite reasonable.
\begin{figure*}[!t]
 \centering 
 \includegraphics[width=0.69\textwidth]{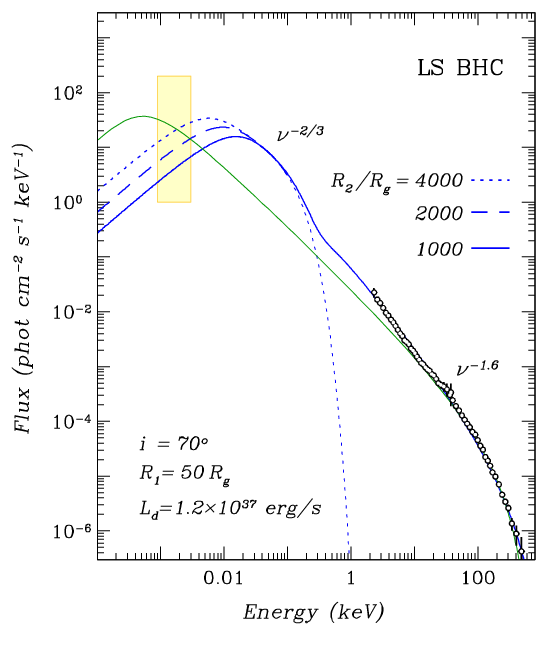}

\caption{\rm The same radiation spectrum of the accretion disk
  near the black hole in its hard state as that on the left
  panel of Fig.~\ref{fig:dbb-spec}, but at the binary
  inclination $i=70\deg$. In the optical and infrared ranges the
  radiation from the high-temperature central disk zone clearly
  dominates.\label{fig:dbb-spec70}}
\end{figure*}

It is important to note that the solid blue line in the figure
indicates the sum of the contributions from the hot and cold
disk zones only to $h\nu\sim 100$ eV; at lower energies it takes
into account only the radiation from the outer region. The
entire radiation spectrum of the inner hot disk zone is
indicated separately by the green line for the convenience of
comparison. The dotted black line shows that the natural
extension of the hard X-ray spectrum as a power law downward
along the energy axis by no means always reproduces the true
optical radiation spectrum of the hot disk region. Nevertheless,
the figure clearly demonstrates that in the optical range, let
alone the near-infrared one, (highlighted by the shaded yellow
rectangular sector) the radiation from the central hot disk zone
can exceed noticeably the contribution of its outer cold
blackbody regions.

The right panel in Fig.~\ref{fig:dbb-spec} presents the spectrum
computed by assuming the accretion rate to be $\dot{M}=2.2\times
10^{-8}\ M_{\odot}\ \mbox{\rm yr}^{-1}$, corresponding to a
total disk luminosity $L_d\simeq 1.06\times10^{38}\, \mbox{erg
  s}^{-1}$ or $\sim 8$\%
of the Eddington luminosity for a black hole with
$M=10\ M_{\odot}$. At such a luminosity the source is usually
already in its soft (two-component) state or during the
transition to it. Assuming that the hot disk region compresses
dramatically, we take the outer radius of this zone to be
$R_1=5\, R_{\rm g}$. According to Eq.~(\ref{eq:lumin}), the
luminosity of this zone should account for 13\% of the total
disk luminosity, i.e., $L_{\rm H}\simeq 1.37\times
10^{37}\ \mbox{erg s}^{-1}$. The selected radiation spectrum of
this zone (green line) was obtained at the same optical depth
$\tau_{\rm T}$ and plasma density $N_{\rm e}$ as those on the
left panel of Fig.~\ref{fig:dbb-spec}, but at a slightly higher
electron temperature, $kT_{\rm e}=70$ keV, and, most
importantly, at a noticeable number of external soft photons
entering the hot layer (in the corona model the temperature of
the underlying surface is $kT_{\rm s}=0.5$ keV and the fraction
of the photons entering the hot layer is $\xi=20$\%). In this
case, the required luminosity of this zone $L_{\rm H}$ given
above is achieved approximately. For comparison, we present the
spectrum of the X-ray Nova Musca 1991 (GRS~1124-68) measured
during its soft spectral state in February 1991 by the GRANAT
observatory (Grebenev et al. 1993) that was properly normalized.
We emphasize again that we by no means tried to accurately
select the parameters to fit this spectrum and provide it only to
illustrate that our view of its formation mechanism is correct.

The radiation spectrum of the outer disk region dominates in its
soft X-ray radiation, which, obviously, is related to a high
accretion rate $\dot{M}$, a small radius $R_1$, and,
accordingly, a high surface temperature $T_1$ of the outer disk
at this radius. The spectrum of the cold disk region is again
presented for three different assumptions about its outer
radius, $R_2=4000,\ 2000,\ \mbox{and}\ 1000\ R_{\rm g}$,
providing a change of its shape at energies $h\nu\la 10$--30 eV
from $\sim\nu^{-2/3}$ (typical for the ``multicolor'' disk
radiation) to $\sim\nu^1$ (Rayleigh-Jeans radiation with a flux
falling toward low energies). In this case, however, there is no
reason to expect a noticeable contribution of the radiation from
the central hot zone to the near-infrared and optical
ranges. This is not only because the luminosity of the outer
disk regions during the soft state of the source ($\simeq 9.2\times
10^{37}\ \mbox{erg s}^{-1}$) is almost two orders of
magnitude higher than that during its hard state, but also because the radiation
shape of the spectrum forming in the hot central zone due to the
Comptonization of soft photons from the outer disk region is
more complex. Given the contribution of external photons, the
X-ray flux and luminosity of the hot zone exceed noticeably
those of the intrinsic radiation from a plasma with the same
temperature and density, compensating for the noticeable
decrease in the size and area of the zone. In contrast, the
optical and infrared radiation of the zone is associated with
the intrinsic radiation of a plasma with a given temperature; it
is completely independent of the presence of external
photons. Because of the decrease in the area of the hot zone,
its flux is much weaker than the OIR flux from such a zone
during the low state of the source.

It should be noted that the radiation flux from the outer disk
regions depends strongly on the binary inclination (see
Eq.~(\ref{eq:dbbflf})), whereas a semitransparent hot central
zone radiates almost isotropically. The case with the binary
inclination $i=0\deg$ is considered in
Fig.~\ref{fig:dbb-spec}. Figure~\ref{fig:dbb-spec70} shows the
radiation spectrum of the same accretion disk (and near the same
black hole) as that on the left panel of
Fig.~\ref{fig:dbb-spec}, but when it it observed from the edge,
with the inclination $i=70\deg$. It can be seen than the optical
radiation from the binary system, let alone the near-infrared
one, in this case is completely determined by the radiation from
the hot plasma in the central accretion disk zone rather than by
the radiation from its outer cold extended regions, as is
commonly thought. The heating of the surface of the outer disk
by X-rays from the central zone (Lyuty and Sunyaev 1976) cannot
affect significantly this conclusion (see Grebenev et al. 2016).

\section*{CONCLUSIONS}
\noindent
Our computations that take into account the bremsstrahlung
processes in a semitransparent plasma and the subsequent
Comptonization of bremsstrahlung photons show the following.

(1)~The intrinsic radiation from the high-temperature plasma in
the central zone of the accretion disk surrounding a black hole
in a binary system can completely explain the characteristic
canonical X-ray spectra (with a photon index $\alpha\sim 1.6$
and an exponential cutoff at high energies $h\nu\ga 50$ keV)
observed from X-ray novae and sources like Cyg~X-1 during their
hard (low) state. No additional external soft photons that could
be involved in Comptonization are required; moreover, their
presence generally leads to a noticeable softening of the
spectrum to a photon index $\alpha\sim 2.2-2.5.$ All of the
previous computations of the Comptonization of seed photons in a
plasma can be incorrect, since they completely ignore the plasma
bremsstrahlung.

(2)~Even if the Comptonization of seed photons in the model case
of a low-density plasma leads to a canonical X-ray spectrum, as
the plasma density increases to realistic values, $N_{\rm
  e}\simeq 10^{16}-10^{19}\ \mbox{\rm cm}^{-3}$, the slope,
shape, normalization, and other characteristics of the spectrum
change radically. As the density increases further to $N_{\rm
  e}\ga 10^{19}\ \mbox{\rm cm}^{-3}$ or higher, the
Comptonization of external photons ceases to affect noticeably
the forming radiation spectrum, which, in this case, is simply
the intrinsic radiation spectrum of the hot plasma (its
bremsstrahlung subjected to Comptonization). Of course, the
aforesaid pertains to the reasonable assumption about the number
of soft photons that can enter the hot zone, for example, from
the outer cold accretion disk regions.

(3)~When photons from the outer cold accretion disk regions
enter the hot region, its X-ray spectrum usually softens and
becomes similar to the spectrum observed from accreting black
holes during their soft or two-component states (with a photon
index $\alpha\sim 2.2-2.5$). An increase in the flux of external
photons is most likely caused by the fact that the inner edge of
the cold disk region approaches the black hole maximally
closely.

(4)~The intrinsic hard X-ray spectra of the hot plasma in the
central accretion disk region extend in an invariable form (with
the same power law) to the optical and infrared ranges,
transforming at $h\nu\la 1$ eV into the Rayleigh-Jeans spectrum
corresponding to the plasma temperature that falls off toward
low energies. The external photons entering the hot disk zone do
not affect noticeably the near-infrared, optical, and
long-wavelength ultraviolet emission from this zone, although,
of course, the compression of this zone itself as the source
passes to its soft state is accompanied by a decrease in the
optical and infrared flux.

(5)~In many cases (at an orbital period of the order of several
hours and a noticeable inclination of the binary system), the
near-infrared, optical, and even ultraviolet radiation observed
from accreting black holes during their hard state is precisely
this radiation from the hight-temperature plasma in the central
disk region rather than the ``multicolor'' blackbody radiation
from its extended cold regions (as is commonly assumed). During
the soft state of such sources the ``multicolor'' radiation from
the outer disk regions almost always dominates in these ranges.

(6)~Obviously, the transition between the spectral states of
accreting black holes is determined by the change in the
accretion rate and the corresponding approach or recession of
the boundary between the central hot and outer cold accretion
disk regions (the truncated disk model). The surface temperature
of the outer disk near this boundary plays an important role in
whether a change in the shape and type of the X-ray spectrum
will be observed in this case. As was revealed during our
computations, external photons of sufficiently low energies,
even when entering the hot disk zone, do not affect noticeably
the forming spectrum of its radiation. Otherwise, we could have
observed these sources in their canonical hard state much more
rarely, since it is impossible to completely rule out the
penetration of external photons into the high-temperature
region.

(7)~The broadband radiation spectra presented in the paper were
obtained without any controversial, let alone doubtful,
assumptions. If a sufficiently dense high-temperature plasma
exists near the black hole (in its accretion disk), and this is
beyond doubt, then it should radiate as described in the paper.
In real accretion disks the temperature, the density, and the
optical depth, of course, should be distributed along the
radius. However, the cases of such distributions and, in
particular, the distributions derived within the two-temperature
accretion disk model by Shapiro et al. (1976) considered in this
paper convincingly show that even those our computations that
were performed for a homogeneous isothermal plasma layer lead to
qualitatively correct and quantitatively reasonable results. In
the presence of photons with sufficiently low energies entering
the hot plasma from outside, an additional radiation component
that changes the general form of the spectrum can appear in its
spectrum because of their subsequent Comptonization, but this is
precisely an additional component.  At a noticeable (realistic)
plasma density, considering the Comptonization of seed photons
alone in it does not lead to physically correct results. The
characteristics of the forming radiation spectrum depend not
only on the temperature and optical depth of the plasma, but
also on its electron density.

The only question with regard to which the results obtained
require a refinement pertains to the nonrelativistic
consideration of the radiative transfer in this paper. In
future, we are going to confirm its results by including the
relativistic corrections to the Kompaneets equation in our
computations or by using other methods. We are also going to
calculate the degree of polarization of the radiation produced
in the accretion disk in much the same way as was done by
Sunyaev and Titarchuk (1985) and Titarchuk et al. (2025). The
discovery of a strong, energy-dependent, polarization of the
X-ray emission from a number of accreting black holes by the
IXPE observatory (see, e.g., Poutanen et al. 2023; Ravi et
al. 2026) has shown a tremendous importance of such
computations. In the case of Comptonization of intrinsic plasma
bremsstrahlung photons, the degree of polarization may turn out
to be different from that obtained when considering the Compton
scattering of seed photons in an accretion disk.

Our results in many respects overturn the long existing views of
the production and properties of the hard X-ray, optical, and
infrared emission from accreting black holes, primarily during
their hard spectral state. In this paper we presented the
broadband observations of some sources, black hole candidates,
in (convincing) comparison with our theoretical calculations. In
future, we are going to expand significantly this part of the
work, understanding that an experiment is the ultimate truth
criterion for any theory.

\vspace{6mm}

\appendix
\section{I. THE ELECTRON DENSITY IN THE DISK}\label{app:A}
\noindent
In our computation we assume a very simple geometry of the
high-temperature plasma cloud --- an infinite layer of the same
thickness with uniform density and temperature
distributions. The results of our computations depend not only
on the obvious (measured in X-ray observations from the shape of
the radiation spectrum) plasma parameters, the transverse
optical depth for Thomson scattering $\tau_{\rm T}$ and the
electron temperature $kT_{\rm e}$, but also on the electron
density $N_{\rm e}$.

It is easy to find the density given the optical depth $\tau_{\rm
  T}$ of the disk:
\begin{equation}\label{eq:Netau}
  N_{\rm{e}} \simeq \frac{\tau_{\rm T}}{2\sigma_{\rm
T}H}\simeq 7.52\times10^{18}\ H_{*}^{-1}\ \tau_{\rm T}\ \mbox{cm}^{-3}.
\end{equation}
Here, $H$ is the disk half-thickness, and $H_*$ is the
half-thickness expressed in km. It can be determined from
the hydrostatic equilibrium equation in the vertical di-
rection that in a viscous disk takes the form (Shakura
and Sunyaev 1973)
\begin{equation}\label{eq:Heq}
  H=\sqrt{2}\ \frac{v_{\rm s}}{v_{\phi}}\ R, \ \mbox{\rm where}
  \ v_{\rm s} 
  \ \mbox{\rm and}\ v_{\phi}=\left(\frac{GM}{R}\right)^{1/2}
\end{equation}
are the speed of sound and the Keplerian gas velocity,
respectively.

\subsection{The Model with the Dominant Gas Pressure}
\noindent
If the gas pressure dominates in the disk region of
interest to us, then $$v_{\rm s}^2=\frac{k(T_{\rm e}+T_{\rm
    p})}{m_{\rm p}}.$$
Accordingly, the disk half-thickness is
\begin{equation}\label{eq:Hgas}
H=\left(\frac{2 k[T_{\rm e}+T_{\rm p}] R^3}{G M\, m_{\rm
    p}}\right)^{1/2}\simeq 3.2\ m_{*} r^{3/2}
kT_{*}^{1/2}\ \mbox{\rm km},
\end{equation}
where, as previously, $r=R/R_0$ ($R_0=3\,R_{\rm g}$),
$m_{*}=M/10\,M_{\odot},$ and $kT_{*}=kT_{\rm e}/50\ \mbox{\rm
  keV}$. We assume here that $T_{\rm e}=T_{\rm p}$. Substituting $H$
into Eq.~(\ref{eq:Netau}), we find the electron density
\begin{equation}\label{eq:Negas}
N_{\rm e}\simeq 2.3\times10^{18}\ \tau_{\rm T}\ r^{-3/2}
kT_{*}^{-1/2} m_{*}^{-1}\ \mbox{\rm cm}^{-3}.
\end{equation}

\subsection{The Model with the Dominant Radiation Pressure}
\noindent
If the radiation pressure dominates in the hot disk
region and the Compton scattering dominates in the
opacity, then
\begin{equation}
  v_{\rm s}^2=\frac{U_{\rm r}}{3\, N_{\rm p}\,m_{\rm p}}=
  \frac{Q_{\rm vis}\ \sigma_{\rm T}\,H}{2\,c\,m_{\rm p}}.
\end{equation}
Here, $U_{\rm r}$ is the radiation density in the central disk
plane, and $Q_{\mathrm vis}$ is the radiation flux from unit disk
surface area associated with viscous energy release
(see Eq.~(\ref{eq:dbbqv})).
Substituting this value of $v_{\rm s}$ into
Eq.~(\ref{eq:Heq}), we obtain for the disk half-thickness
\begin{equation}
  H=\frac{3}{8\pi}\frac{\dot{M}\sigma_{\rm T}\phi(r)}{c\,m_{\rm
      p}}=2.1\ \phi(r)\, \dot{m}_{*}\ \mbox{\rm km}.
\end{equation}
Here, $\dot{m}_{*}=\dot{M}/\dot{M}_{*},$ where
$\dot{M}_{*}=2.1\times 10^{-9}\ M_{\odot}\ \mbox{yr}^{-1}$ is
the accretion rate corresponding to a luminosity
$L_*=\dot{M}_{*}\,c^2/12\simeq 1\times 10^{37}\ \mbox{erg
  s}^{-1}$, $\phi(r)=1-r^{-1/2}$. Substituting $H$ into
Eq.~(\ref{eq:Netau}), we find the electron density
\begin{equation}\label{eq:Nerad}
N_{\rm e}\simeq 3.6\times10^{18}\ \tau_{\rm T}\
\phi(r)^{-1}\ \dot{m}^{-1}_{*}\ \mbox{\rm cm}^{-3}. 
\end{equation}
The radiation pressure dominates in the plasma
at near- or super-Eddington accretion rates, leading
to an intense outflow of material (a ``radiation wind''
from the disk). As a rule (see, e.g., Zhang et al. 2025),
the temperature of such a plasma is $T_{\rm e}\la 10^8$ K, and
the hard radiation being investigated in the paper is
not produced. Equation~(\ref{eq:Nerad}) shows that the electron
density in the hot radiation-dominated disk region
can be, nevertheless, high enough for the effects considered to
influence the formation of the X-ray (diluted blackbody)
spectra.

\subsection{The Two-Temperature Model}
\noindent
This solution was proposed by Eardley et al. (1975) and Shapiro
et al. (1976) to describe the extremely hard X-ray spectra
observed from the source Cyg~X-1 and other black hole candidates
during their hard (low) state. It suggests that the accretion
disk near the black hole is hot, optically thin, and is
maintained in the vertical direction by the pressure of ions
(protons) with a temperature much higher (by a factor of 3--300)
than that of electrons.  Below we provide the plasma parameters
in the two-temperature model. They suggest that the electron
density, the temperature, and the Thomson optical depth remain
comparable to their estimates obtained above for the
single-temperature model.

For the electron temperature we can write
\begin{equation}\label{eq:Tion}
  \left(\frac{T_{\rm p}+T_{\rm e}}{T_{\rm e}}\right)^{5/4}\!(rkT_*)^{3/4}
  =5.0\,\left(\frac{\dot{m}_{*}\phi}{m_{*}}\right)^{1/2}\!
  \left(\frac{A_{\rm ff}}{\alpha}\right).
\end{equation}
Here, $\alpha$ is the parameter that characterizes the viscous
stresses in the disk relative to the pressure and $A_{\rm
  ff}\geq 1$ is the factor that takes into account the
enhancement of the plasma cooling rate compared to the
bremsstrahlung. The real cooling rate is determined by
the Comptonization. Substituting $T_{\rm p}+T_{\rm e}$ from this
expression into Eq.~(\ref{eq:Hgas}), for the disk half-thickness
we find
\begin{equation}
H=4.3\ m_*^{4/5} (A_{\rm ff}/\alpha)^{2/5} \dot{m}_{*}^{1/5}
\phi^{1/5} r^{6/5}\ 
\mbox{km}.
\end{equation}
To eliminate $A_{\rm ff}$ from this expression and to determine
the electron temperature $T_{\rm e}$, we need to take
into account the Coulomb energy exchange between protons and
electrons. This was done by Shapiro et al. (1976), who
consistently showed that the protons have no time to heat the
electrons that efficiently cool down through 
Comptonization. They found the exact dependence of the disk
half-thickness $H$ and
the electron density  $N_{\rm e}$ on the radius $r=R/R_0$:
\begin{equation}\label{eq:2tmodel-ne}
H=10.9\ (m_*/\alpha)^{7/12} \dot{m}_{*}^{5/12} \phi^{5/12} r^{7/8}\
\mbox{km},
\end{equation}
$$
N_{\rm e}=1.6\times10^{18}\ (m_{*}/\alpha)^{-3/4}
\dot{m}^{-1/4}_* \phi^{-1/4}\ r^{-9/8}\ \mbox{cm}^{-3}.\\ \nonumber
$$
The corresponding transverse optical depth of the
disk for Thomson scattering is
\begin{equation}\label{eq:2tmodel-tau}  
\tau_{\rm T}=2.2\ (m_*/\alpha)^{-1/6} \dot{m}_{*}^{1/6}
\phi^{1/6}\ r^{-1/4}.
\end{equation}
Finally, the dependence of the electron, $kT_{\rm e}$, and ion, 
$kT_{\rm p}$, temperatures on the radius $r$ is
\begin{equation}\label{eq:2tmodel-kte}
kT_{\rm e}=110\ (m_*/\alpha)^{1/6} \dot{m}_{*}^{-1/6} \phi^{-1/6}
r^{1/4}\ \mbox{keV},
\end{equation}
\begin{equation}\label{eq:2tmodel-ktp}
kT_{\rm p}=2140\ m_*^{-5/6} \dot{m}_{*}^{5/6} \alpha^{-7/6}
\phi^{5/6} r^{-5/4}\ \mbox{keV}.
\end{equation}
The photon radiation spectra computed in accordance with
these formulas are presented in Fig.~\ref{fig:2tmodel}.

\section{II. CHARACTERISTIC ENERGIES AND ESTIMATES}\label{app:B}
\noindent
The plasma layer ceases to be optically thick for bremsstrahlung
absorption at $\alpha_{\rm ff}\la 1$. Using the expression for
the opacity given when describing Eq.~(\ref{eq:komp}), we find
the corresponding energy
\begin{equation}
h\nu_1=0.28\ N_{18}^{1/2}\ kT_*^{-3/4}\ \mbox{eV}.
\end{equation}
Here, $N_{18}=N_{\rm e}/10^{18}\ \ \mbox{\rm cm}^{-3}$.
The value of $h\nu_1$ obtained takes into account the Gaunt
factor that at these energies is $g\simeq 7.1.$

In reality, given the scattering, the cloud also remains opaque
at higher energies up to $h\nu_2$ determined by the condition
$\tau_{\rm eff}=(\alpha_{\rm ff}/\alpha_{\rm
  T})^{1/2}\,\tau_{\rm T}\simeq 1$. Substituting the opacities,
we find
\begin{equation} 
h\nu_2=0.43\ \tau_{\rm T}\ N_{18}^{1/2}\ kT_*^{-3/4} \ \mbox{eV} 
\end{equation}
Here, we took into account the Gaunt factor $g\simeq 6.47$
corresponding to the energy $h\nu_2$ at $\tau_{\rm T}=2.$

The dimensionless energy $x_1=h\nu/kT_1,$ where $kT_1$ is the
surface temperature of the cold disk at its inner radius $R_1$, for
optical light with $h\nu\simeq 2$~eV used in
Eq.~(\ref{eq:dbbflf}) is very low, $x_1= 1.44\times 10^{-3}\,
r_1^{3/4} m_*^{1/4} (L_{d}/L_{\mathrm c})^{-1/4}\ll 1$. If, in
addition, the outer radius of the disk $R_2$ exceeds
considerably $R_2^*=1.8\times 10^4\ m_{*}^{-1/3}
(L_{d}/L_{\mathrm c})^{1/3}\, R_{\rm g},$ so that $x_2\gg 1$,
then Eq.~(\ref{eq:dbbflf}) transforms to
\begin{equation}\label{eq:dbbflr}
  F_{\nu}\simeq 7.0\times 10^{3}\cos{i}\ \left(\frac{h\nu}{2\ 
    \mbox{\rm eV}}\right)^{-2/3}\times
\end{equation}
$$\times
\left(\frac{M}{10\ M_{\odot}}\right)^{8/3}\!\left(\frac{L_{d}}{L_{\rm c}}
\right)^{2/3}\!\left(\frac{d}{10\ \mbox{\rm
    kpc}}\right)^{-2}\!\mbox{\rm s}^{-1}\,\mbox{\rm cm}^{-2}\,\mbox{\rm keV}^{-1}.  
$$ Here, as previously, $L_{d}$ is the total luminosity of the
accretion disk and $L_{\rm c}$ is the critical Eddington
luminosity. Since the disk surface temperature drops slowly with
radius, $T_s(r)\simeq T_0\,r^{-3/4}$, the regions with large
radii make a major contribution to its optical luminosity
$d\,F_{\nu}\sim B_{\nu}(T_s) R\,d\,R \sim kT_s 
\nu^2\ R\,d\,R\sim R^{1/4}.$ On the right panel of
Fig.~\ref{fig:dbb-spec} solution~(\ref{eq:dbbflr}) is indicated
by the black dotted line. Because of the small presumed disk
radius $R_2$, it passes above the real spectrum of its outer
region in the optical and infrared ranges, but it is useful,
since it shows the asymptotics of the law $\nu^{-2/3}$.

It has been noted above that the Roche lobe in many X-ray novae
and other binary systems containing accreting black holes should
be sufficiently compact, i.e., have a radius $R_L\sim (2-4)
\times 10^3\ P_5^{2/3}\ R_{\rm g}$, where $P_5=P_b/5$ h is the
binary period. The outer radius of the accretion disk should be
even smaller. Therefore, the condition $R_2\gg R_2^*$ can be
fulfilled only for binary systems in a state with a low
accretion rate corresponding to a disk luminosity $L_{d}\la
0.01\ L_{\rm c}.$

\section*{ACKNOWLEDGMENTS}
\noindent
I am grateful to R.A. Sunyaev for the useful discussions of
various aspects of my work and the numerous advice on the
Comptonization theory as well as to S.Yu. Sazonov and
E.M. Churazov for their valuable remarks.

\section*{FUNDING}
\noindent
This study was supported by the ``BASIS'' Foundation for the
Advancement of Theoretical Physics and Mathematics within the
Program ``Leading Scientist (Theoretical Physics)'' (grant
no. 22-1-1-57-1).

\section*{CONFLICT OF INTEREST}
\noindent
The author of this work declares that he has no conflicts of
interest. 

\vspace{6mm}

\vspace{5mm}

\begin{flushright}
{\sl Translated by V. Astakhov}
\end{flushright}
\end{document}